\def\HPpage{
\documentclass[a4paper,12pt]{article}
\addtolength{\topmargin}{-2.5 true cm}
\addtolength{\textheight}{3.5 true cm}
\addtolength{\textwidth}{4 true cm} \addtolength{\hoffset}{-1
true cm} \setlength{\oddsidemargin}{0cm}
\setlength{\evensidemargin}{0cm} }

\def\Izvpage{
\documentclass[a4paper,12pt]{article}
\addtolength{\topmargin}{-1.5 true cm}
\addtolength{\textheight}{1.5 true cm}
\addtolength{\textwidth}{1 true cm} }

\Izvpage

\usepackage{amsmath, amsthm, amsfonts, amssymb}
\usepackage{graphicx}

\theoremstyle{plain}
\newtheorem{theorem}{Theorem}[section]

\newtheorem{lemma}{Lemma}[section]

\newtheorem{corollary}{Corollary}

\theoremstyle{remark}
\newtheorem{remark}{Remark}[section]

\numberwithin{equation}{section}

\newcommand{\sect}[1] 
{\addtocounter{section}{1} 

\medskip

\begin{center}
{\textbf{\large\arabic{section}. #1}}
\end{center}
\setcounter{equation}{0}\setcounter{theorem}{0}\setcounter{lemma}{0}
\setcounter{remark}{0}\setcounter{corollary}{0} \smallskip }

\newcommand{\sectn}[1] 
{ 
\medskip
\begin{center} {\textbf{\large #1}}
\end{center}
\setcounter{equation}{0}\setcounter{theorem}{0}\setcounter{lemma}{0}
\setcounter{remark}{0}\setcounter{corollary}{0} \smallskip }

\def\th{\theta}
\def\Om{\Omega}

\def\e{\varepsilon}
\def\g{\gamma}
\def\G{\Gamma}
\def\l{\lambda}
\def\p{\partial}
\def\D{\Delta}

\def\a{\alpha}
\def\b{\beta}
\def\t{\widetilde}
\def\si{\sigma}

\def\d{\delta}
\def\L{\Lambda}

\def\vs{\varsigma}

\def\h{\widehat}

\newcommand{\PF}[1]
{\noindent\textbf{#1}}

\newcounter{cons} 

\begin{document}

\allowdisplaybreaks

\setcounter{cons}{-1}


\begin{center}

\textbf{\large ASYMPTOTICS AND ESTIMATES FOR EIGENELEMENTS OF
LAPLACIAN WITH FREQUENT NONPERIODIC INTERCHANGE OF BOUNDARY
CONDITIONS}

\bigskip
\bigskip

{\large Denis I. Borisov}\footnotetext[1]{The work was partially
supported by RFBR (Nos. 02-01-00693, 00-15-96038) and Program
''Universities of Russia'' of Ministry of Education of Russia
(UR.04.01.010).}

\end{center}

\begin{quote}
{\small {\em  Bashkir State Pedagogical University, October
Revolution St.,~3a, 450000, Ufa, Russia. E-mail:}
\texttt{BorisovDI@ic.bashedu.ru, BorisovDI@bspu.ru}}
\end{quote}


\begin{abstract}
We consider singular perturbed eigenvalue problem for Laplace
operator in a two-dimensional domain. In the boundary we select
a set depending on a character small parameter and consisting of
a great number of small disjoint parts. On this set the
Dirichlet boundary condition is  imposed while on the rest part
of the boundary we impose the Neumann condition. For the case of
homogenized Neumann or Robin boundary value problem we obtain
highly weak restrictions for distribution and lengths of
boundary Dirichlet parts of  the boundary under those we manage
to get the leading terms of asymptotics expansions for perturbed
eigenelements. We provide explicit formulae for these terms.
Under weaker assumptions we estimate the degrees of convergence
for perturbed eigenvalues.
\end{abstract}

\sectn{Introduction}

The object of this work is to study a two-dimensional boundary
value problem with frequent nonperiodic interchange of type of
boundary conditions. First we describe the formulation of such
problems in general outline. The elliptic equation is considered
in a domain with a boundary smooth enough. In the boundary the
subset consisting of a great number of disjoint parts of small
measure is selected. On this subset the boundary condition of
one type (ex. Dirichlet condition) is imposed while on the rest
part of the boundary the condition of another type (ex. Neumann
condition) is set. The question is: What is  the behaviour of
the solution of  such problem when a number of parts of selected
boundary's subset infinitely decreases while the measure of each
part and distance between neighbouring ones tends to zero. It is
also possible to formulate a problem, where such type of
boundary condition described is imposed not on a whole boundary
but only on its part while on the remaining part one of classic
boundary condition is imposed.

\begin{figure}[htb]
\begin{center}
\noindent
\includegraphics[width=9.85 true cm, height=6.35 true
cm]{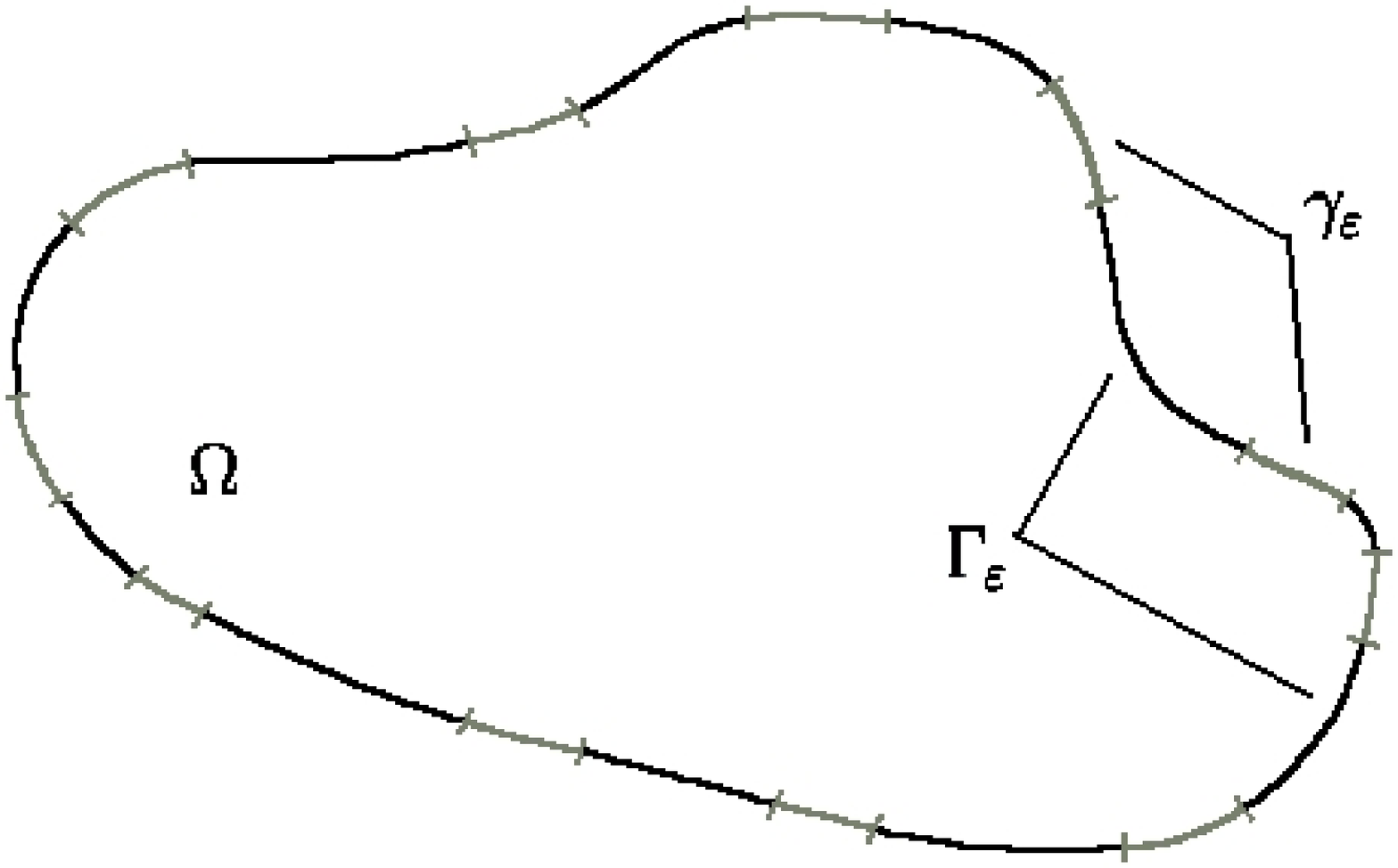}
\end{center}
\end{figure}

The homogenization of problems with frequent interchange of
boundary condition were widely investigated (see, for instance,
\cite{Dl}--\cite{Dv}). The main object of these works was to
determine the limiting (homogenized) problems under minimal set
of constraints for the structure of interchange of boundary
conditions, i.e., for the behaviour of sets with different
boundary conditions. Damlamian and Li Ta-Tsien in \cite{Dl}
considered Laplace equation in a bounded domain with frequent
interchange of boundary conditions. They studied alternation of
Dirichlet and Neumann condition and also the case when the
former was replaced by integral boundary condition. The
homogenized problems were obtained under severe constraints for
the structure of alternation. In papers \cite{ChNov}-\cite{BlCh}
for the problems with the alternation of Dirichlet and Neumann
or Robin conditions the homogenized problems were obtained and
simple conditions determining the dependence of homogenized
problem's type on the structure of alternation were adduced. The
case when the Dirichlet part of boundary had the periodic
structure was investigated in \cite{LP2}--\cite{Ch}. The
convergence in nonperiodic case was studied in \cite{ChNov},
\cite{Fr}, \cite{BlCh}. Barenbaltt, Bell and Crutchfiled
\cite{BBC} and D\'avila  \cite{Dv} considered nonlinear elliptic
equations with frequent interchange of type of boundary
condition. In \cite{BBC} the problem was solved numerically; in
\cite{Dv} the homogenization was studied. The results obtained
in investigating of problems with frequent interchange of
boundary condition (both periodic and nonperiodic) can be
briefly formulated as follows. Under general assumptions the
elliptic boundary value problems with frequent interchange of
boundary condition converge to classic problems. The type of
boundary condition in homogenized problem depends on
relationship between measures of parts of boundary with
different type of boundary condition in the perturbed problem.

The homogenization of boundary value problem close to problem
with interchange of boundary condition was studied in monograph
\cite{Mh}. Here they consider elliptic problems in whole space.
Boundary condition (Dirichlet or Neumann one) was imposed on a
boundary of a set consisting of a great number of  small
disjoint domains located closely each to other. Also it was
considered the case when small domains were replaced by small
curves on those Neumann boundary condition was imposed.
Asymptotics expansions for solutions of problems having such
geometry of boundary condition were constructed by Gadyl'shin in
the paper \cite{Suz}.

Besides the determining of homogenized problems for ones with
frequent interchange also it is important and actual the
question about estimates of degrees of convergence. For periodic
interchange of boundary conditions such estimates were obtained
by Chechkin and Gadyl'shin in \cite{Ch}, \cite{GCh}. Nonperiodic
interchange was studied by Oleinik, Chechkin and Doronina; they
considered the interchange of Dirichlet condition with Robin
condition (or Neumann one as a particular case). The case of
homogenized Dirichlet problem was treated  in \cite{ChOl}, the
case of homogenized Robin problem (or Neumann one as a
particular case) was studied in  \cite{Dr}.

In last years the papers appeared where the asymptotics of
solutions of problems with periodic structure of interchange
were constructed. First of all we stress that this periodicity
was essentially employed. Two-dimensional case is represented by
papers \cite{AA}-\cite{ZhVM}. In these works they considered
interchange of Dirichlet and Neumann conditions. For the circle
under some additional assumptions in  \cite{AA} and \cite{DU}
the complete power asymptotics for eigenelements of Laplace
operator in the case of homogenized Dirichlet or Neumann problem
were obtained. In the paper \cite{MZ} the results of \cite{AA}
were generalized and having assumed only periodicity of
interchange Borisov obtained complete two-parametrical
asymptotics of eigenvalues of Laplace operator converging to
simple limiting eigenvalues. The asymptotics expansions for
associated eigenfunctions were got, too. In papers \cite{TMF}
and \cite{ZhVM} for an arbitrary domain with periodic structure
of interchange they constructed the leading terms of asymptotics
expansions for perturbed eigenelements, corresponding
eigenvalues were assumed to converge to simple limiting
eigenvalues of Neumann or Robin problem.

In papers \cite{F1}-\cite{F3} the authors studied problems for
parabolic equations with frequent interchange of Dirichlet and
Robin condition assuming that measures of parts of the boundary
with different conditions have same smallness order. In
homogenization it led to Dirichlet boundary condition. In
\cite{F1}, \cite{F2} for periodic interchange of boundary
conditions and in \cite{F3} for almost periodic one they
estimated degrees of convergences and constructed first terms of
asymptotic expansions for solutions of the problems studied.

In the present paper we consider eigenvalue problem for Laplace
in an arbitrary two-dimensional domain with frequent and,
generally speaking, nonperiodic interchange of boundary
conditions. We study the interchange of Dirichlet and Neumann
boundary condition. In the problem we extract two character
small parameters governing lengths of Dirichlet and Neumann
parts of boundary. We give highly weak constraints for the set
with Dirichlet condition under those it is possible to construct
leading terms of asymptotics expansions for eigenelements
converging to eigenelements of homogenized Neumann or Robin
problem. These expansions are  simultaneously asymptotical with
respect to both small parameters; for leading terms the explicit
formulae are obtained. For the case of homogenized Neumann
problem we carry out additional studying and show that leading
terms of asymptotics can be obtained under weaker constraints
for the structure of interchange. These asymptotics include
leading terms of asymptotics from \cite{DU}, \cite{TMF},
\cite{ZhVM} as a particular case.

Loosening the constraints for the structure of Dirichlet part,
we obtain double-sided estimates for difference between
perturbed and limiting eigenvalues. The cases of homogenized
Dirichlet, Neumann and Robin problems are considered. These
differences are estimated by infinitesimal having the order of
smallness same with the order of smallness of first terms of
asymptotics for eigenvalues of perturbed problems obtained, of
course, under more severe constraints. Our constraints imposed
to the interchange are more severe  than ones in  \cite{ChOl},
\cite{Dr}. At the same time, the estimates from these works are
rougher than ones proved in this paper.

The results of this paper were announced in \cite{DAN}.

In conclusion of this section we mention that questions on
homogenization and estimates of degree of convergences for
three-dimensional problems with frequent interchange of boundary
condition were studied in \cite{Dl}, \cite{LP2}-\cite{L-P},
\cite{BlCh1}-\cite{Dv}, \cite{ChOl}-\cite{Dr}; asymptotics for
eigenvalues of Laplace operators in cylinder with periodic
frequent interchange of boundary conditions on narrow strips
lying on lateral surface \cite{CR}-\cite{BMs}.  We note also
that in the papers \cite{Ch1}, \cite{Ch2} Chechkin studied
boundary value problem for Poisson equation in $n$-dimensional
layer with frequent periodic interchange of Dirichlet and
Neumann conditions on parts of the boundary shrinking to a
point. It was also assumed in addition that that measures of the
parts of the boundary with different type of boundary condition
have the same smallness order. For the solution of the problem
considered the complete asymptotics expansions was obtained.


\sect{Description of the problem and the main results}

Let $x=(x_1,x_2)$ be Cartesian coordinates, $\Om$ be an
arbitrary bounded simply-connected domain in $\mathbb{R}^2$
having smooth boundary, $s$ be a natural parameter of the curve
$\p\Om$, and $S$ be a length of this curve, $s\in[0,S)$. We will
describe the points of  $\p\Om$ by natural parameter, fixing the
direction of going around  (counterclockwise) and choosing
arbitrary a point in  $\p\Om$ associated with a value $s=0$. For
convenience of presentation we additionally associate the points
corresponding to values of $s$ close to $S$ or to zero with the
values $(s-S)$ and $(S+s)$. We assume $N\gg1$ to be a natural
number, $\e=2N^{-1}$ is a  small positive parameter. For each
value of $N$ we define a set $\g_\e$ in the boundary $\p\Om$
consisting of  $N$ open disjoint connected (cf. fig.). Let us
define the set $\g_\e$ more concretely. For each  $N$ we define
points $x^\e_j\in\p\Om$, $j=0,\ldots,N-1$, associated with
values $s^\e_j\in[0,S)$ of natural parameter, where the distance
between each two neighbouring points measured along the boundary
of the domain $\Om$ is of order $\e$. Next, we introduce two
sets of $N$ functions: $a_j(\e)$ and $b_j(\e)$,
$j=0,\ldots,N-1$, where the functions $a_j$ and $b_j$ are
nonnegative and bounded. The set $\g_\e$ is defined as follows:
\begin{equation*}
\g_\e=\bigcup\limits_{j=0}^{N-1}\g_{\e,j}, \quad \g_{\e,j}=\{x:
-\e a_j(\e)<s-s_j^\e<\e b_j(\e)\},\quad j=0,\ldots,N-1.
\end{equation*}
Without loss of generality the sets $\g_{\e,j}$ are assumed to
be disjoint.
\begin{remark}\label{rm1.7}
We stress that we does not exclude the situation when for some
$\e$ and $j$  function $a_j$ or $b_j$ vanishes. In this case the
point $x_j^\e$ does not belong to the set $\g_{\e,j}$.
\end{remark}

In the paper we consider  singular perturbed eigenvalue problem:
\begin{align}
-&\D\psi_\e=\l_\e\psi_\e,\quad x\in\Om,\label{1.1}
\\
{}&
\begin{aligned}
\psi_\e&=0,\quad x\in\g_\e,
\\
\frac{\p\psi_\e}{\p\nu}&=0,\quad x\in\G_\e,
\end{aligned}\label{1.2}
\end{align}
where $\nu$ is the outward unit normal for the boundary $\p\Om$,
$\G_\e=\p\Om\backslash\overline{\g}_\e$. The object of the paper
is to investigate the behaviour of solutions of the perturbed
problem as $\e\to0$ (or, equivalently, $N\to\infty$).

We set $a_N(\e)\equiv a_0(\e)$, $b_N(\e)\equiv b_0(\e)$,
$s_N^\e\equiv s_0^\e$. Everywhere in the paper the expressions
of the form $f'$ denote the derivations on $s$.

Throughout the paper we suppose the following assumption to be
held.
\begin{enumerate}
\setcounter{enumi}{\value{cons}} \addtocounter{cons}{1}
\def\theenumi{(C\arabic{enumi})}
\item\label{cnA0} There exists a function $\th_\e(s)$, $\th_\e:
[0,S]\to[0,2\pi]$, $\th_\e(0)=0$,  $\th_\e(S)=2\pi$, such that
\begin{equation*}
\th_\e(s_j^\e)=\th_\e(s^\e_0)+\e\pi j, \quad j=0,\ldots,N-1,
\end{equation*}
$\th'_\e\in C^\infty(\p\Om)$, $0<c_1\le \th'_\e(s)\le c_2$,
where $c_1$, $c_2$ are some constants independent on $\e$ and
$s$. The function  $\th_\e(s)$ converges to some function
$\th_0(s)$ in $C^1[0,S]$ as $\e\to0$, $\th'_0\in
C^\infty(\p\Om)$. The norm $\|\th'_\e\|_{C^3(\p\Om)}$ is bounded
on $\e$.
\end{enumerate}
Geometrically the assumption \ref{cnA0} means that the boundary
$\p\Om$ can be smoothly and in one-to-one manner mapped onto
circumference of unit radius such that the set of points
$\{x_j^\e\}$ is mapped to a periodic set of points dividing unit
circumference into $N$ arcs of length $\e\pi$, and also this
transformation may depend on $\e$. The only constraints imposed
to this dependence are convergence of  $\th_\e$ and boundedness
of $\|\th'_\e\|_{C^3(\p\Om)}$.

\begin{remark}\label{rm1.4}
It should be stressed that the nonperiodic structure of
interchange of boundary condition is not generated by
transformation $\th_\e$, i.e., the function  $\th_\e$, generally
speaking, does not map $\g_\e$ into periodic set. For instance,
let $\Om$ be a unit circle with center at the origin,
$x_j^\e=(\cos\e\pi j,\sin\e\pi j)$, $s_j^\e=\e\pi j$,
$a_j(\e)=\e j(1+\e\sin j)/2$, $b_j(\e)=1-\e j/2$. Here the set
$\g_\e$ is a union of arcs having different lengths, laying in
$\p\Om$, moreover, $j$-th arc contains the point
$\{r=1,\th=\e\pi j\}$, ($(r,\th)$ are polar coordinates), but is
not centered with respect to this point. The assumption
\ref{cnA0} for such set holds with
$\th_\e(s)\equiv\th_0(s)\equiv s\equiv\th$, $a^j(\e)=a_j(\e)$,
$b^j(\e)=b_j(\e)$.
\end{remark}

We order the perturbed eigenvalues in ascending order counting
simplicity: $\lambda_\varepsilon^1\le
\lambda_\varepsilon^2\le\ldots\le\lambda_\varepsilon^k\le
\ldots$ Associated eigenfunctions $\psi_\e^k$ are supposed to be
orthonormalized in $L_2(\Om)$.

The first part of the main results of the article are the
estimates for degree of convergence given in formulation of the
following three theorems.

\begin{theorem}\label{th1.2}
Let the assumption \ref{cnA0} and the following ones hold:
\begin{enumerate}
\def\theenumi{(\arabic{enumi})}
\item\label{cnA6} There exists positive bounded function
$\eta=\eta(\e)$  satisfying an equality
\begin{equation}\label{1.3}
\lim\limits_{\e\to0}\e\ln\eta(\e)=0,
\end{equation}
such that estimates
\begin{equation*}
2c_1^{-1}\eta(\e)\le \min\limits_j a_j(\e)+\min\limits_i
b_i(\e),
\end{equation*}
take place where $c_1$ is from \ref{cnA0};

\item\label{cnA7} There exists $\mathsf{d}>0$ such that a H\"older
norm $\|\th'_\e\|_{C^{3+\mathsf{d}}(\p\Om)}$ is bounded on $\e$;
\end{enumerate}

Then eigenvalues  $\l_\e^k$ of the perturbed problem converge to
the eigenvalues $\l_0^k$ (taken in ascending order counting
multiplicity) of the limiting problem
\begin{equation}\label{1.4}
-\D\psi_0=\l_0\psi_0,\quad x\in\Om,\qquad\psi_0=0,\quad
x\in\p\Om,
\end{equation}
and estimates
\begin{equation*}
C_{k,1}\e\ln\sin\eta(\e)-C_{k,2}
\left|\e\ln\eta(\e)\right|^{3/2}\left(\pi/2-\eta(\e)\right)\le\l_\e^k-\l_0^k\le0,
\end{equation*}
hold true, where $C_{k,i}$ are some positive constants
independent on $\e$ and the function $\eta(\e)$ is bounded above
by a number $\pi/2$.
\end{theorem}

\begin{theorem}\label{th1.4}
Let the assumption \ref{cnA0} and the following one hold:
\begin{enumerate}
\def\theenumi{(\arabic{enumi})}
\item\label{cnA5} There exist positive bounded functions
$\eta=\eta(\e)$ and $\eta_0=\eta_0(\e)$  satisfying equalities
$\lim\limits_{\e\to0}\e\ln\eta_0(\e)=0$ and
\begin{equation}\label{1.5}
\lim\limits_{\e\to0}(\e\ln\eta(\e))^{-1}=-A
\end{equation}
with $A=\mathrm{const}>0$, such that estimates
\begin{equation*}
2c_1^{-1}\eta_0\eta\le a_j+b_j\le 2c_2^{-1}\eta,\quad
j=0,\ldots,N-1,
\end{equation*}
take place with  the constants  $c_1$ and $c_2$ from \ref{cnA0}.
\end{enumerate}

Then eigenvalues  $\l_\e^k$ of the perturbed problem converge to
the eigenvalues $\l_0^k$ (taken in ascending order counting
multiplicity) of the limiting problem
\begin{equation}\label{1.6}
-\D\psi_0=\l_0\psi_0,\quad x\in\Om,\qquad\left(\frac{\p}{\p\nu
}+A\th'_0(s)\right)\psi_0=0,\quad x\in\p\Om,
\end{equation}
and estimates
\begin{equation*}
C_{k,1}\mu(\e)+C_{k,2}\e\ln\eta_0(\e)-C_{k,3}\e-C_{k,4}\si(\e)
\le\l_\e^k-\l_0^k\le
C_{k,5}\mu(\e)+C_{k,6}\e^{3/2}+C_{k,7}\si(\e),
\end{equation*}
hold true, where
$\mu=\mu(\e)=-\left(\e\ln\eta(\e)\right)^{-1}-A$,
$\si(\e)=\|\th'_\e-\th'_0\|_{C(\p\Om)}$, $C_{k,i}$ are some
positive constants independent on $\e$ and the function
$\eta_0(\e)$ is bounded above by a unit.
\end{theorem}

\begin{theorem}\label{th1.3}
Let the assumption \ref{cnA0} and the following one hold:
\begin{enumerate}
\def\theenumi{(\arabic{enumi})}
\item\label{cnA4} There exists positive bounded function
$\eta=\eta(\e)$  satisfying  the equality (\ref{1.5}) with $A=0$
such that estimates
\begin{equation*}
a_j+b_j\le 2c_2^{-1}\eta,\quad j=0,\ldots,N-1,
\end{equation*}
take place with the constant $c_2$ from \ref{cnA0}.
\end{enumerate}

Then eigenvalues  $\l_\e^k$ of the perturbed problem converge to
the eigenvalues $\l_0^k$  of the  problem (\ref{1.6}) with $A=0$
and the estimates
\begin{equation*}
0\le\l_\e^k-\l_0^k\le C_k\mu(\e),
\end{equation*}
hold true, where $\mu=\mu(\e)=-\left(\e\ln\eta(\e)\right)^{-1}$,
$C_{k}$ are some positive constants independent on $\e$.
\end{theorem}

Let us outline the geometrical meaning of the hypothesises of
Theorems~\ref{th1.2}-\ref{th1.3}. The assumptions~\ref{cnA6} of
these theorems are posed to the lengths of individual components
of the set  $\g_\e$ and allow the sets $\g_{\e,j}$ to have
lengths of \emph{different} orders. Moreover, the estimate from
assumption~\ref{cnA4} of Theorem~\ref{th1.3} admits the
situation, when for some $\e$ and $j$ the equality
$a_j(\e)+b_j(\e)=0$ holds, i.e., corresponding set $\g_{\e,j}$
is empty and Neumann condition is imposed in a neighbourhood of
the point $x^\e_j$.  It should be noted that the constants in
the assumptions~\ref{cnA6} of Theorems~\ref{th1.2}-\ref{th1.3}
can be arbitrary, however, they can always be chosen in  a shown
way by multiplying the functions $\eta$ and $\eta_0$ by an
appropriate numbers.

The second part of the article's main results is asymptotics
expansions for eigenelements of the perturbed problem. Clear,
the restrictions for the set $\g_\e$ needed for constructing
such expansions should be more severe in comparing with
hypothesises of Theorems~\ref{th1.2}-\ref{th1.3}. One of such
restriction for the set  $\g_\e$ looks as follows:

\begin{enumerate}
\setcounter{enumi}{\value{cons}} \addtocounter{cons}{1}
\def\theenumi{(C\arabic{enumi})}
\item\label{cnA2} There exists positive bounded function
$\eta=\eta(\e)$ such that estimates
\begin{equation*}
c_3\eta(\e)\le a_j(\e)+b_j(\e)\le 2c_2^{-1}\eta(\e),\quad
j=0,\ldots,N-1,
\end{equation*}
hold true where positive constant $c_3$ is independent on $\e$,
$\eta$ and $j$.
\end{enumerate}
Geometrically the assumption \ref{cnA2} means that all sets
$\g_{\e,j}$ have length of order $\e\eta$ as $\e\to0$, that,
however, does not mean the coincidence of these lengths.
Observe, in the right side of the inequality from this
assumption we would have written just some constant  $c_4$.
However, multiplying $\eta(\e)$ by an appropriate number it is
easy to make this constant equal to shown value.

In order to formulate main results of the work about asymptotics
expansions we will need some auxiliary facts and additional
notations.

We continue the function $\th_\e$ to the values  $s\in[-S,2S)$
by a rule $\th_\e(s)=\th_\e(s-kS)+2\pi k$, $s\in[kS,(k+1)S)$,
$k=-1,0,1$. Denote:
\begin{align*}
d_j(\e)&=\frac{a_j(\e)+b_j(\e)}{2\eta(\e)},&
d^j(\e)&=\frac{\th_\e(s_j^\e+\e b_j(\e))-\th_\e(s_j^\e-\e
a_j(\e))}{2\e\eta(\e)},
\\
\d_j(\e)&=d_{j+1}(\e)-d_j(\e), & \d^j(\e)&=d^{j+1}(\e)-d^j(\e).
\end{align*}
Let $\chi(t)$ be an infinitely differentiable cut-off function
equalling to one as  $t<1/4$ and vanishing as $t>3/4$ whose
values belong to a segment $[0,1]$. We introduce one more
function $\mathsf{f}_\e(\th)$:
\begin{equation*}
\mathsf{f}_\e(\th)=d^{j+1}(\e)-\chi\left((\th-\th_\e(s^\e_j))/
(\e\pi)\right)\d^j(\e),
\end{equation*}
as $\e\pi j\le\th-\th_\e(s^0_\e)\le\e\pi(j+1)$,
$j=0,\ldots,N-1$. The eigenvalues of the problem (\ref{1.6}),
like above, are taken in ascending order counting multiplicity:
$\lambda_0^1\le\lambda_0^2\le\ldots\le\lambda_0^k\le\ldots$, and
we orthonormalize associated eigenfunctions $\psi_0^k$ in
$L_2(\Omega)$.

The following proposition has an auxiliary character ant it will
be proved in the second section.

\begin{lemma}\label{lm1.1} Let $\|\th'_\e-\th'_0\|_{C(\p\Om)}\to0$.
Then eigenvalues $\L_0^k=\L_0^k(\mu,\e)$ of a problem
\begin{gather}
-\D\Psi_0^k=\L_0^k\Psi_0^k, \quad x\in\Om, \label{2.6}
\\
\left(\frac{\p}{\p\nu}+(A+\mu)\th'_\e(s)\right)\Psi_0^k=0,\quad
x\in\p\Om,\label{2.25}
\end{gather}
where $A\ge0$, taken in ascending order counting multiplicity
converge to eigenvalues $\l_0^k$ of the problem (\ref{1.6}) as
$(\e,\mu)\to0$. For each fixed value of $\e$ the eigenvalues
$\L_0^k$ and the associated orthonormalized in $L_2(\Om)$
eigenfunctions $\Psi_0^k$ are holomorphic on $\mu$ (latter -- in
$H^1(\Om)$ norm). If $\L_0^k$ is a multiply eigenvalue, then the
associated eigenfunctions can be additionally orthogonalized in
$L_2(\p\Om)$ weighted by $\th'_\e$.
\end{lemma}

Let us formulate the second part of the main results.

\begin{theorem}\label{th1.1}
Suppose the assumptions \ref{cnA0}, \ref{cnA2}, equality
(\ref{1.5}) with $A\ge0$ for the function $\eta$ from \ref{cnA2}
and
\begin{equation}\label{1.7}
\max\limits_{j}|\d_j(\e)|\equiv
\d_*(\e)=o(\e^{1/2}(A+\mu)^{-1}),
\end{equation}
where $\mu=\mu(\e)=-\left(\e\ln\eta(\e)\right)^{-1}-A$, hold.
Then the eigenvalue $\l_\e^k$ of the perturbed problem converge
to eigenvalue $\l_0^k$ of the limiting problem (\ref{1.6}) and
has the asymptotics:
\begin{align}\label{1.8}
{}& \l_\e^k=\L_0^k(\mu,\e)+\e\L_1^k(\mu,\e)+o(\e(A+\mu)),
\\
{}&\L_1^k(\mu,\e)=(A+\mu)^2\int\limits_{\p\Om}
\left(\Psi_0^k(x,\mu,\e)\right)^2\ln\mathsf{f}_\e(\th_\e(s))
\th'_\e(s) \,\mathrm{d}s, \label{1.9}
\end{align}
where $\L_0^k$ and $\Psi_0^k$ meet Lemma~\ref{lm1.1}. The
function $\L_1^k$ is non-positive and holomorphic on $\mu$ for
each fixed $\e$.
\end{theorem}

\begin{remark}\label{rm1.5}
For the case of simple limiting eigenvalue $\l_0^k$ in next
section we will prove in addition that coefficients of Taylor
series in powers of $\mu$ for the functions  $\L_0^k$,
$\Psi_0^k$ are continuous as $\e\to0$, and these functions are
majorized by holomorphic on $\mu$ functions independent on $\e$.
The function $\Psi_0^k$ is majorized in a sense of $H^1(\Om)$
norm. Also it will be shown that  $\Psi_0^k$ converges to
$\psi_0^k$ in $H^1(\Om)$ as $(\e,\mu)\to(0,0)$.
\end{remark}

\begin{remark}\label{rm1.3}
Let us pay attention to the equality~(\ref{1.7}). The quantities
$\d_j$ characterize difference between lengths of two
neighbouring sets $\g_{\e,j+1}$ and $\g_{\e,j}$, so, the
equality (\ref{1.7}) actually means that lengths of two
neighbouring components of the set $\g_\e$ does not differ very
much.
\end{remark}

Along with asymptotics for  $\l_\e^k$ we will prove statements
about asymptotics for associated eigenfunctions $\psi_\e^k$
under the hypothesis of Theorem~\ref{th1.1}. In order to
formulate these statements we have to introduce some additional
notations and that's why it is more convenient to formulate them
in the end of the second section  (see
Theorems~\ref{th2.1},~\ref{th1.5}).

In next theorem we give asymptotics of the perturbed eigenvalues
in the case of breakdown of equality~(\ref{1.7})  and keeping
other assumptions of Theorem~\ref{th1.1}.

\begin{theorem}\label{th1.6}
Suppose the assumptions  \ref{cnA0}, \ref{cnA2} and equality
(\ref{1.5}) with $A\ge0$ for the function $\eta$ from \ref{cnA2}
hold. Then eigenvalue  $\l_\e^k$ of the perturbed problem
converges to the eigenvalue $\l_0^k$ of the limiting problem
(\ref{1.6}) and has the asymptotics:
\begin{equation}\label{1.17}
\l_\e^k=\l_0^k+\mu\int\limits_{\p\Om}
(\psi_0^k(x))^2\th'_0(s)\mathrm{d}s+
O\left(\mu^2+\mu(\si+\e^{3/2})+A(\si+\e^{1/2})\right),
\end{equation}
where in the case of multiply eigenvalue $\l_0^k$ the associated
eigenfunctions are additionally orthogonalized in $L_2(\p\Om)$
weighted by $\th'_0$, $\si=\|\th'_\e-\th'_0\|_{C(\p\Om)}$.
\end{theorem}
The asymptotics (\ref{1.17}) is constructive as $A=0$ and in the
case $A>0$ for  $\si+\e^{1/2}=o(\mu)$.

The statement about asymptotics of eigenfunctions $\psi_\e^k$
under hypothesis of last theorem will be proved in the third
section (see Theorem~\ref{th3.1}).

The  structure of the paper is as follows. In the second section
we prove Theorem~\ref{th1.1} and, under its hypothesis,
Theorems~\ref{th2.1},~\ref{th1.5} about asymptotics of the
perturbed eigenfunctions. The third section is devoted to the
proof of Theorem~\ref{th1.6} and Theorem~\ref{th3.1} about
asymptotics of perturbed eigenfunctions under hypothesis of
Theorem~\ref{th1.6}. In the fourth section we will establish the
correctness of auxiliary statement about asymptotics of
perturbed eigenvalues in the case of limiting Dirichlet problem.
This auxiliary statement will be employed in next section for
the proof of Theorem~\ref{th1.2}. Furthermore, in the fifth
section Theorems~\ref{th1.4},~\ref{th1.3} will be proved.


\sect{Asymptotics for the perturbed eigenelements under
hypothesis of Theorem~\ref{th1.1}}

In this section we will obtain asymptotics for the eigenelements
of the perturbed problem. First we will establish the validity
of some auxiliary statements. We start from Lemma~\ref{lm1.1}.

\PF{\indent Proof of Lemma~\ref{lm1.1}.} Boundary value problem
(\ref{2.6}), (\ref{2.25}) is regular perturbed. Convergence of
eigenelements and  maintained holomorphy on $\mu$ of
eigenelements is easily established by rewriting of (\ref{2.6}),
(\ref{2.25}) to an operator equation and employing then the
results of \cite{Kt}. Keeping all stated properties, the
eigenfunctions $\Psi_0^k$ can be orthonormalized in $L_2(\Om)$.
According to the theorem on diagonalization of two quadratic
forms, the eigenfunctions associated with multiply eigenvalue
can be additionally orthogonalized in  $L_2(\p\Om)$ weighted by
$\th'_\e$. Since $\th'_\e$ is independent on $\mu$, it is clear
that such additional orthogonalization keeps holomorphy on $\mu$
of these eigenfunctions. The proof is complete.

If $\l_0^k$ is a simple eigenvalue of the problem~\ref{1.6},
then exactly one eigenvalue  $\L_0^k$ of the problem
(\ref{2.6}), (\ref{2.25}) converges to it, and associated
eigenfunction $\Psi_0^k$ converges to  $\psi_0^k$ in $H^1(\Om)$.
Represent  $\L_0^k$ and $\Psi_0^k$ as power on $\mu$ series,
substitute them into (\ref{2.6}), (\ref{2.25}) and calculate the
coefficients of the same powers of  $\mu$. The recurrence system
of boundary value problems derived in this way, as it is easy
prove accounting simplicity $\l_0^k$, is uniquely solvable, its
solutions are continuous as $\e\to0$ and can be estimated
uniformly on  $\e$. Last estimates allows to construct
independent on $\e$ and holomorphic on $\mu$ majorants for
$\L_0^k$ and $\Psi_0^k$. Thus, the statement of
Remark~\ref{rm1.5} is proved.

Suppose the assumption \ref{cnA0} holds. We denote:
\begin{equation*}
 a^j(\e)=\left(\th_\e(s_j^\e)-\th_\e(s_j^\e-\e a_j(\e))\right)
/\e, \quad b^j(\e)=\left(\th_\e(s_j^\e+\e
b_j(\e))-\th_\e(s_j^\e)\right)/\e.
\end{equation*}
The functions $a^j$ and $b^j$ describe the image of the set
$\g_{\e,j}$ under mapping  $\th_\e$: length of this image equals
to $\e(a^j+b^j)$, and its end-points associated with angles
$(\th_\e(s^\e_j)- \e a^j(\e))$ and $(\th_\e(s^\e_j)+\e
b^j(\e))$. Suppose that the assumption \ref{cnA2} holds, too. We
set:
\begin{equation*}
\a^j(\e)=\frac{a^j(\e)}{2\eta(\e)},\quad
\b^j=\b^j(\e)=\frac{b^j(\e)}{2\eta(\e)}, \quad
\d^*(\e)=\max\limits_{j}|\d^j(\e)|.
\end{equation*}
Note, that $d^j=\a^j+\b^j$.

\begin{lemma}\label{lm2.6}
Let the assumptions \ref{cnA0} and \ref{cnA2} hold. Then the
estimates
\begin{gather*}
c_1(a_j(\e)+b_j(\e))\le a^j(\e)+b^j(\e)\le c_2(a_j(\e)+b_j(\e)),
\quad \d^*(\e)\le C(\d_*(\e)+\e).
\end{gather*}
are true, where the constant $C$ are independent on  $\e$ and
$\eta$.
\end{lemma}
\PF{Proof.} By Lagrange theorem and the definition of the
functions $a^j$ and $b^j$ we have:
\begin{equation}\label{1.18}
a^j(\e)+b^j(\e)=\th'_\e(M_{j,\e}^{(1)})(a_j(\e)+b_j(\e)),
\end{equation}
where $M_{j,\e}^{(1)}$ is a midpoint belonging to an interval
$(s_j^\e-\e a_j(\e), s_j^\e+\e b_j(\e))$. Employing now the
estimate of the derivation $\th'_\e$ from the assumption
\ref{cnA0}, we arrive at the first inequality from the statement
of the lemma.

Equality (\ref{1.18}) and definition $\d^j(\e)$ imply:
\begin{equation}\label{2.91}
\begin{aligned}
\d^j=&\th'_\e(M_{j+1,\e}^{(1)})(\a_{j+1}+\b_{j+1})-
\th'_\e(M_{j,\e}^{(1)})(\a_{j}+\b_{j})=
\\
{}&=\th'_\e(M_{j+1,\e}^{(1)})\d_j+
(\a_j+\b_j)(\th'_\e(M_{j+1,\e}^{(1)})-\th'_\e(M_{j,\e}^{(1)})).
\end{aligned}
\end{equation}
The quantity
$(\th'_\e(M_{j+1,\e}^{(1)})-\th'_\e(M_{j,\e}^{(1)}))$ in
accordance with Lagrange theorem can be represented in the form:
\begin{equation*}
\th'_\e(M_{j+1,\e}^{(1)})-\th'_\e(M_{j,\e}^{(1)})=
\th''_\e(M_{j,\e}^{(2)}) (M_{j+1,\e}^{(1)}-M_{j,\e}^{(1)}),
\end{equation*}
where, recalling the definition of $M_{j,\e}^{(1)}$, a midpoint
$M_{j,\e}^{(2)}$ lies in an interval $(s_j^\e-\e
a_j(\e),s_{j+1}^\e+\e b_{j+1}(\e))$. In view of last equality
and the assumptions \ref{cnA0} and \ref{cnA2} the second term in
right side of (\ref{2.91}) is estimated as follows:
\begin{equation}\label{2.93}
|(\a_j+\b_j)\th''_\e(M_{j,\e}^{(2)})(M_{j+1,\e}^{(1)}-M_{j,\e}^{(1)})|\le
C\left(s_{j+1}^\e-s_j^\e +\e(b_{j+1}+a_j)\right)\le C\e,
\end{equation}
where constants $C$ are independent on  $\e$, $\eta$ and $j$.
Here we also employed a relationship
\begin{equation*}
c_1|s^\e_{j+1}-s^\e_j|\le
\th_\e(s^\e_{j+1})-\th_\e(s^\e_j)=\e\pi,
\end{equation*}
that is easy to prove. Substitution (\ref{2.93}) into
(\ref{2.91}) and estimate of quantity
$\th'_\e(M_{j+1,\e}^{(1)})$ by the assumption \ref{cnA0} lead us
to a second inequality from the statement of the lemma. The
proof is complete.

The lemma proved in an obvious way yields
\begin{corollary}\label{lm2.6cor1}
Under hypothesis of theorem~\ref{th1.1} the equality
$\d^*(\e)=o(\e^{1/2}(A+\mu)^{-1})$ is true.
\end{corollary}

\begin{corollary}\label{lm2.6cor2}
The function  $\d^*(\e)$ is bounded.
\end{corollary}
\PF{Proof.} It arises from Lemma~\ref{lm2.6} and \ref{cnA2} that
\begin{equation*}
|d^j(\e)|\le c_2\frac{a^j(\e)+b^j(\e)}{2\eta(\e)}\le 1,
\end{equation*}
what implies the boundedness of $\d^*$. The proof is complete.

\PF{\indent Proof of Theorem~\ref{th1.1}.} Convergence of
perturbed eigenvalues to ones of problem (\ref{1.6}) under
assumptions \ref{cnA0}, \ref{cnA2} and equality (\ref{1.5}) can
be easily established, using results and methods of papers
\cite{ChNov}, \cite{Ch}, \cite{Fr}.

Our strategy in proving the asymptotics consists of two main
steps. First we will formally construct the asymptotics for the
eigenelements of the perturbed problem. Second step is to prove
rigorously (to justify) that the asymptotics expansions formally
constructed are really asymptotics of eigenelements. In formal
construction we will use only the boundedness of the function
$\d^*(\e)$ (see Corollary~\ref{lm2.6cor2} of Lemma~\ref{lm2.6}),
while equality (\ref{1.7}) will be employed only in
justification of asymptotics for estimating the errors.

In formal construction we will show in detail only the case of
simple limiting eigenvalue. Such a choose is explained by a
desire to avoid an excessive cumbersomeness of representation,
on the one hand, and, on the other hand, the construction does
not depend essentially on multiplicity of limiting eigenvalue.
Below we will briefly outline the formal construction in the
case of multiply limiting eigenvalue.

Now we proceed to the construction of the asymptotics.  We
suppose $\l_0$ to be a simple eigenvalue of the problem
(\ref{1.6}), $\l_\e$ is an perturbed eigenvalue converging to
$\l_0$, $\psi_\e$ and $\psi_0$ are associated eigenfunctions.
First we will demonstrate the scheme of constructing and
formally obtain first terms of asymptotics. We seek for the
asymptotics of eigenvalue as
\begin{equation}\label{2.2}
\l_\e=\L_0(\mu,\e)+\e\L_1(\mu,\e).
\end{equation}
The asymptotics for  $\psi_\e$ is constructed on the basis of
combination of method of matching asymptotics expansions
\cite{Il}, method of composite expansions \cite{CE} and
multiscaled method \cite{MS}. This asymptotics will be obtained
as a sum of three expansions, namely, outer expansion, boundary
layer and inner expansion. Exterior expansion is constructed as
follows:
\begin{equation}\label{2.3}
\psi_\e^{ex}(x,\mu)=\Psi_0(x,\mu,\e)+\e\Psi_1(x,\mu,\e).
\end{equation}
Using method of composite expansions, we construct the boundary
layer in the form:
\begin{equation}\label{2.4}
\psi_\e^{bl}(\xi,s,\mu)=\e v_1(\xi,s,\mu,\e)+\e^2
v_2(\xi,s,\mu,\e),
\end{equation}
where
$\xi=(\xi_1,\xi_2)=((\th_\e(s)-\th_\e(s^\e_0))/\e,\tau\th'_\e(s)/\e)$
are ''scaled'' variables. Here $(s,\tau)$ are local variables
defined in a neighbourhood of the boundary $\p\Om$, $\tau$ is a
distance from the point to the boundary measured in the
direction of inward normal. Such a definition of variables $\xi$
will be explained below in Remark~\ref{rm2.1}.

Interior expansion will be constructed by the method of matched
asymptotics expansions in small neighbourhoods  of points
$x_j^\e$ in the form:
\begin{equation}\label{2.5}
\psi_\e^{in,j}(\vs^j,s,\mu)=w_{0,0}^{(j)}(\vs^j,s,\mu,\e)+\e
w_{1,0}^{(j)}(\vs^j,s,\mu,\e),
\end{equation}
where $\vs^j=(\vs^j_1,\vs^j_2)=((\xi_1-\pi
j)\eta^{-1},\xi_2\eta^{-1})$.

The aim of the formal construction is to determine the functions
$\L_i$, $\Psi_i$, $v_i$ and $w_{i,0}$.

The equations for the functions $\Psi_0$ and $\Psi_1$ are
derived by standard substitution of (\ref{2.2}) and (\ref{2.3})
into equation (\ref{1.1}) with consequent writing out the
coefficients of the same power of $\e$. Such procedure leads to
the equation (\ref{2.6}) for the function $\Psi_0$ with
$\L_0^k=\L_0$ and to the following equation for $\Psi_1$:
\begin{equation}
(\D+\L_0)\Psi_1=-\L_1\Psi_0, \quad x\in\Om. \label{2.7}
\end{equation}
The boundary condition for the functions $\Psi_0$ and $\Psi_1$
will be deduced later in constructing of the boundary layer and
inner expansion's coefficients.

Let us determine the functions $v_i$. First we should obtain the
problems for them, in order to make it one needs to rewrite the
Laplace operator in variables $(s,\tau)$:
\begin{equation*}
\Delta_x=\frac{1}{\mathsf{H}}\left(\frac{\partial}{\partial\tau}
\left(\mathsf{H}\frac{\partial}{\partial\tau}
\right)+\frac{\partial}{\partial
s}\left(\frac{1}{\mathsf{H}}\frac{\partial}{\partial
s}\right)\right),\qquad
\mathsf{H}=\mathsf{H}(s,\tau)=1+\tau\mathsf{k}(s),
\end{equation*}
$\mathsf{k}=\mathsf{k}(s)=\left(\boldsymbol{r}''(s),
\nu(s)\right)_{\mathbb{R}^2}$, $\nu=\nu(s)$, $\boldsymbol{r}(s)$
is a two-dimensional vector-function describing the curve
$\partial\Om$, $\mathsf{k}\in C^\infty(\p\Om)$. Now we
substitute  (\ref{2.2}), (\ref{2.3}) and the expression for
Laplace operator in variables $(s,\tau)$ in (\ref{1.1}), pass to
the variables  $\xi$ and write out the coefficients of leading
powers of  $\e$. This implies the equations for functions $v_1$
and $v_2$:
\begin{gather}
 \D_\xi v_1=0,\quad \xi_2>0,\label{2.9}
\\
\begin{aligned}
\D_\xi v_2=&-\frac{\th''_\e}{(\th'_\e)^2}
\left(\frac{\p}{\p\xi_1}+ 2\xi_2\frac{\p^2}{\p\xi_1\p\xi_2}
\right) v_1-\\
{}&-\frac{\mathsf{k}}{\th'_\e}\left(\frac{\p}{\p\xi_2}-
2\xi_2\frac{\p^2}{\p\xi_1^2}\right)v_1-\frac{2}{\th'_\e}\frac{\p^2
v_1}{\p\xi_1\p s},\quad\xi_2>0.\end{aligned}\label{2.14}
\end{gather}
In accordance with method of composite expansions, the sum of
functions  $\psi_\e^{ex}$ and $\psi_\e^{bl}$ is to satisfy to
homogeneous boundary condition everywhere in $\p\Om$ except
points $x_j^\e$:
\begin{equation*}
\frac{\p}{\p\nu}\psi_\e^{ex}-\frac{\p}{\p\tau}\psi_\e^{bl}=0,\qquad
x\in\p\Om,\quad x\not=x_j^\e.
\end{equation*}
Now we rewrite second term in last equality to the variables
$\xi$ and replace the functions $\psi_\e^{ex}$ and
$\psi_\e^{bl}$ by right sides of the equalities (\ref{2.3}) and
(\ref{2.4}), after that we calculate the coefficient of the
leading power of $\e$ that is set equalling to zero. As a
result, we have the boundary conditions for the functions
$v_i$:
\begin{align}\label{2.10}
{}&\frac{\p v_1}{\p\xi_2}=\frac{1}{\th'_\e}\Psi_0^\nu,\quad
\xi\in\G^0,
\\
{}&\frac{\p v_2}{\p\xi_2}=\frac{1}{\th'_\e}\Psi_1^\nu,\quad
\xi\in\G^0,\label{2.15}
\end{align}
where $\G^0=\{\xi: \xi_2=0, \xi_1\not=\e\pi j,
j\in\mathbb{Z}\}$,
\begin{align*}
\Psi_i^\nu=\Psi_i^\nu(s,\mu,\e)=\frac{\p}{\p\nu}\Psi_i(x,\mu,\e),\quad
x\in\p\Om.
\end{align*}
\begin{remark}\label{rm2.1}
It follows from the definition of the set $\G^0$ that the
problems for the functions $v_i$ are periodic on  the variable
$\xi_1$ what will be essentially used in solving of the boundary
value problems (\ref{2.9})--(\ref{2.15}). One can easily check
that the periodicity of $\G^0$ is a direct implication of the
assumption \ref{cnA0} (namely, of the equation
$\th_\e(s_j^\e)=\th_\e(s^\e_0)+\e\pi j$) and the definition of
the variable  $\xi_1$ given above, what explains the indicated
definition of the variable $\xi_1$. The variable $\xi_2$ was
selected so that to obtain Poisson equations for the functions
$v_1$ and $v_2$.
\end{remark}
In accordance with method of composite expansions, we are to
seek exponentially decaying as $\xi_2\to+\infty$ solutions to
the problems (\ref{2.9}), (\ref{2.10}) and (\ref{2.14}),
(\ref{2.15}). In constructing of boundary layer we additionally
employ the multiscaled method: the variable  $s$ plays the slow
time's role. We also notice that the boundary layer ''controls''
only Neumann boundary condition, exactly because of this we have
passed to a formal limit as $\eta\to0$ in deducing the boundary
condition (\ref{2.10}). Possibly, Dirichlet boundary condition
is seemed to be satisfied simultaneously by a suitable choose of
the functions $v_1$ and $v_2$. However, this way leads to
unsolvable problems for the functions $v_1$ and $v_2$.

We indicate by $\mathcal{V}_0$ the space of $\pi$-periodic on
the variable  $\xi_1$ functions belonging
$C^\infty(\{\xi:\xi_2>0\}\cup\G^0)$ and decaying exponentially
with all their derivatives as $\xi_2\to+\infty$ uniformly on
$\xi_1$.

Let us construct the solution of the problem (\ref{2.9}),
(\ref{2.10}). We stress that this problem contains the variable
$s$ as a parameter. Consider a function
\begin{equation}\label{2.105}
X(\xi)=\mathrm{Re}\,\ln\sin z+\ln 2-\xi_2,
\end{equation}
$z=\xi_1+\mathrm{i}\xi_2$ is a complex variable. By direct
calculations one can check that $X\in \mathcal{V}_0$ is even on
$\xi_1$ harmonic function meeting a boundary condition
\begin{equation*}
\frac{\p X}{\p\xi_2}=-1,\quad\xi\in\G^0.
\end{equation*}
The representation
\begin{equation}\label{2.12}
X(\xi)=\ln\rho+\ln 2-\xi_2+\t X(\xi),\quad\t
X(\xi)=O(\rho^2),\quad \xi\to\xi^{(j)},
\end{equation}
holds, where $\rho=|\xi-\xi^{(j)}|$, $\xi^{(j)}=(\pi j,0)$,
$j\in\mathbb{Z}$, $\t X(\xi)\in C^\infty(\{\xi: \xi_2\ge0\})$.
Bearing in mind all the facts counted, we conclude that the
function $v_1$ has the form:
\begin{equation*}
v_1(\xi,s,\mu)=-\frac{1}{\th'_\e(s)}\Psi_0^\nu(s,\mu,\e)X(\xi).
\end{equation*}
The solutions for the problem (\ref{2.14}), (\ref{2.15}) can be
constructed explicitly, too. By direct calculations we check
that the function
\begin{equation*}
\t v_2=\frac{\Psi^\nu_0}{2(\th'_\e)^2}\xi_2^2 \left(
\frac{\th''_\e}{\th'_\e}\frac{\p X}{\p\xi_1}+\mathsf{k}\frac{\p
X}{\p\xi_2}\right)-\frac{1}{\th'_\e}
\left(\frac{\Psi_0^\nu}{\th'_\e}\right)'\int\limits_{\xi_2}^{+\infty}
t\frac{\p}{\p\xi_1}X(\xi_1,t)\,\mathrm{d}t
\end{equation*}
is a solution of equation (\ref{2.14}) satisfying  homogeneous
Neumann condition on $\G^0$. The function $\t
v_2\in\mathcal{V}_0$ has the following (differentiable)
asymptotics as $\xi\to\xi^{(j)}$:
\begin{equation}\label{2.16}
\t v_2=O(\rho\ln\rho).
\end{equation}
Taking into account all the described properties of the function
$\t v_2$ and the properties of the function $X$, we arrive at
the formula for the function $v_2$:
\begin{equation}\label{2.17}
v_2(\xi,s,\mu,\e)=\t
v_2(\xi,s,\mu,\e)-\frac{1}{\th'_\e(s)}\Psi_1^\nu(s,\mu,\e)X(\xi).
\end{equation}

As it follows from  (\ref{2.12})--(\ref{2.17}), the functions
$v_i$ have logarithmic singularities in neighbourhoods of points
$\xi^{(j)}$, or, equivalently, in neighbourhoods of points
$x_j^\e $. Moreover, the sum of the outer expansion and boundary
layer does not satisfy (even asymptotically) to Dirichlet
boundary condition on $\g_\e$. That's why in neighbourhoods of
the points $x_j^\e $ we introduce new ''scaled'' variables
$\vs^j$, and the asymptotics of the eigenfunction is constructed
as $\psi^{in}_\e$ by the method of matched asymptotics
expansions. The using of term ''scaled variables'' for $\vs^j$
is correct, since owing to the equality (\ref{1.5}) the function
$\eta$ is of the form:
\begin{equation}\label{2.1}
\eta(\e)=\exp\left(-\frac{1}{\e(A+\mu(\e))}\right),
\end{equation}
where $\mu(\e)$ is defined in a statement of the theorem being
proved. Thus, $\eta(\e)$ is exponentially small in comparing
with $\e$.

First we carry out the matching procedure in a neighbourhood of
the point $x_j^\e$. For the sake of brevity we denote
$\vs=\vs^j$. Clear, the asymptotics
\begin{equation}\label{2.18}
\psi_\e^{ex}(x)=\sum\limits_{i=0}^1\e^i
\left(\Psi_i^D(s,\mu,\e)-\tau\Psi_i^\nu(s,\mu,\e)\right)+O(\tau^2),
\end{equation}
holds true  as $\tau\to0$ where $\Psi_i^D$ indicates the values
of the functions $\Psi_i$ as $x\in\p\Om$, and the variable $s$
ranges  in a small neighbourhood of value $s_j^\e$. Bearing in
mind the asymptotics (\ref{2.12}), (\ref{2.16}) and the formulae
for $v_i$ we get that, as $\xi\to\xi^{(j)}$,
\begin{equation}\label{2.19}
\begin{aligned}
\psi_\e^{bl}(\xi,s,\mu)=-\e(\ln\rho+\ln
2-\xi_2)\sum\limits_{i=0}^1\e^i\frac{\Psi_i^\nu(s,\mu,\e)}
{\th'_\e(s)}+O(\rho\ln\rho).
\end{aligned}
\end{equation}
Let us rewrite the asymptotics (\ref{2.18}), (\ref{2.19}) in the
variables  $\vs$ and take into account that due to (\ref{2.1})
the equality $\e\ln\eta(\e)=-(A+\mu)^{-1}$ is valid. Hence, we
have that for $\frac{1}{4}\eta^{1/4}<\rho<\frac{3}{4}\eta^{1/4}$
(or, equivalently, for
$\frac{1}{4}\eta^{-3/4}<|\vs|<\frac{3}{4}\eta^{-3/4}$)
\begin{align}\label{2.20}
{}&\psi_\e^{ex,\mu}(x)+\psi_\e^{bl}(\xi,s,\mu)=W_{0,0}(s,\mu,\e)+\e
W_{1,0}(\vs,s,\mu,\e)+O(\e^2\ln|\vs|),
\\
{}&W_{0,0}(s,\mu,\e)=\Psi_0^D(s,\mu,\e)+\frac{\Psi_0^\nu(s,\mu,\e)}
{(A+\mu)\th'_\e(s)}, \label{2.21}
\\
{}&W_{1,0}(\vs,s,\mu,\e)=-
\frac{\Psi_0^\nu(s,\mu,\e)}{\th'_\e(s)}(\ln|\vs|+\ln
2)+\Psi_1^D(s,\mu)+\frac{\Psi_1^\nu(s,\mu,\e)}
{(A+\mu)\th'_\e(s)}.\label{2.22}
\end{align}
In accordance with method of matched asymptotics expansions it
arises from (\ref{2.20}) that the functions $w_{i,0}^{(j)}$ must
have the following asymptotics at infinity:
\begin{equation}\label{2.23}
w_{i,0}^{(j)}=W_{i,0}+o(1),\quad \vs\to\infty.
\end{equation}
The problems for the functions $w_{i,0}$ are deduced by standard
substitution of (\ref{2.2}) and (\ref{2.5}) into boundary value
problem (\ref{1.1}), (\ref{1.2}) and by writing out  the
coefficients of leading powers of $\e$ and $\eta$:
\begin{equation}\label{2.24}
\begin{aligned}
\D_\vs w_{i,0}^{(j)}&=0,\quad &\vs_2>0,
\\
w_{i,0}^{(j)}&=0,\quad &\vs\in\g^1_j,
\\
\frac{\p}{\p\vs_2}w_{i,0}^{(j)}&=0,\quad &\vs\in\G^1_j.
\end{aligned}
\end{equation}
Here $\g^1_j$ is an interval $(-2\a_j, 2\b_j)$ lying in the axis
$O\vs_1$, and $\G^1_j$ is a complement of the closure of
$\g^1_j$ with respect to a line $\vs_2=0$.

The problem (\ref{2.24}) has no nontrivial solutions bounded at
infinity,  therefore, in view of (\ref{2.21}), (\ref{2.23}),
\begin{equation*}
w_{0,0}^{(j)}=0.
\end{equation*}
This equality and the asymptotics (\ref{2.21}), (\ref{2.23})
yield the boundary condition  (\ref{2.25}) for the function
$\Psi_0$. The eigenelements $\L_0$ and $\Psi_0$ obey
Lemma~\ref{lm1.1}. The smoothness of domain's boundary and of
the function $\th'_\e(s)$ allows us to maintain that the
function $\Psi_0$ is infinitely differentiable on the variables
$x$.

Let us determine the function $w_{1,0}$. Let
\begin{equation}\label{2.106}
Y^{(j)}(\vs,\e)=\mathrm{Re}\,\ln\left(y+\sqrt{y^2-1}\right),
\end{equation}
where $y=(\vs_1+\mathrm{i}\vs_2+\a^j-\b^j)/(\a^j+\b^j)$ is a
complex variable. It is easy to establish that $Y^{(j)}\in
\mathcal{W}$, where
\begin{equation*}
\mathcal{W}\equiv C^\infty(\{\vs:\vs_2\ge0, \vs\not=(-2\a^j,0),
\vs\not=(2\b^j,0)\})\cap H^1(\{\vs: \vs_2>0,|\vs|<5\}).
\end{equation*}
The function $Y^{(j)}$ is a solution of the problem (\ref{2.24})
having the following asymptotics at infinity:
\begin{equation}\label{2.26}
Y^{(j)}=\ln|\vs|+\ln
2-\ln(\a^j+\b^j)+(\a^j-\b^j)\vs_1|\vs|^{-2}+O(|\vs|^{-2}),\quad
\vs\to\infty.
\end{equation}
Owing to the properties $Y^{(j)}$ stated the function
$w_{1,0}^{(j)}$ is of the form:
\begin{equation*}
w_{1,0}^{(j)}(\vs,s,\mu,\e)=-\frac{\Psi_0^\nu(s,\mu,\e)}
{\th'_\e(s)}Y^{(j)}(\vs,\e).
\end{equation*}
It is obvious that $w_{1,0}^{(j)}\in\mathcal{W}$. Now we write
out the asymptotics of the function $w_{1,0}^{(j)}$ at infinity
(see (\ref{2.26})) and compare it with (\ref{2.22}),
(\ref{2.23}). As a result we arrive at the equality
\begin{equation}\label{2.28}
\frac{\Psi_0^\nu(s,\mu,\e)}{\th'_\e(s)}\ln(\a^j(\e)+\b^j(\e))=
\Psi_1^D(s,\mu,\e)+
\frac{\Psi_1^\nu(s,\mu,\e)}{(A+\mu)\th'_\e(s)}.
\end{equation}
This equality actually is a boundary condition for the function
$\Psi_1$. We just should correctly define the right side of this
condition bearing in mind that, generally speaking, the
quantities $\ln(\a^j+\b^j)$ depend on index $j$ and the
parameter $\e$. As it has already been mentioned above, the
variable $s$ in the equality (\ref{2.28}) ranges in a small (of
order $O(\e\eta^{1/4})$) neighbourhood of point $s_j^\e$.
Therefore, to satisfy the equality (\ref{2.28}) it is sufficient
to construct the function equalling to  $(\a^j+\b^j)$ in these
neighbourhoods of the points $s_j^\e$ and then replace
$(\a^j+\b^j)$ by this function in (\ref{2.28}). The function
$\mathsf{f}_\e(\th)$, as it is easy to prove, is infinitely
differentiable on $\th$ and equals to $d^j=\a^j+\b^j$ for
$|\th-\th_\e(s^\e_j)|\le \e\pi/4$. That's why as the function
the sum $(\a^j+\b^j)$ in (\ref{2.28}) is replaced to we take
$\mathsf{f}_\e(\th_\e(s))$ what immediately implies the boundary
condition for $\Psi_1$:
\begin{equation*}
\left(\frac{\p}{\p\nu}+(A+\mu)\th'_\e(s)\right)\Psi_1=
(A+\mu)\Psi_0^\nu(s,\mu,\e)\ln\mathsf{f}_\e(\th_\e(s)),\quad
x\in\p\Om.
\end{equation*}
Now we take into account that
$\Psi_0^\nu=-(A+\mu)\th'_\e\Psi_0^D$, and finally we have:
\begin{equation}\label{2.29}
\left(\frac{\p}{\p\nu}+(A+\mu)\th'_\e(s)\right)\Psi_1=
-(A+\mu)^2\Psi_0^D(s,\mu,\e)\th'_\e(s)
\ln\mathsf{f}_\e(\th_\e(s)), \quad x\in\p\Om.
\end{equation}
Problem (\ref{2.7}), (\ref{2.29}) is solvable under suitable
choice of  $\L_1$. The solvability condition is deduced in a
standard way, by multiplying equation  (\ref{2.7}) by $\Psi_0$
and integrating by parts with employing the boundary condition
(\ref{2.29}). Bearing in mind the normalization for $\Psi_0$,
this condition implies formula (\ref{1.9}) for the leading term
of the asymptotics. It follows from Lemma~\ref{lm2.6}, the
assumption \ref{cnA2} and the definition of $\mathsf{f}_\e$
that:
\begin{equation}\label{2.104}
c_1 c_3/2\le\mathsf{f}_\e(\th)\le 1,
\end{equation}
what due to formula (\ref{1.9}) gives nonpositiveness of $\L_1$.
The maintained holomorphy on $\mu$ of $\L_1$ is an implication
of the corresponding properties of $\Psi_1$, boundedness of
$\th'_\e$ and $\mathsf{f}_\e(\th)$ and the estimate for the norm
$\|\Psi\|_{L_2(\p\Om)}$ by $\|\Psi\|_{H^1(\Om)}$.

The function $\Psi_1$ is defined up to an additive term
$C\Psi_0$, $C=\mathrm{const}$; we eliminate this arbitrariness
by assuming  $\Psi_1$ to be orthogonal  to $\Psi_0$ in
$L_2(\Om)$. The function $\ln\mathsf{f}_\e(\th_\e(s))$ is
smooth, that's why it is easy to show that $\Psi_1\in
C^\infty(\overline{\Om})$. Moreover,  the function $\Psi_1$ is
holomorphic on $\mu$ in $H^1(\Om)$ norm for each fixed value of
$\e$ \cite{Kt}. Using the simplicity of $\l_0$, one can
establish that coefficients of Taylor series in powers of $\mu$
for $\Psi_1$ are continuous as $\e\to0$, and $\Psi_1$ is
majorized by holomorphic on $\mu$ function independent on $\e$.

The constructing done allowed to determine first terms of
asymptotics expansions for  $\l_\e$ and $\psi_\e$ (formally, of
course). Now we should prove that the asymptotics constructed do
provide the asymptotics for $\l_\e$ and $\psi_\e$. In order to
make such a justification we need to prove first that the
asymptotics constructed satisfy to the perturbed problem up to
sufficiently small discrepancy. Exactly the proof of this
statement will be our aim in this step. To guarantee the
smallness of discrepancy needed we have to construct additional
terms in asymptotics expansions for  $\l_\e$ and $\psi_\e$.

We have to construct one more term in the outer expansion:
\begin{equation}
\psi_\e^{ex}(x,\mu)=\Psi_0(x,\mu,\e)+\e\Psi_1(x,\mu,\e)+\e^2
\Psi_2(x,\mu,\e).\label{2.61}
\end{equation}

In boundary layer it is should be constructed two additional
terms; as a result the boundary layer reads as follows:
\begin{equation}\label{2.34}
\psi_\e^{bl}(\xi,s,\mu)=\sum\limits_{i=1}^4\e^i
v_i(\xi,s,\mu,\e).
\end{equation}
With regard to the equality $w^{(j)}_{0,0}=0$ and additional
terms the inner expansion becomes:
\begin{equation}\label{2.35}
\psi_\e^{in,j}(\vs^j,s,\mu)=\sum\limits_{i=1}^3\e^i
w_{i,0}^{(j)}(\vs^j,s,\mu,\e)+\eta\sum\limits_{i=1}^4\e^i
w_{i,1}^{(j)}(\vs^j,s,\mu,\e).
\end{equation}

First terms of outer expansion (\ref{2.61}) are known, we just
need to determine the function $\Psi_2$. In what follows this
function will be employed only for matching of additional terms
of inner  expansion. Like before, this matching will affect only
the boundary condition of $\Psi_2$, hence, we have an
arbitrariness in choosing the equation for $\Psi_2$, since its
form does not influence very much on the estimate of
discrepancy. We choose the equation for $\Psi_2$ so that to
guarantee the solvability and to simplify the calculations. Both
these aims are achieved by the following choice:
\begin{equation}
(\D-1)\Psi_2=-\L_1\Psi_1,\quad x\in\Om. \label{2.62}
\end{equation}

Additional terms of boundary layer are defined as follows:
\begin{align*}
v_3=&\frac{\Psi^\nu_1}{2(\th'_\e)^2}\xi_2^2 \left(
\frac{\th''_\e}{\th'_\e}\frac{\p X}{\p\xi_1}+\mathsf{k}\frac{\p
X}{\p\xi_2}\right)-\frac{1}{\th'_\e}
\left(\frac{\Psi_1^\nu}{\th'_\e}\right)'\int\limits_{\xi_2}^{+\infty}
t\frac{\p}{\p\xi_1}X(\xi_1,t)\,\mathrm{d}t+
\\
{}&+\mathsf{a}\int\limits_{\xi_2}^{+\infty} t
X(\xi_1,t)\,\mathrm{d}t-\frac{1}{\th'_\e}\Psi_2^\nu X,
\\
v_4=&\frac{\Psi^\nu_2}{2(\th'_\e)^2}\xi_2^2 \left(
\frac{\th''_\e}{\th'_\e}\frac{\p X}{\p\xi_1}+\mathsf{k}\frac{\p
X}{\p\xi_2}\right)-\frac{1}{\th'_\e}
\left(\frac{\Psi_2^\nu}{\th'_\e}\right)'\int\limits_{\xi_2}^{+\infty}
t\frac{\p}{\p\xi_1}X(\xi_1,t)\,\mathrm{d}t,
\end{align*}
where $\mathsf{a}=\mathsf{a}(s,\mu,\e)$ is a some function that
will be determined below, $\Psi_2^\nu$ is a value of normal
derivation for $\Psi_2$ on the boundary $\p\Om$. It is easy to
check that $v_3, v_4\in\mathcal{V}_0$.

In order to match asymptotics expansions and to determine inner
expansion, one needs the following differentiable asymptotics
held as $\rho\to0$:
\begin{equation*}
\begin{aligned}
v_1(\xi,s,\mu,\e)=&-\frac{\Psi_0^\nu}{\th'_\e} \left(\ln\rho+\ln
2-\xi_2\right)+O(\rho^2),
\\
v_2(\xi,s,\mu,\e)=&-\frac{\Psi_1^\nu}{\th'_\e} \left(\ln\rho+\ln
2-\xi_2\right)+
\\
{}&+V_\e(\xi-\xi^{(j)},s,\Psi_0^\nu(s,\mu,\e))+O(\rho^2),
\\
v_3(\xi,s,\mu,\e)=&-\frac{\Psi_2^\nu}{\th'_\e} \left(\ln\rho+
\ln 2-\xi_2\right)-\frac{\zeta(3)}{4}\mathsf{a}(s,\mu,\e)+
\\
{}&+V_\e(\xi-\xi^{(j)},s,\Psi_1^\nu(s,\mu,\e))+O(\rho^2\ln\rho),
\\
v_4(\xi,s,\mu,\e)=&V_\e(\xi-\xi^{(j)},s,\Psi_2^\nu(s,\mu,\e))+
O(\rho^2),
\end{aligned}
\end{equation*}
where $\zeta(t)$ is Riemann zeta function, and it is indicated
\begin{align*}
V_\e(\xi,s,\Psi(s))=&\frac{\Psi(s)}{2(\th'_\e(s))^2}\frac{\xi_2^2}{|\xi|^2}
\left(\frac{\th''_\e(s)}{\th'_\e(s)}\xi_1+\mathsf{k}(s)\xi_2\right)+
\\
{}&+\frac{1}{2\th'_\e(s)}\left(\frac{\Psi(s)}{\th'_\e(s)}\right)'
\xi_1\left(\ln|\xi|+\ln 2-1\right).
\end{align*}
The coefficients of the outer expansion satisfy the
relationships
\begin{equation*}
\Psi_i=\Psi_i^D-\tau\Psi_i^\nu+O(\tau^2),\quad \tau\to0,\quad
i=0,1,2
\end{equation*}
in an neighbourhood of the boundary $\p\Om$. Rewriting now the
asymptotics of the functions  $v_i$ and $\Psi_j$ given above to
the variables  $\vs$ in view of the equality (\ref{2.1}) we
obtain that for
$\frac{1}{4}\eta^{1/4}<\rho<\frac{3}{4}\eta^{1/4}$
($\frac{1}{4}\eta^{-3/4}<|\vs|<\frac{3}{4}\eta^{-3/4}$) the
equality
\begin{align}
{}&
\begin{aligned}
\psi_\e^{ex}(x,\mu)+\psi_\e^{bl}(\xi,s,\mu)=&\sum\limits_{k=1}^3
\e^k W_{k,0}(\vs,s,\mu,\e)+
\\
{}&+\eta\sum\limits_{k=1}^4\e^k
W_{k,1}(\vs,s,\mu\,\e)+O(\eta^2|\vs|^2\ln|\vs|),
\end{aligned}\label{2.41}
\\
{}&
\begin{aligned}
{}& W_{2,0}=-\frac{\Psi_1^\nu}{\th'_\e}\left(\ln|\vs|+\ln
2\right)+\Psi_2^D+\frac{\Psi_2^\nu}{(A+\mu)\th'_\e},
\\
{}& W_{3,0}=-\frac{\Psi_2^\nu}{\th'_\e}\left(\ln|\vs|+\ln
2\right)-\frac{\zeta(3)}{4}\mathsf{a},
\end{aligned}\label{2.42}
\\
{}& W_{1,1}=-\frac{1}{2(A+\mu)\th'_\e}
\left(\frac{\Psi_0^\nu}{\th'_\e}\right)'\vs_1, \nonumber
\\
{}&W_{k,1}=V_\e(\vs,s,\Psi_{k-2}^\nu)-\frac{1}{2(A+\mu)\th'_\e}
\left(\frac{\Psi_{k-1}^\nu}{\th'_\e}\right)'\vs_1,\quad
k=2,3,\nonumber
\\
{}&W_{4,1}=V_\e(\vs,s,\Psi_2^\nu),\nonumber
\end{align}
holds. Here $\Psi_2^D=\Psi_2^D(s,\mu,\e)=\Psi_2(x,\mu,\e)$,
$x\in\p\Om$. Thus, the function $w_{k,i}^{(j)}$  should meet the
asymptotics
\begin{equation}\label{2.46}
w_{k,i}^{(j)}=W_{k,i}+o(|\vs|^{i}),\quad \vs\to\infty.
\end{equation}
Like before, the problems for the coefficients are deduced by
the substitution of (\ref{2.2}) and (\ref{2.35}) into
(\ref{1.1}), (\ref{1.2}) and writing out the coefficients of
leading powers of $\e$ and $\eta$:
\begin{equation}\label{2.47}
\begin{aligned}
\D_\vs w_{k,1}^{(j)}=&-\frac{\th''_\e}{(\th'_\e)^2}
\left(\frac{\p}{\p\vs_1}+ 2\vs_2\frac{\p^2}{\p\vs_1\p\vs_2}
\right) w_{k-1,0}^{(j)}-\\
{}&-\frac{\mathsf{k}}{\th'_\e}\left(\frac{\p}{\p\vs_2}-
2\vs_2\frac{\p^2}{\p\vs_1^2}\right)w_{k-1,0}^{(j)}-
\frac{2}{\th'_\e}\frac{\p^2}{\p\vs_1\p
s}w_{k-1,0}^{(j)},\quad\vs_2>0,
\\
w_{k,1}^{(j)}=&0,\quad \vs\in\g^1_j,\qquad
\frac{\p}{\p\vs_2}w_{k,1}^{(j)}=0, \quad\vs\in\G^1_j,
\end{aligned}
\end{equation}
$k=2,3,4$; and for  $w_{1,1}^{(j)}$ and $w_{k,0}^{(j)}$, $k=2,3$
we obtain the same problem (\ref{2.24}) as for $w_{1,0}^{(j)}$.
We define the functions $w_{k,0}^{(j)}$, $k=2,3$, as follows:
\begin{equation}\label{2.48}
w_{k,0}^{(j)}=-\frac{\Psi_{k-1}^\nu}{\th'_\e} Y^{(j)}.
\end{equation}
The belongings  $w_{k,0}^{(j)}\in\mathcal{W}$ take place. Now we
calculate the asymptotics for the functions $w_{k,0}^{(j)}$ (see
(\ref{2.26}), (\ref{2.48})) and compare them with the
asymptotics (\ref{2.46}), (\ref{2.42}). This procedure gives two
equalities:
\begin{align*}
\frac{\Psi_1^\nu(s,\mu,\e)}{\th'_\e(s)}\ln(\a^j(\e)+\b^j(\e))&=
\Psi_2^D(s,\mu,\e)+
\frac{\Psi_2^\nu(s,\mu,\e)}{(A+\mu)\th'_\e(s)},
\\
\frac{\Psi_2^\nu(s,\mu,\e)}{\th'_\e(s)}\ln(\a^j(\e)+\b^j(\e))&=
-\frac{\zeta(3)}{4}\mathsf{a}(s,\mu,\e).
\end{align*}
The former leads  us to a boundary condition for $\Psi_2$:
\begin{equation}\label{2.64}
\left(\frac{\p}{\p\nu}+(A+\mu)\th'_\e(s)\right)\Psi_2=
(A+\mu)\Psi_1^\nu(s,\mu,\e)\ln\mathsf{f}_\e(\th_\e(s)),\quad
x\in\p\Om,
\end{equation}
while the latter determines the function $\mathsf{a}$:
\begin{equation*}
\mathsf{a}(s,\mu,\e)=-\frac{4}{\zeta(3)\th'_\e(s)}
\Psi_2^\nu(s,\mu,\e)\ln \mathsf{f}_\e(\th_\e(s)).
\end{equation*}

Boundary value problem (\ref{2.62}), (\ref{2.64}) is uniquely
solvable. The right sides of the equation in (\ref{2.62}) and of
the boundary condition (\ref{2.64}) contain smooth on $x$ and
$s$ functions, thus, $\Psi_2\in C^\infty(\overline{\Om})$.

Now we return to the construction of the inner expansion. It is
easy to check that the function
\begin{equation*}
Y_1^{(j)}(\vs,\e)=\left(\a^j(\e)+\b^j(\e)\right)\mathrm{Re}\,
\sqrt{y^2-1}
\end{equation*}
belongs to $\mathcal{W}$ and is a solution of the boundary value
problem (\ref{2.24}) with the asymptotics
\begin{equation}\label{2.49}
Y_1^{(j)}(\vs,\e)=\vs_1+O(|\vs|^{-1}),\quad \vs\to\infty.
\end{equation}
In view of this asymptotics and other mentioned properties of
the function $Y_1^{(j)}$ the function $w_{1,1}^{(j)}$ is given
by
\begin{equation*}
w_{1,1}^{(j)}=-\frac{1}{2(A+\mu)\th'_\e}
\left(\frac{\Psi_0^\nu}{\th'_\e}\right)'Y_1^{(j)}.
\end{equation*}
By direct calculations one can establish that the function
\begin{equation*}
\widetilde{w}_{2,1}=\frac{\Psi_0^\nu}{2(\th'_\e)^2}\vs_2^2\left(
\frac{\th''_\e}{\th'_\e}\frac{\p}{\p\vs_1}+\mathsf{k}\frac{\p}{\p\vs_2}
\right)Y^{(j)}+\frac{1}{\th'_\e}
\left(\frac{\Psi_0^\nu}{\th'_\e}\right)'\vs_1 Y^{(j)},
\end{equation*}
belonging to $\mathcal{W}$ is a solution of the problem
(\ref{2.47}) and satisfies to the following asymptotics at
infinity ($\vs\to\infty$):
\begin{equation*}
\widetilde{w}^{(j)}_{2,1}=\frac{\Psi_0^\nu}{2(\th'_\e)^2}\frac{\vs_2^2}
{|\vs|^2}\left(\frac{\th''_\e}{\th'_\e}\vs_1+\mathsf{k}\vs_2\right)
+\frac{1}{\th'_\e}\left(\frac{\Psi_0^\nu}{\th'_\e}\right)'\vs_1
\left(\ln|\vs|+\ln 2-\ln(\a^j+\b^j)\right)+O(1).
\end{equation*}
To get the function $w_{2,1}^{(j)}$ needed we should add
harmonic function $Y_1^{(j)}$ with an suitable factor to
$\widetilde{w}_{2,1}^{(j)}$ so that the asymptotics of
$w_{2,1}^{(j)}$  to contain the needed coefficient of $\vs_1$.
Such a factor is a function:
\begin{equation*}
\frac{1}{\th'_\e}\left(\frac{\Psi_0^\nu}{\th'_\e}\right)'
(\ln(\a^j+\b^j)-1) -\frac{1}{2(A+\mu)\th'_\e}
\left(\frac{\Psi_1^\nu}{\th'_\e}\right)';
\end{equation*}
i.e.,
\begin{equation*}
w_{2,1}^{(j)}=\widetilde{w}^{(j)}_{2,1}+\frac{1}{\th'_\e}
\left(\left(\frac{\Psi_0^\nu}{\th'_\e}\right)'
\left(\ln(\a^j+\b^j)-1\right) -\frac{1}{2(A+\mu)}
\left(\frac{\Psi_1^\nu}{\th'_\e}\right)'\right)Y^{(j)}_1.
\end{equation*}
The functions $w_{3,1}^{(j)}$ and $w_{4,1}^{(j)}$ are determined
similarly:
\begin{equation*}
\begin{aligned}
w_{3,1}^{(j)}=&\frac{\Psi_1^\nu}{2(\th'_\e)^2}\vs_2^2\left(
\frac{\th''_\e}{\th'_\e}\frac{\p}{\p\vs_1}+
\mathsf{k}\frac{\p}{\p\vs_2} \right)Y^{(j)}+\frac{1}{\th'_\e}
\left(\frac{\Psi_1^\nu}{\th'_\e}\right)'\vs_1 Y^{(j)}+
\\
{}&+\frac{1}{\th'_\e}\left(\left(
\frac{\Psi_1^\nu}{\th'_\e}\right)'\left(\ln(\a^j+\b^j)-1\right)
-\frac{1}{2(A+\mu)}
\left(\frac{\Psi_2^\nu}{\th'_\e}\right)'\right) Y^{(j)}_1;
\\
w_{4,1}^{(j)}=&\frac{\Psi_2^\nu}{2(\th'_\e)^2}\vs_2^2\left(
\frac{\th''_\e}{\th'_\e}\frac{\p}{\p\vs_1}+
\mathsf{k}\frac{\p}{\p\vs_2} \right)Y^{(j)}+
\\
{}&+\frac{1}{\th'_\e} \left(\frac{\Psi_2^\nu}{\th'_\e}\right)'
\left(\vs_1 Y^{(j)}+
\left(\ln(\a^j+\b^j)-1\right)Y^{(j)}_1\right).
\end{aligned}
\end{equation*}
Clear, $w_{k,1}^{(j)}\in\mathcal{W}$. Employing asymptotics
(\ref{2.26}) and (\ref{2.49}), we see that as $\vs\to\infty$
\begin{equation}\label{2.54}
\begin{aligned}
w_{k,i}^{(j)}=W_{k,i}+O(|\vs|^{i-1}).
\end{aligned}
\end{equation}
The formal constructing of outer expansion (\ref{2.61}),
boundary layer  (\ref{2.34}) and inner expansion (\ref{2.35}) is
finished.

Next four auxiliary lemmas will be used in proving that the
eigenelements' asymptotics formally constructed is a formal
asymptotics solution of the perturbed problem.

We denote $\Om^{bl}=\{x: 0<\tau<c_0\}$ where $c_0$ is a some
small fixed number so that in a domain $\Om^{bl}$ the
coordinates $(s,\tau)$ are defined correctly and the function
$\mathsf{H}(s,\tau)$ has no zeroes. Throughout in what follows
we will employ the symbol $C$ for nonspecific constants
independent on $\e$ and $\mu$.

\begin{lemma}\label{lm2.3}
Suppose $F=F(x,\mu,\e)$ and $f=f(s,\mu,\e)$ are infinitely
differentiable on $x$ and $s$ functions,
$\mathsf{a}_0=\mathsf{a}_0(\mu,\e)$ is a some function uniformly
bounded on $\e$ and $\mu$, and norms $\|f\|_{C(\p\Om)}$,
$\|F\|_{C(\overline{\Om})}$ and $\|F\|_{C^k(\overline{\Om_1})}$,
$\Om_1\Subset\Om$ is an arbitrary subdomain, $k\in\mathbb{N}$,
are uniformly bounded on $\e$ and $\mu$. If the boundary value
problem
\begin{equation}\label{2.67}
(\D+\mathsf{a}_0)u=F,\quad x\in\Om,\qquad
\left(\frac{\p}{\p\nu}+(A+\mu)\th'_\e\right)u=f,\quad x\in\p\Om.
\end{equation}
has a solution whose $H^1(\Om)$ norm is uniformly bounded on
$\e$ and $\mu$, then for this solution uniform on $\e$ and $\mu$
estimates hold:
\begin{align*}
{}&\|u\|_{C^1(\Om)}\le
C(\|f\|_{C^1(\p\Om)}+1),\quad\|u\|_{C^k(\overline{\Om}_1)}\le C,
\\
{}&\|u^\nu\|_{C(\p\Om)}\le C(A+\mu+\|f\|_{C(\p\Om)}),
\\
{}&\|u^\nu\|_{C^i(\p\Om)}\le C(\|f\|_{C^i(\p\Om)}+1),\quad
i=1,2,
\\
{}&\|u^\nu\|_{C^3(\p\Om)}\le
C(\|F\|_{C^1(\p\Om)}+\|f\|_{C^3(\p\Om)}+1),
\end{align*}
where $u^\nu=u^\nu(s,\mu,\e)=\frac{\displaystyle\p
u}{\displaystyle\p\nu}(x,\mu,\e)$, $x\in\p\Om$,
$k\in\mathbb{Z}$.
\end{lemma}
\PF{Proof.} The smoothness $f$ and $F$ allows us to maintain
that the solution $u$ for the problem (\ref{2.67}) is infinitely
differentiable on $x$. Moreover, by the boundedness of the norm
$\|u\|_{H^1(\Om)}$ for each couple of strongly inner subdomains
$\Om_1\Subset\Om_2\Subset\Om$ we have
\begin{equation*}
\|u\|_{H^{k+2}(\Om_1)}\le C\left(\|F\|_{H^k(\Om_2)}+1\right)\le
C,\quad k\in\mathbb{N}.
\end{equation*}
Last inequalities and embedding theorems ($C^k\subset H^{k+2}$)
imply that estimates
\begin{equation}\label{2.70}
\|u\|_{C^k(\overline{\Om}_1)}\le C.
\end{equation}
take place. In a domain $\Om^{bl}$ we change the function $u$:
\begin{equation*}
v(x,\mu,\e)=u(x,\mu,\e)\mathrm{e}^{-(A+\mu)\th'_\e(s)\tau}
(\mathsf{a}_1-\mathsf{a}_2\tau^2),
\end{equation*}
where $\mathsf{a}_1$ and $\mathsf{a}_2$ are some positive
numbers. Owing to  (\ref{2.67}) the function $v$ is a solution
of the problem:
\begin{gather*}
\left(\D_x+\mathsf{a}_2\frac{\p}{\p x_1}+
\mathsf{a}_4\frac{\p}{\p x_2}+\mathsf{a}_5\right)v\equiv L_1
v=\widetilde{F},\quad x\in\Om^{bl},
\\
v=\mathsf{a}_6,\quad \tau=c_0,\qquad\frac{\p
v}{\p\tau}=-\mathsf{a}_1 f,\quad \tau=0,
\end{gather*}
where $\widetilde{F}=\mathrm{e}^{-(A+\mu)\th'_\e(s)\tau}
(\mathsf{a}_1-\mathsf{a}_2\tau^2)F$, the functions
$\mathsf{a}_i=\mathsf{a}_i(x,\mu,\e)$, $i=3,4,5$,
$\mathsf{a}_6=\mathsf{a}_6(s,\mu,\e)$ are  smooth on spatial
variables and holomorphic on $\mu$, and also
$\|\mathsf{a}_6\|_{C(\{\tau=c_0\})}\le C$ (see (\ref{2.70})),
$\|\mathsf{a}_i\|_{C^1(\overline{\Om}^{bl})}\le C$. The
functions $\mathsf{a}_i$, $i=3,4,5$, can be easily got
explicitly, we don't adduce here these explicit formulae and
just note that by a suitable choice of constant $\mathsf{a}_2$,
$\mathsf{a}_3$ and constant  $c_0$ from the definition of
$\Om^{bl}$ one can always achieve inequalities
$\mathsf{a}_1-\mathsf{a}_2 c_0^2>0$, $\mathsf{a}_5\le C<0$ for
$x\in\overline{\Om}^{bl}$. Then for the operator $L_1$ and each
function $V\in C^2(\overline{\Om}^{bl})$ the statement holds: if
\begin{equation*}
L_1 V<0,\quad x\in\overline{\Om}^{bl},\qquad
V>0,\quad\tau=c_0,\qquad \frac{\p V}{\p\tau}<0,\quad\tau=0,
\end{equation*}
then $V>0$. Indeed, assuming a contrary, at a point of minimum
in $\overline{\Om}^{bl}$ the function $V$ is negative, $\D
V\ge0$, $\nabla_x V=0$, i.e., at this point $L_1 V>0$. Clear,
this point of minimum lies strongly inside the domain
$\Om^{bl}$; the contradiction obtained proves the statement. Now
we take a ''barrier'' function $(\mathsf{a}_7-\mathsf{a}_8
\tau-\mathsf{a}_9\tau^2)$, $\mathsf{a}_7$,
$\mathsf{a}_8$,$\mathsf{a}_9$  are positive constant, and apply
this statement to the functions $V=(\mathsf{a}_7-\mathsf{a}_8
\tau-\mathsf{a}_9\tau^2)\pm v$, each time choosing the constants
$\mathsf{a}_i$ in a suitable way. As a result we have an
estimate
\begin{align*}
\|u\|_{C(\overline{\Om}^{bl})}&\le
C\|v\|_{C(\overline{\Om}^{bl})}\le C\left(
\|\widetilde{F}\|_{C(\overline{\Om}^{bl})}+\|f\|_{C(\p\Om)}+
\|\mathsf{a}_6\|_{C(\{\tau=c_0\})} \right)\le
\\
{}&\le C\left( \|F\|_{C(\overline{\Om}^{bl})}+\|f\|_{C(\p\Om)}+
\|u\|_{C(\Om\backslash\Om^{bl})}\right)\le C.
\end{align*}
Combining last inequality with (\ref{2.70}), we finally get
\begin{equation}\label{2.71}
\|u\|_{C(\overline{\Om})}\le C.
\end{equation}
In \cite[Chapter 3, \S\ 3, Theorem 3.1]{Ld} the estimate is
given, by that, dividing the equation and boundary condition in
(\ref{2.67}) to sufficiently great fixed number and taking into
account the smoothness $u$, we obtain:
\begin{equation}\label{2.72}
\|u\|_{C^2(\overline{\Om})}\le
C\left(\|F\|_{C(\overline{\Om})}+\|f\|_{C^1(\p\Om)}+
\|u\|_{C(\overline{\Om})}\right).
\end{equation}
It follows from (\ref{2.71}) and (\ref{2.72}) that
\begin{equation}\label{2.73}
\|u\|_{C^2(\overline{\Om})}\le
C\left(\|f\|_{C^1(\p\Om)}+1\right),
\end{equation}
what, in particular, gives needed estimate for
$\|u\|_{C^1(\overline{\Om})}$. Due to boundary condition
$u^\nu=-(A+\mu)\th'_\e u+f$, hence, using (\ref{2.71}) and
(\ref{2.73}) and bearing in mind the boundedness of
$\|\th'_\e\|_{C^2(\p\Om)}$, we derive the estimate for the
quantities $\|u^\nu\|_{C^i(\p\Om)}$, $i=0,1,2$, given in the
statement of the lemma. Let us estimate
$\|u^\nu\|_{C^3(\p\Om)}$. For $x\in\Om^{bl}$ we differentiate
the problem  (\ref{2.67}) on $s$. Then we have, that the
function $U=\frac{\p u}{\p s}$ is a solution of the boundary
value problem:
\begin{align*}
{}&\left(\D_x+\frac{\p}{\p s
}\left(\mathsf{H}^{-2}\right)'\frac{\p}{\p s} +
\mathsf{a}_0\right)U=\frac{\p F}{\p
s}-\frac{\mathsf{k}'}{\mathsf{H}^2}\frac{\p u}{\p\tau}\equiv F_1
,\quad x\in\Om^{bl},
\\
{}&\left(\frac{\p}{\p\nu}+(A+\mu)\th'_\e\right)U=f'-(A+\mu)\th''_\e
U\equiv f_1,\quad x\in\p\Om,
\\
{}&\phantom{\Bigg(}\frac{\p U}{\p\tau}=\frac{\p^2 u}{\p
s\p\tau},\quad \tau=c.
\end{align*}
For such problem, leaning for \cite[Chapter 3, \S\ 3, Theorem
3.1]{Ld}, we can write the estimate of (\ref{2.72}) kind; here
it is of the form:
\begin{equation}\label{2.74}
\|U\|_{C^2(\overline{\Om}^{bl})}\le
C\left(\|F_1\|_{C(\overline{\Om}^{bl})}+\|f_1\|_{C^1(\p\Om)}+
\left\|\frac{\p^2 u}{\p
s\p\tau}\right\|_{C^1(\{\tau=c_0\})}+\|U\|_{C(\overline{\Om}^{bl})}\right).
\end{equation}
The quantity $\left\|\frac{\p^2 u}{\p
s\p\tau}\right\|_{C^1(\{\tau=c_0\})}$ is estimated above by some
constant $C$ due to (\ref{2.70}); the sum of other three
summands can be estimated by (\ref{2.73}):
\begin{equation*}
\|F_1\|_{C(\overline{\Om}^{bl})}+\|f_1\|_{C^1(\p\Om)}+\|U\|_{C(\overline{\Om}^{bl})}
\le C\left(\|F\|_{C^1(\overline{\Om}^{bl})}
+\|f\|_{C^2(\p\Om)}+1\right).
\end{equation*}
Substituting the estimate obtained into (\ref{2.74}), we arrive
at the inequality
\begin{equation*}
\left\|\frac{\p u}{\p s}\right\|_{C^2(\overline{\Om}^{bl})}\le
C\left(\|F\|_{C^1(\overline{\Om})} +\|f\|_{C^2(\p\Om)}+1\right),
\end{equation*}
from what, the equality $u^\nu=-(A+\mu)\th'_\e u+f$ and the
boundedness of $\|\th'_\e\|_{C^3(\p\Om)}$  the estimate for
$\|u^\nu\|_{C^3(\p\Om)}$ follows. The proof is complete.

\begin{lemma}\label{lm2.1} The functions $\Psi_1$ and
$\L_1$ are represented in the form:
\begin{equation}\label{2.75}
\Psi_1(x,\mu,\e)=(A+\mu)^2\widetilde{\Psi}_1(x,\mu,\e), \quad
\L_1(\mu,\e)=(A+\mu)^2\widetilde{\L}_1(\mu,\e),
\end{equation}
where $\widetilde{\Psi}_1$ is infinitely differentiable on $x$,
$\widetilde{\Psi}_1$ and $\widetilde{\L}_1$ are holomorphic on
$\mu$ for each fixed value of $\e$. The uniform on $\e$ and
$\mu$ estimates ($i=1,2,3$)
\begin{equation}\label{2.76}
\begin{aligned}
{}&|\L_0|\le C,\phantom{(A+\mu)^2}\quad \|\Psi_0\|_{H^1(\Om)}\le
C,\phantom{(A+\mu)\|}\quad \|\Psi_0^\nu\|_{C^3(\p\Om)}\le
C(A+\mu),
\\
{}&|\L_1|\le C(A+\mu)^2,\quad \|\Psi_1\|_{H^1(\Om)}\le
C(A+\mu)^2, \quad \|\Psi_2\|_{H^1(\Om)}\le C(A+\mu)^3,
\\
{}&\|\Psi_1^\nu \|_{C(\p\Om)}\le C(A+\mu)^2, \quad
 \|\Psi_1^\nu \|_{C^i(\p\Om)}\le
C(A+\mu)^2(\e^{-i}\d^*(\e)+1),
\\
{}&\|\Psi_2^\nu\|_{C(\p\Om)}\le
C(A+\mu)^3,\quad\|\Psi_2^\nu\|_{C^i(\p\Om)}\le
C(A+\mu)^3(\e^{-i}\d^*(\e)+1).\phantom{+\mu)^2\scriptstyle\p}
\end{aligned}
\end{equation}
hold true.
\end{lemma}
\PF{Proof.} The proof of representations (\ref{2.75}) is very
simple. Indeed, the representation for $\L_1$ is a direct
implication of (\ref{1.9}). Employing this representation and
presence of the factor $(A+\mu)^2$ in the boundary condition
(\ref{2.29}), we arrive at the needed representation for
$\Psi_1$.

The proof of the estimates for $\L_0$ and $\Psi_0$ from
(\ref{2.76}) is elementary. The boundedness of $\L_0$ follows
from the convergence $\L_0\to\l_0$. Since
$\|\Psi_0\|_{L_2(\Om)}=1$, multiplying equation (\ref{2.6}) by
$\Psi_0$ and integrating once by parts, owing to boundedness of
$\L_0$ and $\|\th'_\e\|_{C(\p\Om)}$ we get the needed estimate
for the norm $\|\Psi_0\|_{H^1(\Om)}$. Now, applying
Lemma~\ref{lm2.3} to the problem for the function $\Psi_0$, we
obtain the estimate for $\|\Psi_0^\nu\|_{C^3(\p\Om)}$, and also,
$\|\Psi_0\|_{C^1(\overline{\Om})}\le C$,
$\|\Psi_0\|_{C^k(\overline{\Om}_1)}\le C$ for each subdomain
$\Om_1\Subset\Om$.

The estimate for  $\L_1$ arises from the proven estimates for
$\Psi_0$, boundedness of the function $\th'_\e$ and
$\mathsf{f}_\e(\th_\e)$ and the formula (\ref{1.9}).

We  prove the inequalities for  $\Psi_1$ and $\Psi_2$ from
(\ref{2.76}) on the basis of Lemma~\ref{lm2.3}, too. Since
$\Psi_1$ is orthogonal to $\Psi_0$ in $L_2(\Om)$, and the
quantities $\L_0$ and $\L_1$ are bounded, an uniform estimate
\begin{equation*}
\|\Psi_1\|_{H^1(\Om)}\le C(A+\mu)^2\left(\|\Psi_0\|_{L_2(\Om)}+
\|\Psi_0^D\th'_\e\ln
\mathsf{f}_\e(\th_\e)\|_{H^1(\Om)}\right)\le C(A+\mu)^2
\end{equation*}
takes place. The right side of the equation (\ref{2.7}) and the
boundary condition (\ref{2.29}) obey to hypothesis of
Lemma~\ref{lm2.3}. We also note that the estimating of the
derivatives of boundary condition  (\ref{2.29}) actually reduces
to the estimating of derivatives of (bounded) function
$\mathsf{f}_\e(\th_\e(s))$, since the derivatives of  $\th'_\e$
are estimated by assumption \ref{cnA0}, while the estimates for
the derivatives of $\Psi_0^D$ are deduced from the estimates for
$\Psi_0^\nu$  proved  already and the equality
$\Psi_0^\nu=(A+\mu)\th'_\e\Psi_0^D$. Obviously, the derivatives
$\mathsf{f}_\e(\th_\e)$ are estimated as follows
\begin{equation*}
\|\mathsf{f}_\e(\th_\e(s))\|_{C^i(\p\Om)}\le
C\left(\e^{-i}\d^*(\e)+1\right),\quad i=1,2,3.
\end{equation*}
Using this obvious fact and applying Lemma~\ref{lm2.3} to the
problem for  $\t\Psi_1$, we arrive at the estimates for $\Psi_1$
from (\ref{2.76}). Besides, Lemma~\ref{lm2.3} implies
inequalities
\begin{align*}
\|\Psi_1\|_{C^1(\overline{\Om})}\le
C(A+\mu)^2(\e^{-1}\d^*(\e)+1),\quad
\|\Psi_1\|_{C^k(\overline{\Om}_1)}\le C,
\end{align*}
for all $k\in\mathbb{Z}_+$ and all $\Om_1\Subset\Om$. By obvious
inequality
\begin{equation*}
\|\Psi_2\|_{H^1(\Om)}\le
C\left(|\L_1|\|\Psi_1\|_{L_2(\Om)}+(A+\mu)
\|\Psi_1^\nu\|_{L_2(\p\Om)}\right),
\end{equation*}
and estimates for $\Psi_1$ and $\L_1$ proved already we get the
needed estimate for the norm $\|\Psi_2\|_{H^1(\Om)}$.
Representing $\Psi_2$ as $\Psi_2=(A+\mu)^3\t\Psi_2$ and applying
Lemma~\ref{lm2.3} to $\t\Psi_2$, we obtain other estimates for
$\Psi_2$ from (\ref{2.76}). The proof is complete.

We denote $\h\l_\e=\L_0(\mu,\e)+\e\L_1(\mu,\e)$,
$\Pi^{(j)}=\{\xi: |\xi_1-\pi j|<\pi/2, \xi_2>0\}$,
$\Pi^{(j)}_\eta=\Pi^{(j)}\cap\{\xi:
4|\xi-\xi^{(j)}|>\eta^{1/4}\}$, $\Om_\eta^{bl}=\Om^{bl}\cap\{x:
\xi\in\Pi^{(j)}_\eta, j=0,\ldots, N-1\}$.

\begin{lemma}\label{lm2.2}
For boundary layer (\ref{2.34}) uniform estimates
\begin{align*}
{}& \|\psi_\e^{bl}\|_{L_2(\Om^{bl})}\le C\e^{3/2}(A+\mu),\quad
\|\psi_\e^{bl}-\e v_1-\e^2(v_2-\t v_2)\|_{H^1(\Om^{bl}_\eta)}\le
C\e^{3/2}(A+\mu)^2,
\\
{}& \|(\D_x+\widehat{\l}_\e)\psi_\e^{bl}\|_{L_2(\Om^{bl})}\le
C\left(\e^{3/2}(A+\mu)+\e^{1/2}\d^*(\e)(A+\mu)^2\right)
\end{align*}
hold as  $\e\to0$
\end{lemma}

\PF{Proof.} Everywhere in the proof, not stressing it
additionally, we will employ the fact that
\begin{equation*}
\xi_2^{i+k}\frac{\p^i X}{\p\xi_2^i},\quad \xi_2^{i+k}\frac{\p^i
X}{\p\xi_1\p\xi_2^{i-1}}\in
L_2(\Pi^{(j)})\cap\mathcal{V}_0,\qquad i,k\in\mathbb{Z}, \quad
i,k\ge0.
\end{equation*}
An estimate
\begin{equation*}
\|\psi_\e^{bl}\|_{L_2(\Om^{bl})}\le
\e\left(\sum\limits_{j=0}^{N-1}
\|\psi_\e^{bl}\|^2_{L_2(\Pi^{(j)})}\right)^{1/2}
\end{equation*}
is true. We estimate the norms
$\|\psi_\e^{bl}\|_{L_2(\Pi^{(j)})}$ using explicit form of the
functions $v_i$ and estimates from Lemma~\ref{lm2.1}:
\begin{equation*}
\|\psi_\e^{bl}\|_{L_2(\Pi^{(j)})}\le C\e(A+\mu).
\end{equation*}
Last two estimates and the equality $N=2\e^{-1}$ yield first
inequality from the statement of the lemma. Second inequality
can be proved by analogy on the basis of  explicit form of the
functions $v_i$ and Lemma~\ref{lm2.1}. For the sake of brevity
we denote: $F_\e=(\D_x+\widehat{\l}_\e)\psi_\e^{bl}$. Employing
explicit form of the functions $v_i$ we calculate:
\begin{align*}
\mathsf{H}^3 F_\e=&\e\sum\limits_{i=1}^3\xi_2^i
\left(\mathsf{c}_{0,i}\frac{\p^i X}{\p\xi_2^i}+
\mathsf{c}_{1,i-1}\frac{\p^i X}{\p\xi_1\p\xi_2^{i-1}}
\right)+\e\mathsf{c}_{0,0} X+
\\
{}&+\e^2\mathsf{c}_{1,-1}\int\limits_{\xi_2}^{+\infty}
t\frac{\p}{\p\xi_1}X(\xi_1,t)\,\mathrm{d}t+
\e\mathsf{c}_{0,-1}\int\limits_{\xi_2}^{+\infty} t
X(\xi_1,t)\,\mathrm{d}t.
\end{align*}
Here $\mathsf{c}_{i,k}=\mathsf{c}_{i,k}(\xi_2,s,\e,\mu)$ are
polynomials on $\xi_2$ whose coefficients depends on other
variables and owing to Lemma~\ref{lm2.1} can be estimated above
by a quantity
 $C\left((A+\mu)+\e^{-1}\d^*(\e)(A+\mu)^2\right)$, where $C$ are
independent on $\e$, $\mu$, $s$. Using these estimates for
coefficients of polynomials  $\mathsf{c}_{i,k}$ and the form of
the function $\mathsf{H}^3 F_\e$, we persuade to
\begin{equation*}
\|\mathsf{H}^3 F_\e\|_{L_2(\Pi^{(j)})}\le
C\left(\e(A+\mu)+\d^*(\e)(A+\mu)^2\right),
\end{equation*}
where $C$ is independent on $\e$, $\mu$, $s$ and $j$. Since for
$x\in\overline{\Om}^{bl}$ the function $\mathsf{H}$ does not
vanish it follows that
\begin{equation*}
\| F_\e\|_{L_2(\Om^{bl})}\le C \|\mathsf{H}^3
F_\e\|_{L_2(\Om^{bl})} \le C \e \left(\sum\limits_{j=0}^{N-1}
\|\mathsf{H}^3 F_\e\|^2_{L_2(\Pi^{(j)})}\right)^{1/2},
\end{equation*}
what with the estimates for the norms
$\|\mathsf{H}^3F_\e\|_{L_2(\Pi^{(j)})}$ obtained already gives
third inequality from the statement of Lemma. The proof is
complete.

We denote $\Om^{in}_j=\{x: 4\eta^{3/4}|\vs^j|<3\}$, $\Om^{mat}_j
=\{x: 1<4|\vs^j|\eta^{3/4}<3, j=0,\ldots, N-1\}$.

By analogy with Lemma~\ref{lm2.2} one can establish the validity
of following statement.
\begin{lemma}\label{lm2.4}
For the  inner expansions (\ref{2.35}) uniform on $\e$, $\mu$
and $\eta$ estimates
\begin{align*}
{}&\|\psi_\e^{in,j}\|_{L_2(\Om^{in}_j)}\le C\eta^{1/5}, \quad
 \|(\D_x+\widehat{\l}_\e)\psi_\e^{in,j}\|_{L_2(\Om^{in}_j)}\le
C\eta^{1/5},
\\
{}&  \|\psi_\e^{in,j}-\e
w^{(j)}_{1,0}-\e^2w^{(j)}_{2,0}\|_{H^1(\Om^{in}_j)}\le
C\e^2(A+\mu)^{5/2}
\end{align*}
take place as  $\e\to0$.
\end{lemma}

\noindent Let
\begin{align*}
\h\psi_\e(x)=\left(\psi^{ex}_\e(x,\mu)+\chi(\tau/
c_0)\psi_\e^{bl}(\xi,s,\mu)\right)\chi_\e(x)+
\sum\limits_{j=0}^{N-1}\chi(|\vs^j|\eta^{3/4})
\psi_\e^{in,j}(\vs^j,s,\mu),
\end{align*}
where $\psi_\e^{ex}$, $\psi_\e^{bl}$ and $\psi_\e^{in,j}$ are
from (\ref{2.61})--(\ref{2.35}),
\begin{equation*}
\chi_\e(x)=1-\sum\limits_{j=0}^{N-1}\chi(|\vs^j|\eta^{3/4}).
\end{equation*}
In next statement we will prove that formally constructed
asymptotics $\h\l_\e$ and  $\h\psi_\e$ are formal asymptotics
solutions of the perturbed problem.
\begin{lemma}\label{lm2.5}
The functions $\h\psi_\e\in
C^\infty\left(\Om\cup\g_\e\cup\G_\e\right)\cap H^1(\Om)$ and
$\h\l_\e$ satisfy boundary value problem
\begin{equation}\label{2.79}
-\D u_\e=\l u_\e+f,\quad x\in\Om,\qquad u_\e=0,\quad
x\in\g_\e,\qquad\frac{\p u_\e}{\p\nu}=0,\quad x\in\G_\e,
\end{equation}
with $u_\e=\h\psi_\e$, $\l=\h\l_\e$ and $f=f_\e$, where for
$f_\e$ the uniform estimate holds:
\begin{equation}
\|f_\e\|_{L_2(\Om)}\le
C\left(\e^{3/2}(A+\mu(\e))+\e^{1/2}\d^*(\e)
(A+\mu(\e))^2\right).\label{2.96}
\end{equation}
The function $\h\l_\e$ converges to $\l_0$ as $\e\to0$, and for
$\h\psi_\e$ the relationship
$\|\h\psi_\e-\Psi_0\|_{L_2(\Om)}=o(1)$ holds true.
\end{lemma}
\PF{Proof.} The maintained smoothness of the function
$\h\psi_\e$ is obvious. Convergence of  $\h\l_\e$ to $\l_0$
follows from Lemmas~\ref{lm1.1}~and~\ref{lm2.1}. The
relationship $\|\h\psi_\e-\Psi_0\|_{L_2(\Om)}=o(1)$ is a direct
implication of Lemmas~\ref{lm2.1}-\ref{lm2.4}. Let us check the
boundary conditions from (\ref{2.79}). Vanishing of the function
$\h\psi_\e$ on $\g_\e$ arises from vanishing of  $\chi_\e$ on
$\g_\e$ and of $\psi_\e^{in,j}$ on $\g^1_j$. It is easy to check
that for $x\in\G_\e$
\begin{align*}
{}&\frac{\p\h\psi_\e}{\p\nu}=\left(\frac{\p\psi_\e^{ex}}{\p\nu}-
\frac{\th'_\e}{\e}\frac{\p\psi_\e^{bl}}{\p\xi_2}\bigg|_{\xi\in\G^0}
\right)\chi_\e(x)-\frac{\th'_\e}{\e\eta}\sum\limits_{j=0}^{N-1}
\chi(|\vs^j|\eta^{3/4})\frac{\p\psi_\e^{in,j}}{\p\vs_2}
\bigg|_{\vs\in\G^1_j}=
\\
{}&=
\chi_\e(x)\sum\limits_{i=0}^2\e^i\left(\Psi_i^\nu-\th'_\e\frac{\p
v_{i+1}}{\p\xi_2}\bigg|_{\xi\in\G^0}\right)=0.
\end{align*}
Let us estimate the function $f_\e$. Clear, it is of the form:
\begin{align*}
{}&f_\e=-(\D_x+\h\l_\e)\h\psi_\e=-\sum\limits_{i=1}^5
f^{(i)}_\e,
\\
{}&f_\e^{(1)}=\chi_\e(\D_x+\h\l_\e)\psi_\e^{ex},
\\
{}&f_\e^{(2)}=\chi_\e\chi(\tau/c_0)(\D_x+\h\l_\e)\psi_\e^{bl},
\\
{}&f_\e^{(3)}=\chi_\e\left( 2\left(\nabla_x\psi_\e^{bl},
\nabla_x
\chi(\tau/c_0)\right)+\psi_\e^{bl}\D_x\chi(\tau/c_0)\right),
\\
{}&f_\e^{(4)}=\sum\limits_{j=0}^{N-1}\chi(|\vs^j|\eta^{3/4})
(\D_x+\h\l_\e)\psi_\e^{in,j},
\\
{}&f_\e^{(5)}=\sum\limits_{j=0}^{N-1}\left(
2\left(\nabla_x\psi_\e^{mat,j}, \nabla_x \chi(|\vs^j|\eta^{3/4})
\right)+\psi_\e^{mat,j}\D_x\chi(|\vs^j|\eta^{3/4})\right),
\\
{}&\psi_\e^{mat,j}=\psi_\e^{in,j}-\psi_\e^{ex}-\psi_\e^{bl}.
\end{align*}
Employing equations (\ref{2.6}), (\ref{2.7}) and (\ref{2.62}) we
see that
\begin{equation*}
(\D_x+\h\l_\e)\psi_\e^{ex}=\e^2(\L_0+1+\e\L_1)\Psi_2,
\end{equation*}
thus, by Lemma~\ref{lm2.1},
\begin{equation*}
\|f_\e^{(1)}\|_{L_2(\Om)}\le C\e^2(A+\mu)^3.
\end{equation*}
The function $f_\e^{(2)}$ is estimated by Lemma~\ref{lm2.2}:
\begin{equation*}
\|f_\e^{(2)}\|_{L_2(\Om)}\le \|f_\e^{(2)}\|_{L_2(\Om^{bl})}\le
C\left(\e^{3/2}(A+\mu)+\e^{1/2}\d^*(\e)(A+\mu)^2\right).
\end{equation*}
The functions $v_i$ and, therefore, $\psi_\e^{bl}$ decay
exponentially, and integrating of $f_\e^{(3)}$ over $\Om$ due to
definition $\chi$ actually reduces to integrating over a domain
$\{x: \frac{c_0\th'_\e}{4\e}\le\xi_2\le \frac{3
c_0\th'_\e}{4\e}\}$, thus
\begin{equation*}
\|f_\e^{(3)}\|_{L_2(\Om)}\le
C(A+\mu)\mathrm{e}^{-1/\e^{\mathsf{b}}},
\end{equation*}
where $\mathsf{b}>0$ is a some fixed number. Next,  we estimate
the function $f_\e^{(4)}$ on the basis of Lemma~\ref{lm2.4}:
\begin{equation*}
\|f_\e^{(4)}\|_{L_2(\Om)}\le C\sum\limits_{j=0}^{N-1}
\|(\D_x+\h\l_\e)\psi_\e^{in,j}\|_{L_2(\Om^{in}_j)}\le
C\eta^{1/6}.
\end{equation*}
By the matching carried out (see (\ref{2.41}), (\ref{2.54})) the
function $\psi_\e^{mat,j}$ for $\eta^{-3/4}\le4|\vs^j|\le
3\eta^{-3/4}$ has a differentiable asymptotics:
\begin{equation}\label{2.85}
\psi_\e^{mat,j}=O(\eta^2|\vs|^2|\ln|\vs||+\e|\vs|^{-1}+\e\eta),
\end{equation}
using that we estimate $f_\e^{(5)}$:
\begin{equation*}
\|f_\e^{(5)}\|_{L_2(\Om)}\le C\sum\limits_{j=0}^{N-1}
\|f_\e^{(5)}\|_{L_2(\Om^{mat}_j)}\le C\eta^{1/5}.
\end{equation*}
Collecting the estimates obtained for $f^{(i)}_\e$  we arrive at
the statement of the lemma. The proof is complete.

Now we proceed to the justification of the asymptotics. By
analogy with \cite{AA}, \cite{DU}, \cite{BDU}, \cite{BMs},
\cite{MYu} it can be shown that for $\l$ close to $p$-multiply
limiting eigenvalue  $\l_0$, for the solution of the problem
(\ref{2.79}) with $f\in L_2(\Om)$ the representation
\begin{equation}\label{2.81}
u_\e=\sum\limits_{k=q}^{q+p-1}\frac{\psi_\e^{k}}{\l_\e^{k}-\l}
\int\limits_{\Om}f\psi_\e^{k}\,\mathrm{d}x+\t u_\e,
\end{equation}
takes place, where, recall, $\l_\e^{k}$, $k=q,\ldots,q+p-1$, are
perturbed eigenvalues converging to  $\l_0$, $\psi_\e^{k}$ are
associated orthonormalized in $L_2(\Om)$ eigenfunctions, $\t
u_\e$ is a holomorphic on $\l$ in $H^1(\Om)$-norm function
orthogonal to all $\psi_\e^k$, $k=q,\ldots,q+p-1$, in
$L_2(\Om)$; for the function $\t u_\e$ a uniform on $\e$, $\mu$,
$\l$ and $f$ estimate
\begin{equation}\label{2.82}
\|\t u_\e\|_{H^1(\Om)}\le C\|f\|_{L_2(\Om)}
\end{equation}
is valid. In our case $\l_0$ is a simple eigenvalue. We set
$u_\e=\h\psi_\e$, $\l=\h\l_\e$ and $f=f_\e$. Then from
Lemma~\ref{lm2.5} and (\ref{2.81}), (\ref{2.82}) we obtain
\begin{equation}\label{2.83}
\begin{aligned}
{}&\h\psi_\e=\frac{\psi_\e}{\l_\e-\h\l_\e}
\int\limits_{\Om}f_\e\psi_\e\,\mathrm{d}x+\t u_\e,
\\
{}&\|\t u_\e\|_{H^1(\Om)}\le
C\left(\e^{3/2}(A+\mu)+\e^{1/2}\d^*(\e)(A+\mu)^2\right).
\end{aligned}
\end{equation}
Since $\|\h\psi_\e-\t
u_\e\|_{L_2(\Om)}=\|\Psi_0\|_{L_2(\Om)}(1+o(1))=1+o(1)$ (see
Lemma~\ref{lm2.5}), by (\ref{2.83}) we have:
\begin{equation*}
C\le \frac{\|f_\e\|_{L_2(\Om)}}{|\l_\e-\h\l_\e|}\; \Rightarrow\;
|\l_\e-\h\l_\e|\le
C\left(\e^{3/2}(A+\mu)+\e^{1/2}\d^*(\e)(A+\mu)^2\right).
\end{equation*}
The Corollary~\ref{lm2.6cor1} of Lemma~\ref{lm2.6} allows us to
replace the function  $\d^*(\e)$ by $o(\e^{1/2}(A+\mu)^{-1})$ in
last estimate, i.e., the asymptotics (\ref{1.8}) is correct.

In general, the case of $p$-multiply eigenvalue
$\l_0=\l_0^q=\ldots=\l_0^{q+p-1}$ is proved similarly. In
constructing the multiplicity of  $\l_0$ becomes apparent in the
fact that by same scheme we simultaneously construct several
asymptotics corresponding to eigenvalues $\l_\e^{k}$ converging
to  $\l_0$. Besides, the multiplicity becomes apparent in
solving the problem (\ref{2.6}), (\ref{2.25}), that has several
eigenvalues $\L_0^k$, converging to  $\L_0$, and, of course,
several eigenfunctions $\Psi_0^k$. These eigenfunctions are
assumed to meet Lemma~\ref{lm1.1}. In particular, the
orthogonality of $\Psi_0^k$ in $L_2(\p\Om)$ weighted by
$\th'_\e$ is exactly a solvability condition of the problems for
$\Psi_1^k$, those again are chosen to be orthogonal to
$\Psi_0^k$.  All other arguments of formal constructing hold
true, including Lemmas~\ref{lm2.1}-\ref{lm2.4}. Thus, as a
result of formal constructing we have functions $\h\l_\e^k$ and
$\h\psi_\e^k$, $k=q,\ldots,p+q-1$, those are defined as
$\h\l_\e$ and $\h\psi_\e$ with replacement $\L_0$ by $\L_0^k$
and $\Psi_0$ by $\Psi_0^k$. For $\h\l_\e^k$ and $\h\psi_\e^k$
Lemma~\ref{lm2.5} is valid. By $f_\e^k$ we denote right sides of
equations from (\ref{2.79}) with $u_\e=\h\psi_\e^k$,
$\l=\h\l_\e^k$.

Now we apply the representation (\ref{2.81}) to the functions
$\h\psi_\e^k$:
\begin{equation}\label{2.88}
\begin{aligned}
{}&\h\psi^k_\e=
\sum\limits_{i=q}^{q+p-1}\mathsf{b}^\e_{ki}\psi_\e^{i}+\t
u^k_\e,\quad \mathsf{b}^\e_{ki}=\frac{1}{\l_\e^{i}-\h\l^k_\e}
\int\limits_{\Om}f_\e^k\psi_\e^{i}\,\mathrm{d}x,
\\
{}&\|\t u^k_\e\|_{H^1(\Om)}\le C
\left(\e^{3/2}(A+\mu)+\e^{1/2}\d^*(\e)(A+\mu)^2\right).
\end{aligned}
\end{equation}
Last estimate for $\t u_\e^k$ arises from (\ref{2.82}) and
Lemma~\ref{lm2.5}. By (\ref{2.88}) and the orthogonality of $\t
u^k_\e$ to the functions $\psi_\e^k$ we get the assertions
\begin{equation}\label{2.101}
\mathsf{b}_{ki}^\e=\left(\h\psi_\e^k,\psi_\e^i\right)_{L_2(\Om)},
\end{equation}
those imply boundedness of the quantities $\mathsf{b}_{ki}^\e$.
Let us prove the asymptotics (\ref{1.8}) for the eigenvalues
$\l_\e^k$, $k=q,\ldots,q+p-1$. Assume a contrary, namely,
suppose there exists a subsequence  $\e_m$, on that for some of
eigenvalues $\l_\e^k$, $k=q,\ldots,q+p-1$, the asymptotics
(\ref{1.8}) are wrong, and  for $k=q,\ldots,q+p-1$, $i\in
I\not=\emptyset$
\begin{equation}\label{2.89}
|\l^i_{\e_m}-\h\l^k_{\e_m}|\ge
m(\e_m^{3/2}(A+\mu)+\e_m^{1/2}\d^*(\e_m)(A+\mu)^2),
\end{equation}
where $I\subseteq\{q,\ldots,q+p-1\}$ a subset of indices of
eigenvalues not satisfying to asymptotics (\ref{1.8}). By
estimate for the functions $f^k_\e$, the formulae for
$\mathsf{b}_{ki}^\e$ from (\ref{2.88}) and the inequalities
(\ref{2.89}) we deduce that
\begin{equation}\label{2.90}
\mathsf{b}^{\e_m}_{ki}\xrightarrow[m\to\infty]{}0,\quad
k=q,\ldots,q+p-1,\quad i\in I.
\end{equation}
Bearing in mind the boundedness $\mathsf{b}^{\e_m}_{ki}$ and
extracting a subsequence form $\e_m$ if it is needed, we assume
that $\mathsf{b}^{\e_m}_{ki}\to\mathsf{b}^0_{ki}$, where due to
(\ref{2.90}) the equalities $\mathsf{b}^0_{ki}=0$ are true for
$k=q,\ldots,q+p-1$, $k\in I$. By numbers
$\mathsf{b}^{\e_m}_{ki}$ we compose $p$ vectors
$\mathsf{b}^{\e_m}_{k}$ by a rule: as components of vector
$\mathsf{b}^{\e_m}_{k}$ we take consequently the numbers
$\mathsf{b}^{\e_m}_{ki}$, where index $i$ ranges in
$q,\ldots,q+p-1$ and does not takes values from the set $I$. In
a similar way we compose $p$ vectors $\mathsf{b}^0_{k}$ from
numbers $\mathsf{b}^0_{ki}$. The dimension of the vectors
composed are equal to $(p-|I|)<p$. Now multiply in $L_2(\Om)$
the representations (\ref{2.88}) for  $\h\psi_\e^k$ each to
other for all values of $k$ and take in account the equalities
$\|\h\psi_\e^k-\Psi_0^k\|_{L_2(\Om)}=o(1)$, the estimates for
$\t u^k_\e$ and orthonormality of the functions $\Psi_0^k$ and
$\psi_\e^k$. Then we get that
\begin{equation*}
(\mathsf{b}^0_k,\mathsf{b}^0_j)_{L_2(\Om)}=
\lim\limits_{m\to\infty}
(\mathsf{b}^{\e_m}_k,\mathsf{b}^{\e_m}_j)_{L_2(\Om)}=\d_{kj},
\quad k,j=q,\ldots,q+p-1,
\end{equation*}
where $\d_{kj}$ is a Kronecker delta, i.e., $\mathsf{b}^0_k$
make up a system of  $p$ orthonormalized $(p-|I|)$-dimensional
vectors. The contradiction obtained proves the estimates
\begin{equation*}
|\l_\e^k-\h\l_\e^k|\le
C\left(\e^{3/2}(A+\mu)+\e^{1/2}\d^*(\e)(A+\mu)^2\right),
\end{equation*}
what owing to the equality (\ref{1.7}), Lemma~\ref{lm2.6}, and
Corollary~\ref{lm2.6cor1} of this lemma leads us to the
asymptotics (\ref{1.8}) in the case of multiply eigenvalue
$\l_0$. The proof of Theorem~\ref{th1.1} is complete.

Let us clear up the asymptotics behaviour of perturbed
eigenfunctions under hypothesis of Theorem~\ref{th1.1}. Under
assumptions  \ref{cnA0}, \ref{cnA2} and equality (\ref{1.5})
with $A\ge0$ by analogy with papers \cite{ChNov}, \cite{Ch},
\cite{Fr} one can establish following facts. If $\l_0^k$ is a
simple eigenvalue of problem~\ref{1.6}, and $\psi_0^k$ is an
associated eigenfunction, then the eigenfunction $\psi_\e^k$
converges to $\psi_0^k$. If $\l_0=\l_0^q=\ldots\l_0^{q+p-1}$ is
$p$-multiply eigenvalue and $\l_\e^k\to\l_0^k$,
$k=q,\ldots,q+p-1$, then for each associated eigenfunction
$\psi_0^k$, $k=q,\ldots,q+p-1$, there exists a linear
combination of eigenfunctions  $\psi_\e^l$, $l=q,\ldots,q+p-1$
converging to $\psi_0^k$. This convergence is strong in
$L_2(\Om)$ and weak in  $H^1(\Om)$ if limiting problem is the
Robin one ($A>0$) and it is strong in $H^1(\Om)$ if limiting
problem is the Neumann one ($A=0$).

We will keep the notations of the proof of Theorem~\ref{th1.1}.
Let $\l_0$ be a simple eigenvalue. It arises from
Lemma~\ref{lm2.5} and Remark~\ref{rm1.5} that $\h\psi_\e$
converges to  $\psi_0$ in $L_2(\Om)$.

Multiplying (\ref{2.83}) by $\psi_\e$ in $L_2(\Om)$, owing to
Lemmas~\ref{lm2.1}-\ref{lm2.4} we see that
\begin{equation*}
\frac{1}{\l_\e-\h\l_\e}
\int\limits_{\Om}f_\e\psi_\e\,\mathrm{d}x=
(\h\psi_\e,\psi_\e)_{L_2(\Om)}=
(\Psi_0+\e\Psi_1,\psi_\e)_{L_2(\Om)}+O(\e^{3/2}(A+\mu)).
\end{equation*}
From last assertion, denoting
\begin{equation}\label{2.100}
\t\psi_\e=(\Psi_0+\e\Psi_1,\psi_\e)_{L_2(\Om)}\psi_\e,
\end{equation}
and from (\ref{2.83}), (\ref{1.7}) and Corollary~\ref{lm2.6cor1}
of Lemma~\ref{lm2.6} we derive that
\begin{equation*}
\|\t\psi_\e-\h\psi_\e\|_{L_2(\Om)}\le C
\left(\e^{3/2}(A+\mu)+\e^{1/2}\d^*(\e)(A+\mu)^2\right)
=o(\e(A+\mu)),
\end{equation*}
what, Remark~\ref{rm1.5} and Lemma~\ref{lm2.5} imply that the
perturbed eigenfunction $\t\psi_\e$, associated with $\l_\e$,
converges to $\psi_0$ in $L_2(\Om)$ and due to
Lemmas~\ref{lm2.1}-\ref{lm2.4} and the matching carried out has
the following asymptotics in $H^1(\Om)$-norm:
\begin{equation}
\begin{aligned}
\t\psi_\e(x)=&\left(\Psi_0(x,\mu,\e)+\e\Psi_1(x,\mu,\e)-
\frac{\chi(\tau/c_0)}{\th'_\e(s)}\sum\limits_{l=0}^1\e^{l+1}
\Psi^\nu_l(s,\mu,\e)X(\xi)\right)\chi_\e(x)-
\\
{}&-\sum\limits_{j=0}^{N-1}
\frac{\chi(|\vs^j|\eta^{3/4})}{\th'_\e(s)}
\sum\limits_{l=0}^1\e^{l+1}
\Psi^\nu_l(s,\mu,\e)Y^{(j)}(\vs^j,\e)+o(\e(A+\mu)).
\end{aligned}\label{2.87}
\end{equation}

Let $\l_0=\l_0^q=\ldots=\l_0^{q+p-1}$ be a $p$-multiply
eigenvalue. Let us calculate the coefficients of linear
combination of perturbed eigenfunctions converging to
$\psi_0^q,\ldots,\psi_0^{q+p-1}$ and the asymptotics for them.
First we will prove an auxiliary lemma.

\begin{lemma}\label{lm2.7}
In $H^1(\Om)$ a convergence holds:
\begin{equation*}
\sum\limits_{l=q}^{q+p-1}(\psi_0^k,\Psi_0^l)_{L_2(\Om)}
\Psi_0^l\to\psi_0^k.
\end{equation*}
\end{lemma}
\PF{Proof.} The eigenfunctions $\Psi_0^k$, $k=q,\ldots,q+p-1$,
converge to eigenfunction $\psi_0^k$ of the problem (\ref{1.6})
in such sense that for each eigenfunction $\psi_0^k$,
$k=q,\ldots,q+p-1$, there exists a linear combination of
eigenfunctions $\Psi_0^l$, $l=q,\ldots,q+p-1$, converging to
$\psi_0^k$ in $H^1(\Om)$ (\cite{Kt}):
\begin{equation*}
\sum\limits_{l=q}^{q+p-1}\mathsf{b}_{lk}\Psi_0^l=
\psi_0^k(1+o(1)).
\end{equation*}
Multiplying this equality by $\Psi_0^i$ in $L_2(\Om)$, we have:
$\mathsf{b}_{ik}=(\psi_0^k,\Psi_0^i)_{L_2(\Om)}(1+o(1))$, what
proves the lemma. The proof is complete.

It follows from formulae (\ref{2.101}) and
Lemmas~\ref{lm2.1}-\ref{lm2.4} that
\begin{equation}\label{2.102}
\mathsf{b}_{ki}^\e=(\Psi_0+\e\Psi_1,\psi_\e^i)_{L_2(\Om)}+
O(\e^{3/2}(A+\mu)),
\end{equation}
moreover, last assertions hold under the assumption of
boundedness of function $\d^*(\e)$. Using these assertions and
(\ref{2.88}), we derive an estimate:
\begin{equation}\label{2.99}
\Big\|\sum\limits_{i=q}^{q+p-1}
(\Psi_0^l+\e\Psi_1^l,\psi_\e^i)_{L_2(\Om)}\psi_\e^{i}
-\h\psi_\e^l\Big\|_{H^1(\Om)}=
O(\e^{3/2}(A+\mu)+\e^{1/2}\d^*(\e)(A+\mu)^2),
\end{equation}
from that, Lemma~\ref{lm2.7} and the estimate
$\|\h\psi_\e^l-\Psi_0^l\|_{L_2(\Om)}=o(1)$ (see
Lemma~\ref{lm2.5}) it follows the convergence in $L_2(\Om)$:
\begin{equation}\label{2.103}
\t\psi_\e^k\equiv\sum\limits_{l=q}^{q+p-1}
(\psi_0^k,\Psi_0^l)_{L_2(\Om)} \sum\limits_{i=q}^{q+p-1}
(\Psi_0^l+\e\Psi_1^l,\psi_\e^i)_{L_2(\Om)}\psi_\e^{i}
\to\psi_0^k,
\end{equation}
i.e., $\t\psi_\e^k$ is a linear combination of the perturbed
eigenfunctions, converging to $\psi_0^k$ in $L_2(\Om)$. On the
other hand, by (\ref{2.99}) for $\t\psi_\e^k$ the estimate
\begin{equation*}
\Big\|\t\psi_\e^k-\sum\limits_{l=q}^{q+p-1}
(\psi_0^k,\Psi_0^i)_{L_2(\Om)}\h\psi_\e^i\Big\|_{H^1(\Om)}
=O(\e^{3/2}(A+\mu)+\e^{1/2}\d^*(\e)(A+\mu)^2),
\end{equation*}
takes place, from that, equality (\ref{1.7}) and
Corollary~\ref{lm2.6cor1} of Lemma~\ref{lm2.6} it arises that
the asymptotics for $\t\psi_\e^k$ in $H^1(\Om)$ has the
following form:
\begin{equation}\label{2.84}
\begin{aligned}
\t\psi_\e^k(x)=&\sum\limits_{i=q}^{q+p-1}
(\psi_0^k,\Psi_0^i)_{L_2(\Om)}\Bigg(
\Big(\Psi_0^i(x,\mu,\e)+\e\Psi_1^i(x,\mu,\e)-
\\
{}&-\frac{\chi(\tau/c_0)}{\th'_\e(s)}\sum\limits_{l=0}^1\e^{l+1}
\Psi^{i,\nu}_l(s,\mu,\e)X(\xi)\Big)\chi_\e(x)-
\\
{}&-\sum\limits_{j=0}^{N-1}
\frac{\chi(|\vs^j|\eta^{3/4})}{\th'_\e(s)}
\sum\limits_{l=0}^1\e^{l+1}
\Psi^{i,\nu}_l(s,\mu,\e)Y^{(j)}(\vs^j,\e)\Bigg)+o(\e(A+\mu)).
\end{aligned}
\end{equation}
Thus, we have proved

\begin{theorem}\label{th2.1}
Suppose the hypothesis of Theorem~\ref{th1.1} takes place. If
$\l_0=\l_0^k$ is a simple eigenvalue of the problem (\ref{1.6}),
then eigenfunction $\t\psi_\e$ from (\ref{2.100}) with
$\psi_\e=\psi_\e^k$  converges to $\psi_0^k$ in $L_2(\Om)$-norm
and  has the asymptotics (\ref{2.87}) in $H^1(\Om)$, where
$\Psi_1$ is a solution of problem (\ref{2.7}), (\ref{2.29}) with
$\Psi_0=\Psi_0^k$, $\Psi_l^\nu=\Psi_l^{k,\nu}$ are values of
normal derivatives of functions $\Psi_l^k$ on $\p\Om$, $X$ and
$Y^{(j)}$ are defined by equalities (\ref{2.105}) and
(\ref{2.106}). If $\l_0=\l_0^q=\ldots=\l_0^{q+p-1}$ is a
$p$-multiply eigenvalue of the problem (\ref{1.6}), then for
each associated eigenfunction $\psi_0^k$, $k=q,\ldots,q+p-1$,
there exists a linear combination (\ref{2.103}) of the perturbed
eigenfunctions, converging to $\psi_0^k$ in $L_2(\Om)$ norm and
having in  $H^1(\Om)$ the asymptotics (\ref{2.84}).
\end{theorem}

From Theorem~\ref{th1.1}~and~\ref{th2.1} it follows the validity
of next statement.

\begin{lemma}\label{lm2.8}
Suppose the assumptions \ref{cnA0}, \ref{cnA2} and the equality
(\ref{1.5}) with $A\ge0$ for the function $\eta$ from \ref{cnA2}
hold. Then the remainders in the asymptotics (\ref{1.8}),
(\ref{2.87}) and (\ref{2.84}) are of order
$O(\e^{3/2}(A+\mu)+\e^{1/2}\d^*(\e)(A+\mu)^2)$.
\end{lemma}

If $A=0$, i.e., the limiting problem is the Neumann one, the
statement of Theorem~\ref{th2.1} can be strengthened as follows.

\begin{theorem}\label{th1.5}
Suppose the assumptions \ref{cnA0}, \ref{cnA2} and the equality
(\ref{1.5}) with $A=0$  for the function $\eta$ from \ref{cnA2}
hold. Then in the case of simple limiting eigenvalue --
eigenfunction $\t\psi_\e^k$ from (\ref{2.100}) and in the case
of multiply limiting eigenvalue  -- the linear combination of
eigenfunctions $\t\psi_\e^k$ from (\ref{2.103}) converges to the
limiting eigenfunction $\psi_0^k$ in $H^1(\Om)$.
\end{theorem}

\PF{Proof.} Let us prove, that the equality
\begin{equation}\label{3.1}
\|\h\psi_\e^k-\Psi_0^k\|_{H^1(\Om)}=o(1)
\end{equation}
holds for all $k$ as $\e\to0$. Since
\begin{equation*}
\|\h\psi_\e^k-\Psi_0^k\|^2_{H^1(\Om)}=
\|\nabla_x(\h\psi_\e^k-\Psi_0^k)\|^2_{L_2(\Om)}+
\|\h\psi_\e^k-\Psi_0^k\|^2_{L_2(\Om)},
\end{equation*}
and also, last term tends to zero as $\e\to0$ by
Lemma~\ref{lm2.5}, it remains to estimate the gradient's norm.
Taking into account the form of $\h\psi_\e^k$, the gradient's
norm $(\h\psi_\e^k-\Psi_0^k)$ is estimated as follows:
\begin{equation}\label{3.2}
\begin{aligned}
\|&\nabla_x(\h\psi_\e^k-\Psi_0^k)\|_{L_2(\Om)}^2  \le
2\|\nabla_x(\psi_\e^{ex}-\Psi_0^k)\|_{L_2(\Om)}^2+
\sum\limits_{j=0}^{N-1} \|\nabla_x
\psi_\e^{in,j}\|_{L_2(\Om^{in}_j)}^2+
\\
{}&+2\|\nabla_x\left(\psi_\e^{bl}
\chi(\tau/c_0)\right)\|_{L_2(\Om_\eta^{bl})}^2+2\sum\limits_{j=0}^{N-1}
\|\nabla_x\left(\psi_\e^{mat,j}\chi(|\vs^j|\eta^{3/4})\right)
\|_{L_2(\Om^{mat}_j)}^2,
\end{aligned}
\end{equation}
where $\psi_\e^{ex}$, $\psi_\e^{bl}$,  $\psi_\e^{in,j}$ and
$\psi_\e^{mat,j}$ are the functions defined in formal
constructing in the  proof of Theorem~\ref{th1.1} and associated
with $\psi_0^k$.

In view of Lemma~\ref{lm2.1} and the definition $\psi_\e^{ex}$
we have:
\begin{equation}\label{3.3}
\|\nabla_x(\psi_\e^{ex}-\Psi_0^k)\|_{L_2(\Om)}^2\le C\e^2\mu^4.
\end{equation}
It is easy to see
\begin{equation*}
\|\nabla_x\left(\psi_\e^{bl}\chi(\tau/c_0)
\right)\|_{L_2(\Om_\eta^{bl})}^2\le
C\left(\|\nabla_x\psi_\e^{bl}\|_{L_2(\Om_\eta^{bl}\cap\Om^{bl})}^2+
\|\psi_\e^{bl}\|_{L_2(\Om^{bl})}^2\right).
\end{equation*}
Second term in the right side of the inequality obtained is
estimated above by  $C\e^{3/2}\mu$ (see Lemma~\ref{lm2.2}). By
direct calculations with employing explicit form of the
functions $v_i$, the boundedness of function $\d^*(\e)$,
Lemma~\ref{lm2.1} and the equality $N=2\e^{-1}$ one can check
that
\begin{align*}
\|\nabla_x\psi_\e^{bl}\|_{L_2(\Om_\eta^{bl}\cap\Om^{bl})}^2 \le
C\sum\limits_{j=0}^{N-1}
\left(\|\nabla_\xi\psi_\e^{bl}\|_{L_2(\Pi^{(j)}_\eta)}^2+\e^2
\left\|\frac{\p}{\p
s}\psi_\e^{bl}\right\|_{L_2(\Pi^{(j)})}^2\right)\le C\mu.
\end{align*}
Thus,
\begin{equation}\label{3.4}
\|\nabla_x\left(\psi_\e^{bl}
\chi(\tau/c_0)\right)\|_{L_2(\Om_\eta^{bl})}^2 \le C\mu.
\end{equation}
It follows from explicit form of the functions $\psi_\e^{in,j}$
that
\begin{equation}\label{3.5}
\sum\limits_{j=0}^{N-1} \|\nabla_x
\psi_\e^{in,j}\|_{L_2(\Om^{in}_j)}^2\le C\mu.
\end{equation}
Using the asymptotics (\ref{2.85}), we prove that
\begin{equation}\label{3.6}
\sum\limits_{j=0}^{N-1}
\|\nabla_x\left(\psi_\e^{mat,j}\chi(|\vs^j|\eta^{3/4})\right)
\|_{L_2(\Om^{mat}_j)}^2\le C\eta^{1/5}.
\end{equation}
Collecting (\ref{3.2})--(\ref{3.6}), we get (\ref{3.1}). We
stress that convergence (\ref{3.1}) was proved without using the
equality (\ref{1.7}) and holds true for each bounded function
$\d^*(\e)$.

Let $\psi_0$ be associated with simple eigenvalue. Owing to
convergence (\ref{3.1}) and $\Psi_0\xrightarrow{H^1(\Om)}\psi_0$
(see Remark~\ref{rm1.5}) we conclude that $\h\psi_\e$ converges
to $\psi_0$ strongly in $H^1(\Om)$. Therefore, by
Lemma~\ref{lm2.8} the eigenfunction $\t\psi_\e$ from
(\ref{2.100}) satisfies an equality
\begin{equation*}
\|\t\psi_\e-\h\psi_\e\|_{H^1(\Om)}=o(1),
\end{equation*}
from what it follows that eigenfunction $\psi_\e$ converges to
$\psi_0$ in $H^1(\Om)$ norm.

Let $\psi_0$ be associated with $p$-multiply eigenvalue
$\l_0=\l_0^q=\ldots=\l_0^{q+p-1}$, that the eigenfunctions
$\psi_0^k$, $k=q,\ldots,q+p-1$ are associated with. For the
functions $\h\psi_\e^k$ and $\Psi_0^k$ the relationships
(\ref{3.1}) hold. These relationships, Lemma~\ref{lm2.7} and
(\ref{2.99}) yield that the linear combination  $\t\psi_\e^k$ of
perturbed eigenfunctions from (\ref{2.103}) converges to
$\psi_0^k$ in $H^1(\Om)$. The proof is complete.

\sect{Asymptotics for the perturbed eigenelements under
hypothesis of Theorem~\ref{th1.6}.}

In this section we will obtain the asymptotics for  the
perturbed eigenelements in the case of breakdown of equality
(\ref{1.7}) of Theorem~\ref{th1.1}. First we will prove
Theorem~\ref{th1.6} about asymptotics for eigenvalues, and then
we will establish Theorem~\ref{th3.1} about asymptotics for
associated eigenfunctions. Everywhere in the section, if it is
not said specially, we keep the notations of the previous
section.

\smallskip
\PF{\indent Proof of Theorem~\ref{th1.6}.} In proving we lean on
the boundedness of the function $\d^*(\e)$ established in
Corollary~\ref{lm2.6cor2} of Lemma~\ref{lm2.6}. In Appendix we
will show that eigenvalues of problem (\ref{2.6}), (\ref{2.25})
satisfy following asymptotics formulae
\begin{equation}\label{3.17}
\L_0^k(\mu,\e)=\l_0^k+\mu\int\limits_{\p\Om}(\psi_0^k)^2\th'_0
\,\mathrm{d}s+O(\mu^2+(A+\mu)\si),
\end{equation}
where in the case of multiply eigenvalue  $\l_0^k$ the
associated eigenfunctions $\psi_0^k$ are additionally assumed to
be orthogonal in $L_2(\p\Om)$  weighted by  $\th'_0$,
$\si=\si(\e)=\|\th'_\e-\th'_0\|_{C(\p\Om)}=o(1)$. From
Lemmas~\ref{lm2.1},~\ref{lm2.8} and Corollary~\ref{lm2.6cor2} of
Lemma~\ref{lm2.6} it follows that
$|\l_\e^k-\L_0^k|=O(\e^{3/2}(A+\mu)+\e^{1/2}(A+\mu)^2)$, what by
asymptotics (\ref{3.17}) implies the correctness of the theorem.
The proof is complete.

Let us derive the asymptotics for the perturbed eigenfunctions
under hypothesis of Theorem~\ref{th1.6}. We start from the case
of simple eigenvalue  $\l_0$. Assertion  (\ref{2.100}) and
Lemma~\ref{lm2.1} imply that perturbed eigenfunction
\begin{equation}\label{3.18}
\t\psi_\e=(\Psi_0,\psi_\e)_{L_2(\Om)}\psi_\e,
\end{equation}
associated with $\l_\e\xrightarrow[\e\to0]{}\l_0$, satisfies an
estimate
\begin{equation*}
\|\t\psi_\e-\h\psi_\e\|_{H^1(\Om)}=O(\e^{3/2}(A+\mu)+
\e^{1/2}(A+\mu)^2).
\end{equation*}
By direct calculations and employing
Lemmas~\ref{lm2.1}-\ref{lm2.4} and the results of matching
procedure made in the previous section one can check see that
$H^1(\Om)$-norm of the function
\begin{equation*}
\left(\e\Psi_1-\e^2\frac{\chi(\tau/c_0)}{\th'_\e} \Psi^\nu_1 X
\right)\chi_\e-\e^2\sum\limits_{j=0}^{N-1}
\frac{\chi(|\vs^j|\eta^{3/4})}{\th'_\e} \Psi^\nu_1 Y^{(j)}
\end{equation*}
is of order $O(\e^{1/2}(A+\mu))$. Hence, the function
$\t\psi_\e$ from (\ref{3.18}) converges to $\psi_0$ in
$L_2(\Om)$ and has the following asymptotics in $H^1(\Om)$:
\begin{equation}\label{3.15}
\begin{aligned}
\t\psi_\e(x)=&\left(\Psi_0(x,\mu,\e)+\e
\frac{\chi(\tau/c_0)}{\th'_\e(s)}
\Psi^\nu_0(s,\mu,\e)X(\xi)\right)\chi_\e(x)-
\\
{}&+\e\sum\limits_{j=0}^{N-1}\frac{\chi(|\vs^j|\eta^{3/4})}
{\th'_\e(s)} \Psi^\nu_0(s,\mu,\e)Y^{(j)}(\vs^j,\e)
+O(\e^{1/2}(A+\mu)).
\end{aligned}
\end{equation}

Now we proceed to the case of $p$-multiply eigenvalue
$\l_0=\l_0^q=\ldots=\l_0^{q+p-1}$. Due to (\ref{2.103}) and
Lemmas~\ref{lm2.1},~\ref{lm2.8} we see that a linear combination
of perturbed eigenfunctions
\begin{equation}
\t\psi_\e^k\equiv\sum\limits_{l=q}^{q+p-1}
(\psi_0^k,\Psi_0^l)_{L_2(\Om)} \sum\limits_{i=q}^{q+p-1}
(\Psi_0^l,\psi_\e^i)_{L_2(\Om)}\psi_\e^{i}\label{3.19}
\end{equation}
converges to $L_2(\Om)$ in $\psi_0^k$, $k=q,\ldots,q+p-1$ and
its asymptotics in $H^1(\Om)$ reads as follows:
\begin{equation}\label{3.16}
\begin{aligned}
\t\psi_\e^k(x)=\sum\limits_{i=q}^{q+p-1}
{}&(\psi_0^k,\Psi_0^i)_{L_2(\Om)}\Bigg(\Big(\Psi_0^i(x,\mu,\e)+\e
\frac{\chi(\tau/c_0)}{\th'_\e(s)}
\Psi^{i,\nu}_0(s,\mu,\e)X(\xi)\Big)\chi_\e(x)-
\\
{}&+\e\sum\limits_{j=0}^{N-1}\frac{\chi(|\vs^j|\eta^{3/4})}
{\th'_\e(s)} \Psi^{i,\nu}(s,\mu,\e)Y^{(j)}(\vs^j,\e)\Bigg)
+O(\e^{1/2}(A+\mu)).
\end{aligned}
\end{equation}
Similar to the case of simple limit eigenvalue, $H^1(\Om)$-norm
of neglected terms of $\h\psi_\e^l$ is of order
$O(\e^{1/2}(A+\mu))$.

Lemmas~\ref{lm2.1} and Theorem~\ref{th1.5} yield that
Theorem~\ref{th1.5} takes place for the functions (\ref{3.18}),
(\ref{3.19}), too.

Thus, we have proved
\begin{theorem}\label{th3.1}
Suppose the hypothesis of Theorem~\ref{th1.5} holds. If
$\l_0=\l_0^k$ is a simple eigenvalue of problem the (\ref{1.6}),
then the eigenfunction $\t\psi_\e$ from (\ref{3.18}) with
$\psi_\e=\psi_\e^k$, $\Psi_0=\Psi_0^k$, converges to $\psi_0^k$
in $L_2(\Om)$ as $A\ge0$ and in $H^1(\Om)$ as $A=0$ and has in a
sense of $H^1(\Om)$-norm the asymptotics (\ref{3.15}). If
$\l_0=\l_0^q=\ldots=\l_0^{q+p-1}$ is a $p$-multiply eigenvalue,
then for each associated eigenfunction  $\psi_0^k$,
$k=q,\ldots,k+p-1$, there exists a linear combination
(\ref{3.19}), converging to $\psi_0^k$ in $L_2(\Om)$ as $A\ge0$
and in $H^1(\Om)$ as $A=0$  having asymptotics  (\ref{3.16}) in
$H^1(\Om)$-norm. In asymptotics (\ref{3.15}), (\ref{3.16}) the
notations of Theorem~\ref{th2.1} are used.
\end{theorem}

\sect{Auxiliary statement}

In this section we will prove an auxiliary statement that will
be employed in next section in the proof of Theorem~\ref{th1.2}.
Let us formulate this lemma.

\begin{lemma}\label{lm5.1} Suppose the assumptions \ref{cnA0}
and \ref{cnA2} hold, the function $\eta(\e)$ from \ref{cnA2} is
bounded above by a number $\pi/2$ and satisfies the equality
(\ref{1.3}), and for each $i$, $j$ and $\e$ the equalities
$a^j(\e)+b^j(\e)=2\eta(\e)$, $a^i(\e)=a^j(\e)$,
$b^i(\e)=b^j(\e)$ take place. Suppose also that there exists a
fixed number $\mathsf{d}>0$ for that H\"older norm
$\|\th'_\e\|_{C^{3+\mathsf{d}}(\p\Om)}$ is bounded on $\e$. Then
the perturbed eigenvalue $\l_\e^k$ converges to the eigenvalue
$\l_0^k$ of limiting problem (\ref{1.4}) and has the asymptotics
\begin{equation*}
\l_\e^k=\l_0^k+\e\ln\sin\eta(\e)\int\limits_{\p\Om}
\left(\frac{\p\psi_0^k}{\p\nu}\right)^2\frac{\mathrm{d}s}
{\th'_\e}+
O\left(\e^{3/2}\left(\left|\ln\eta(\e)\right|^{3/2}+1\right)
\left(\frac{\pi}{2}-\eta(\e)\right)\right).
\end{equation*}
\end{lemma}

\PF{Proof.} The convergence of eigenvalues is established by
analogy with papers~\cite{ChNov}, \cite{Ch}, \cite{Fr}. We will
prove the asymptotics by the scheme employed in the second
section. As before, first we will formally construct asymptotics
and after we will justify them. It should be noted that formal
construction of the asymptotics that will be used in general
coincide with the scheme proposed in \cite{AA}, \cite{GDan}. The
difference is a more general formulation of the problem
considered here, the renunciation of additional assumptions made
in \cite{AA}, \cite{GDan}, and the estimate for the error with
respect two both parameters $\e$ and $\eta$. We will consider in
detail only the case of simple limiting eigenvalue;  the case of
multiply limiting eigenvalue is established by analogy.

Let $\l_0$ be a simple eigenvalue of limiting problem
(\ref{1.4}), $\psi_0$ be the associated eigenfunction
(normalized in $L_2(\Om)$), $\l_\e$ be the perturbed eigenvalue
converging to $\l_0$.

We seek for the asymptotics of $\l_\e$ as follows:
\begin{equation}\label{5.1}
\l_\e=\l_0+\e\l_1(\e)\ln\sin\eta,
\end{equation}
and the asymptotics for associated eigenfunction is constructed
as a sum of an outer expansion and a boundary layer:
\begin{align}
\psi_\e(x)&=\psi_\e^{ex}(x,\eta)+\chi(\tau/c_0)
\psi_\e^{bl}(\xi,s,\eta),\nonumber
\\
\psi_\e^{ex}(x,\eta)&=\psi_0(x)+\e\psi_1(x,\e)\ln\sin\eta,
\label{5.2}
\\
\psi_\e^{bl}(\xi,s,\eta)&=\e v_1(\xi,s,\e,\eta)+ \e^2
v_2(\xi,s,\e,\eta),\label{5.3}
\end{align}
where $\xi=(\xi_1,\xi_2)$,
$\xi_1=(\th_\e(s)-\th_\e(s^\e_0))/\e-(b^j(\e)-a^j(\e))/2$,
$\xi_2=\tau\th'_\e(s)/\e$. Observe, here it is possible to carry
out the construction of asymptotics  without employing method of
matched asymptotics expansions.

We substitute  (\ref{5.1}) and (\ref{5.2}) into equation
(\ref{1.1}) and write out the coefficient of $\e\ln\sin\eta$:
\begin{equation}\label{5.4}
(\D+\l_0)\psi_1=-\l_1\psi_0,\quad x\in\Om.
\end{equation}
Substitution (\ref{5.1}) and (\ref{5.3}) into equation
(\ref{1.1}) lead us to the equation (\ref{2.9}) and (\ref{2.14})
for the functions $v_1$ and $v_2$. Boundary conditions for these
functions are derived from the claim the sum of (\ref{5.2}) and
(\ref{5.3}) to satisfy  both boundary conditions in (\ref{1.2}):
\begin{gather}
v_1=-\psi_1^D\ln\sin\eta, \quad\xi\in\g^\eta,\qquad \frac{\p
v_1}{\p\xi_2}=\frac{1}{\th'_\e}\psi_0^\nu,\quad\xi\in\G^\eta,
\label{5.5}
\\
\frac{\p v_2}{\p\xi_2}=\frac{1}{\th'_\e}\psi_1^\nu\ln\sin\eta,
\quad\xi\in\G^\eta, \label{5.6}
\end{gather}
where $\g^\eta$ is a union of intervals $(\pi j-\eta,\pi
j+\eta)$, $j\in\mathbb{Z}$, lying in the axis $O\xi_1$, and
$\G^\eta$ is a complement of $\overline{\g}^\eta$ on the axis
$O\xi_1$, $\psi_1^D$ and $\psi_i^\nu$ are values of the
functions $\psi_i$ and their normal derivatives on the boundary
$\p\Om$. Problem (\ref{2.9}), (\ref{5.5}) is solved explicitly:
\begin{align}\label{5.7}
{}&
v_1(\xi,s,\e,\eta)=-\frac{1}{\th'_\e(s)}\psi_0^\nu(s)X_\eta(\xi),
\\
{}& X_\eta(\xi)=\mathrm{Re}\,\ln\left(\sin z+\sqrt{\sin^2
z-\sin^2 \eta}\right)-\xi_2.\nonumber
\end{align}
It is easy to check that $X_\eta\in\mathcal{V}_\eta\cap
H^1(\Pi^{(j)})$ is even on $\xi_1$ harmonic function, where
$\mathcal{V}_\eta$ denotes the space of $\pi$-periodic on the
variable $\xi_1$ functions decaying exponentially as
$\xi_2\to+\infty$ uniformly on $\xi_1$ with all their
derivatives and belonging to $C^\infty(\{\xi:
\xi_2>0\}\cup\g^\eta\cup\G^\eta)$. The function $X_\eta$ obeys
boundary condition
\begin{equation}\label{5.8}
X_\eta(\xi)=\ln\sin\eta,\quad \xi\in\g^\eta,\qquad \frac{\p
X}{\p\xi_2}=-1,\quad \xi\in\G^\eta.
\end{equation}
The function $v_1$ defined by the equality (\ref{5.7}) due to
(\ref{5.8}) meets the boundary condition
\begin{equation*}
v_1=-\frac{1}{\th'_\e}\psi_0^\nu\ln\sin\eta,\quad\xi\in\g^\eta,
\end{equation*}
comparing that with (\ref{5.5}), we obtain the boundary
condition for $\psi_1$:
\begin{equation}\label{5.9}
\psi_1=\frac{1}{\th'_\e}\frac{\p\psi_0}{\p\nu},\quad x\in\p\Om.
\end{equation}
The solvability condition of boundary value problem (\ref{5.4}),
(\ref{5.9}) gives the formula for $\l_1$:
\begin{equation}\label{5.10}
\l_1=\int\limits_{\p\Om}\left(\frac{\p\psi_0}{\p\nu}\right)^2
\frac{\mathrm{d}s}{\th'_\e(s)}.
\end{equation}
The function $\psi_1$ is chosen to be orthogonal to $\psi_0$ in
$L_2(\Om)$. The function $v_2$ is defined as follows:
\begin{equation}\label{5.25}
v_2=\frac{\psi^\nu_0}{2(\th'_\e)^2}\xi_2^2 \left(
\frac{\th''_\e}{\th'_\e}\frac{\p
X_\eta}{\p\xi_1}+\mathsf{k}\frac{\p
X_\eta}{\p\xi_2}\right)-\frac{2}{\th'_\e}
\left(\frac{\psi_0^\nu}{\th'_\e}\right)'v_2^{odd}-
\frac{1}{\th'_\e}\psi_1^\nu\ln\sin\eta X_\eta,
\end{equation}
where $v_2^{odd}$ is an exponentially decaying solution for the
boundary value problem
\begin{equation}\label{5.11}
\D_\xi v_2^{odd}=\frac{\p X_\eta}{\p\xi_1},\quad \xi_2>0,\qquad
v_2^{odd}=0,\quad \xi\in\g^\eta,\qquad \frac{\p
v_2^{odd}}{\p\xi_2}=0,\quad \xi\in\G^\eta.
\end{equation}
The solution for problem (\ref{5.11}) exists; this existence and
also its evenness on $\xi_1$ and belonging to
$\mathcal{V}_\eta\cap H^1(\Pi^{(j)})$ were proved in \cite{AA}.

For justification of the asymptotics constructed formally  we
will use following lemmas.

\begin{lemma}\label{lm5.2}
The properties takes place:
\begin{enumerate}
\renewcommand{\theenumi}{(\arabic{enumi})}
\item\label{lm5.2_3} for integer $m\ge0$ the inequalities
\begin{equation*}
\|\xi_2^m X_\eta\|_{L_2(\Pi^{(j)})}\le
C\left(\frac{\pi}{2}-\eta\right)^2
\left(\left|\ln\left(\frac{\pi}{2}-
\eta\right)\right|^{1/2}+1\right)
\end{equation*}
are true, where constants  $C$ are independent on $\eta$.

\item\label{lm5.2_2} for integer $m,p\ge0$ the estimates
\begin{align*}
{}&\Big\|\xi_2^m\nabla_\xi\frac{\p^m X_\eta}{\p\xi_2^m}
\Big\|_{L_2(\Pi^{(j)})}\le C|\ln\sin\eta|^{1/2},
\\
{}&\Big\|\xi_2^{m+p+1}\nabla_\xi\frac{\p^m X_\eta}{\p\xi_2^m}
\Big\|_{L_2(\Pi^{(j)})}\le C\left(\frac{\pi}{2}-\eta\right)^2
\left(\left|\ln\left(\frac{\pi}{2}-
\eta\right)\right|^{1/2}+1\right),
\end{align*}
take place, where constants  $C$ are independent on $\eta$.
\end{enumerate}
\end{lemma}

\PF{Proof.} First we prove the statement of item \ref{lm5.2_3}
for $m=0$. It was shown in \cite[\S 3]{MZ} that
$\|X_\eta\|_{L_2(\Pi^{(j)})}$ is continuous on
$\eta\in[0,\pi/2]$ function. To prove the estimate needed it is
sufficient to clear up the behaviour of this function as
$\eta\to\pi/2$. It is easy to see that the function
\begin{equation*}
X_\eta^1(\xi)=-\frac{1}{2}\xi_2\int\limits_{\xi_2}^{+\infty}
X_\eta(\xi_1,t)\,\mathrm{d}t
\end{equation*}
is even on $\xi_1$, belong to $\mathcal{V}_\eta$ and is a
solution for the equation $\D_\xi X_\eta^1(\xi)=X_\eta$ in a
domain $\xi_2>0$ satisfying boundary conditions:
\begin{equation*}
X_\eta^1=0,\quad \frac{\p
X_\eta^1}{\p\xi_2}=-\frac{1}{2}\int\limits_0^{+\infty}
X_\eta(\xi_1,t)\,\mathrm{d}t,\quad \xi_2=0.
\end{equation*}
Using these properties of the functions $X_\eta^1$ and $X_\eta$
and the equality
\begin{equation*}
\int\limits_{\Pi^{(j)}}X^2_\eta\,\mathrm{d}\xi=
\int\limits_{\Pi^{(j)}}(X_\eta+\xi_2-\ln\sin\eta)
X_\eta\,\mathrm{d}\xi
\end{equation*}
proved in \cite[\S 3]{MZ} and integrating by parts we have:
\begin{equation}\label{5.12}
\begin{aligned}
\int\limits_{\Pi^{(j)}}&X^2_\eta\,\mathrm{d}\xi=
\int\limits_{\Pi^{(j)}}(X_\eta+\xi_2-\ln\sin\eta)\D_\xi
X^1_\eta\,\mathrm{d}\xi=\\&=\int\limits_\eta^{\pi/2}\left(
X_\eta(\xi_1,0)-\ln\sin\eta\right)\int\limits_{0}^{+\infty}
X_\eta(\xi_1,t)\,\mathrm{d}t\,\mathrm{d}\xi_1.
\end{aligned}
\end{equation}
Since as $\xi_1\in(\eta,\pi/2]$
\begin{equation*}
\frac{d^2}{d\xi_1^2}\int\limits_{0}^{+\infty}
X_\eta(\xi_1,t)\,\mathrm{d}t=-\int\limits_{0}^{+\infty}
\frac{\p^2}{\p t^2}X_\eta(\xi_1,t)\,\mathrm{d}t=-1,
\end{equation*}
due to evenness and  $\pi$-periodicity of $X_\eta$ on $\xi_1$ we
get:
\begin{equation}\label{5.13}
\int\limits_{0}^{+\infty} X_\eta(\xi_1,t)\,\mathrm{d}t=
-\frac{1}{2}\left(\xi_1-\frac{\pi}{2}\right)^2+\int\limits_{0}^{+\infty}
X_\eta\left(\frac{\pi}{2},t\right)\,\mathrm{d}t.
\end{equation}
Applying the estimate  $|\ln(1+a)|\le a$, $a\ge0$, to the
integrand function
\begin{equation*}
X_\eta\left(\frac{\pi}{2},t\right)=\ln\left( 1+\frac{
\mathrm{e}^{-2t}-1+\sqrt{(1-\mathrm{e}^{-2t})^2+4\mathrm{e}^{-2t}
\cos^2\eta}}{2}\right)
\end{equation*}
in the right side of the equality (\ref{5.13}) and integrating
the integral obtained we deduce an assertion ($\eta\to\pi/2$):
\begin{equation}\label{5.14}
\int\limits_{0}^{+\infty}
X_\eta\left(\frac{\pi}{2},t\right)\,\mathrm{d}t
=
O\left(\eta_1^2\ln\eta_1\right),
\end{equation}
where $\eta_1=\pi/2-\eta$. In \cite{AA} it was proved that:
\begin{equation}\label{5.15}
\int\limits_{\gamma^\eta\cap\overline{\Pi}}\frac{\partial
X_\eta}{\partial\xi_2}\,d\xi_1=\pi-2\eta,\quad
\int\limits_{\Gamma^\eta\cap\overline{\Pi}}
X_\eta\,d\xi_1=-2\eta\ln\sin\eta.
\end{equation}
Substituting  (\ref{5.13})--(\ref{5.15}) into (\ref{5.12}), we
arrive at equalities ($\eta\to\pi/2$):
\begin{align*}
\int\limits_{\Pi^{(j)}}&X^2\,\mathrm{d}\xi=-\frac{1}{2}
\int\limits_{\eta}^{\pi/2}\ln\left(\frac{\sin\xi_1}{\sin\eta}
+\sqrt{\frac{\sin^2\xi_1}{\sin^2\eta}-1}
\right)\left(\xi_1-\frac{\pi}{2}\right)^2\,\mathrm{d}\xi_1+
\\
{}&+\frac{1}{2}\int\limits_{0}^{+\infty}X_\eta
\left(\frac{\pi}{2},\xi_2\right)\,\mathrm{d}\xi_2\,\int\limits_{\eta}^{\pi/2}\left(X_\eta(\xi_1,0)
-\ln\sin\eta\right)\mathrm{d}\xi_1=
\\
{}&=-\frac{1}{2}\int\limits_{0}^{\eta_1}t^2\left(\ln\left(\cos
t+\sqrt{\cos^2
t-\sin^2\eta}\right)-\ln\sin\eta\right)\,\mathrm{d}t
+O(\eta_1^4\ln\eta_1)=\\&=O(\eta_1^4\ln\eta_1).
\end{align*}
In calculations the change $t=\pi/2-\xi_1$ has been done. The
estimate for  $\|X_\eta\|^2_{L_2(\Pi^{(j)})}$ obtained  and the
continuity of this function on $\eta\in[0,\pi/2]$ imply the
statement of item \ref{lm5.2_3} for $m=0$.

It follows from explicit form of $X$, its infinitely
differentiability $(\xi,\eta)$ for $\xi_2\ge1$, continuity on
$(\xi,\eta)\in\{\xi: \xi_2>0\}\times(0,\pi/2]$ and exponential
decaying as $\xi_2\to+\infty$ that for $m\ge1$ the quantity
$\|\xi_2^m X_\eta\|_{L_2(\Pi^{(j)})}$ is continuous on
$\eta\in(0,\pi/2]$ function, and the estimate:
\begin{equation*}
\|\xi_2^m X_\eta\|_{L_2(\Pi^{(j)}\cap\{\xi: \xi_2>1\})}\le
C\eta_1^2,
\end{equation*}
holds, where constant $C$ is independent on $\eta$. Then by an
inequality
\begin{equation*}
\|\xi_2^m X_\eta\|_{L_2(\Pi^{(j)}\cap\{\xi: \xi_2<1\})}<
\|X_\eta\|_{L_2(\Pi^{(j)})}\le C\eta_1^2
\left(|\ln\eta_1|^{1/2}+1\right)
\end{equation*}
and the statement of item \ref{lm5.2_3} for $m=0$ we derive that
this item takes place for  $m>0$, too.

Let us integrate by parts in the equality
$\int\limits_{\Pi}X_\eta\Delta_\xi X_\eta\,d\xi=0$; as a result
we have:
\begin{align*}
\int\limits_\Pi\left|\nabla_\xi
X_\eta\right|^2\,d\xi=-\ln\sin\eta
\int\limits_{\gamma^\eta\cap\overline{\Pi}^{(j)}}\frac{\partial
X_\eta}{\partial\xi_2}\,d\xi_1+
\int\limits_{\Gamma^\eta\cap\overline{\Pi}^{(j)}}X_\eta\,d\xi_1,
\end{align*}
from what and (\ref{5.15}) it arises:
\begin{equation}\label{5.17}
\|\nabla_\xi
X_\eta\|_{L_2(\Pi^{(j)})}=\pi^{1/2}|\ln\sin\eta|^{1/2}.
\end{equation}
The chain of equalities ($m\ge0$, $p\ge0$, $m+p\ge1$,
$m,p\in\mathbb{Z}$):
\begin{align*}
0=&\int\limits_{\Pi^{(j)}}\xi_2^{2(m+p)}\frac{\partial^m
X_\eta}{\partial\xi_2^m}\Delta_\xi \frac{\partial^m
X_\eta}{\partial\xi_2^m}\,d\xi=-\int\limits_{\Pi^{(j)}}
\xi_2^{2(m+p)}\Big|\nabla_\xi\frac{\partial^m
X_\eta}{\partial\xi_2^m}\Big|^2\,d\xi-
\\
{}&-2(m+p)\int\limits_{\Pi^{(j)}}\xi_2^{2(m+p)-1}\frac{\partial^m
X_\eta}{\partial\xi_2^m}\frac{\partial^{m+1}
X_\eta}{\partial\xi_2^{m+1}}\,d\xi=-\Big\|
\xi_2^{m+p}\nabla_\xi\frac{\partial^m
X_\eta}{\partial\xi_2^m}\Big\|_{L_2(\Pi^{(j)})}^2+
\\
{}&+ (m+p)(2(m+p)-1)\Big\|\xi_2^{m+p-1}\frac{\partial^m
X_\eta}{\partial\xi_2^m}\Big\|_{L_2(\Pi^{(j)})}^2
\end{align*}
gives the formulae:
\begin{equation}\label{5.18}
\Big\|\xi_2^{m+p}\nabla_\xi\frac{\partial^m
X_\eta}{\partial\xi_2^m}\Big\|_{L_2(\Pi^{(j)})}
=\sqrt{(m+p)(2(m+p)-1)}\Big\|\xi_2^{m+p-1}\frac{\partial^m
X_\eta}{\partial\xi_2^m}\Big\|_{L_2(\Pi^{(j)})}.
\end{equation}
Employing these formulae for $p=0$, $m\ge1$ and with $p\ge1$,
$m\ge0$, we get estimates
\begin{align*}
{}&\Big\|\xi_2^m\nabla_\xi\frac{\p^m X_\eta}{\p\xi_2^m}
\Big\|_{L_2(\Pi^{(j)})}\le C \|\nabla_\xi
X_\eta\|_{L_2(\Pi^{(j)})},
\\
{}&\Big\|\xi_2^{m+p+1}\nabla_\xi\frac{\p^m X_\eta}{\p\xi_2^m}
\Big\|_{L_2(\Pi^{(j)})}\le C \|X_\eta\|_{L_2(\Pi^{(j)})},
\end{align*}
from those, the item \ref{lm5.2_3} and the   equality
(\ref{5.17}) it follows the statement of item \ref{lm5.2_2}. The
proof is complete.

\begin{lemma}\label{lm5.3}
The function $v_2^{odd}$ satisfies estimates:
\begin{align*}
{}& \|\xi_2^p v_2^{odd}\|_{L_2(\Pi^{(j)})}\le
C|\ln\sin\eta|^{1/2},
\\
{}& \|\xi_2^p \nabla_\xi v_2^{odd}\|_{L_2(\Pi^{(j)})}\le
C|\ln\sin\eta|^{1/2},
\\
{}& \Big\|\xi_2^p \nabla_\xi \frac{\p}{\p\xi_2}
v_2^{odd}\Big\|_{L_2(\Pi^{(j)})}\le C|\ln\sin\eta|^{1/2},
\end{align*}
where $p\ge0$, $p\in\mathbb{Z}$, and constants $C$ are
independent on $\eta$.
\end{lemma}
\PF{Proof.} Let $v\in\mathcal{V}_\eta\cap H^1(\Pi^{(j)})$ be an
odd on $\xi_1$ function that is a solution of a boundary value
problem
\begin{equation}\label{5.19}
\D_\xi v=f,\quad \xi_2>0,\qquad v=0,\quad \xi\in\g^\eta,\qquad
\frac{\p v}{\p\xi_2}=0,\quad \xi\in\G^\eta,
\end{equation}
where $f\in\mathcal{V}_\eta\cap L_2(\Pi^{(j)})$ is odd on
$\xi_1$. Since $v\in\mathcal{V}_\eta$ is odd on $\xi_1$, it
follows that $v=0$ as $\xi_1=\pi k/2$, $k\in\mathbb{Z}$.
Therefore,
\begin{equation*}
v(\xi)=\int\limits_{-\pi/2+\pi j}^{\xi_1} \frac{\p v}{\p
t}(t,\xi_2)\,\mathrm{d}t,
\end{equation*}
from what owing to Cauchy-Schwarz-Bunyakovskii inequality we
derive an estimate:
\begin{equation*}
|v(\xi)|^2\le \pi \int\limits_{-\pi/2+\pi j}^{\pi/2+\pi j}
\left|\frac{\p v}{\p\xi_1}(\xi)\right|^2\,\mathrm{d}\xi_1,
\end{equation*}
employing that, we finally get:
\begin{equation}\label{5.20}
\|v\|_{L_2(\Pi^{(j)})}\le\pi\|\nabla_\xi v\|_{L_2(\Pi^{(j)})}.
\end{equation}
We multiply equation in (\ref{5.19}) by $v$ and integrate by
parts once:
\begin{equation*}
\|\nabla_\xi v\|_{L_2(\Pi^{(j)})}=-\int\limits_{\Pi^{(j)}} v
f\,\mathrm{d}\xi,
\end{equation*}
what by Cauchy-Schwarz-Bunyakovskii inequality and estimate
(\ref{5.20}) gives:
\begin{equation}\label{5.21}
\|v\|_{L_2(\Pi^{(j)})}\le\pi^2\|f\|_{L_2(\Pi^{(j)})},\quad
\|\nabla v\|_{L_2(\Pi^{(j)})}\le\pi\|f\|_{L_2(\Pi^{(j)})}.
\end{equation}
Applying estimates (\ref{5.21}) to the solution of problem
(\ref{5.11}) and bearing in mind Lemma~\ref{lm5.2}, we obtain
uniform on $\eta$ estimates:
\begin{equation}\label{5.22}
\begin{aligned}
{}& \|v_2^{odd}\|_{L_2(\Pi^{(j)})}\le C\|\nabla_\xi
X_\eta\|_{L_2(\Pi^{(j)})}\le C|\ln\sin\eta|^{1/2},\\ &
\|\nabla_\xi v_2^{odd}\|_{L_2(\Pi^{(j)})}\le
C|\ln\sin\eta|^{1/2}.
\end{aligned}
\end{equation}
Next, the functions $\xi_2^p v_2^{odd}$ are solutions to problem
(\ref{5.19}), where $\g_\eta$ coincides with axis $O\xi_1$;
right sides are
\begin{equation*}
f=p(p-1)\xi_2^{p-2} v_2^{odd}+2p\xi_2^{p-1}\frac{\p
v_2^{odd}}{\p\xi_2}+\xi_2^p\frac{\p X_\eta}{\p\xi_1},
\end{equation*}
thus, applying estimates (\ref{5.21}) to $\xi_2^p v_2^{odd}$
accounting (\ref{5.22}) and Lemma~\ref{lm5.2}, we have:
\begin{equation}\label{5.23}
\begin{aligned}
\|\xi_2 v_2^{odd}\|_{L_2(\Pi^{(j)})}\le C&\left(\|\nabla
v_2^{odd}\|_{L_2(\Pi^{(j)})}+\|\nabla X_\eta\|_{L_2(\Pi^{(j)})}
\right)\le C|\ln\sin\eta|^{1/2},
\\
\|\xi_2^p v_2^{odd}\|_{L_2(\Pi^{(j)})}\le C&\left(\|\xi_2^{p-2}
v_2^{odd}\|_{L_2(\Pi^{(j)})}+ \|\xi_2^{p-1}\nabla
v_2^{odd}\|_{L_2(\Pi^{(j)})}+\right.\\ &\phantom{\left(
\|\xi_2^{p-2}
v_2^{odd}\|_{L_2(\Pi^{(j)})}\right.}\left.+\|\xi_2^p\nabla_\xi
X_\eta\|_{L_2(\Pi^{(j)})} \right),\quad p\ge2.
\end{aligned}
\end{equation}
Integrating by parts in equalities ($m\ge 1$, $m\in\mathbb{Z}$)
\begin{align*}
 \int\limits_{\Pi^{(j)}}\xi_2^{2m}v_2^{odd}\frac{\p
X_\eta}{\p\xi_1}\, \mathrm{d}\xi&=
\int\limits_{\Pi^{(j)}}\xi_2^{2m}v_2^{odd}\D_\xi v_2^{odd}\,
\mathrm{d}\xi,
\\
 \int\limits_{\Pi^{(j)}}\xi_2^{2(m+1)}\frac{\p
v_2^{odd}}{\p\xi_2}\frac{\p^2 X_\eta}{\p\xi_1\p\xi_2}\,
\mathrm{d}\xi&= \int\limits_{\Pi^{(j)}}\xi_2^{2(m+1)}\frac{\p
v_2^{odd}}{\p\xi_2}\D_\xi \frac{\p v_2^{odd}}{\p\xi_2}\,
\mathrm{d}\xi,
\end{align*}
by analogy with how (\ref{5.18}) was deduced, we derive
inequalities:
\begin{align*}
{}&\|\xi_2^m\nabla_\xi v_2^{odd}\|_{L_2(\Pi^{(j)})}\le C
\left(\|\xi_2^{m-1}v_2^{odd}\|_{L_2(\Pi^{(j)})}+
\Big\|\xi_2^{m+1}\frac{\p
X_\eta}{\p\xi_1}\Big\|_{L_2(\Pi^{(j)})}\right),
\\
{}&\Big\|\xi_2^{m+1}\nabla_\xi \frac{\p
v_2^{odd}}{\p\xi_2}\Big\|_{L_2(\Pi^{(j)})}\le C
\left(\|\xi_2^{m}\nabla_\xi v_2^{odd}\|_{L_2(\Pi^{(j)})}+
\Big\|\xi_2^{m+2}\frac{\p^2
X_\eta}{\p\xi_1\p\xi_2}\Big\|_{L_2(\Pi^{(j)})}\right).
\end{align*}
The inequalities obtained, Lemma~\ref{lm5.2} and estimates
(\ref{5.22}), (\ref{5.23}) by induction prove the lemma. The
proof is complete.

\begin{lemma}\label{lm5.4} For each $R>0$ and integer $m\ge3$
the uniform on $R$  and $\eta$ estimates ($k=0,1,2$)
\begin{align*}
{}& \|X_\eta\|_{L_2(\Pi^{(j)}\cup\{\xi:\xi_2>R\})}\le C
R^{-m}\left(\frac{\pi}{2}-\eta\right)^2 \left( \left|\ln\left(
\frac{\pi}{2}-\eta\right)\right|+1\right),
\\
{}&\|\xi_2^k\nabla_\xi
X_\eta\|_{L_2(\Pi^{(j)}\cup\{\xi:\xi_2>R\})}\le C
R^{-m}\left(\frac{\pi}{2}-\eta\right)^2 \left( \left|\ln\left(
\frac{\pi}{2}-\eta\right)\right|+1\right),
\\
{}&\Big\|\xi_2^{k+1}\nabla_\xi \frac{\p
X_\eta}{\p\xi_2}\Big\|_{L_2(\Pi^{(j)}\cup\{\xi:\xi_2>R\})} \le C
R^{-m}\left(\frac{\pi}{2}-\eta\right)^2 \left( \left|\ln\left(
\frac{\pi}{2}-\eta\right)\right|+1\right),
\\
{}&\|v_2^{odd}\|_{L_2(\Pi^{(j)}\cup\{\xi:\xi_2>R\})} \le C
R^{-m} \left|\ln\sin\eta\right|^{1/2},
\\
{}&\|\nabla_\xi
v_2^{odd}\|_{L_2(\Pi^{(j)}\cup\{\xi:\xi_2>R\})}\le C
R^{-m}\left|\ln\sin\eta\right|^{1/2}.
\end{align*}
take place.
\end{lemma}

\PF{Proof.} By Cauchy-Schwarz-Bunyakovskii inequality each
function $v\in\mathcal{V}_\eta$ for $\xi_2\ge R$ obeys
\begin{equation*}
|v(\xi)|=\left|\int\limits_{\xi_2}^{+\infty}\frac{\p v}{\p t
}(\xi_1,t)\,\mathrm{d}t\right|\le
\frac{1}{\sqrt{2m-3}}\xi_2^{-m+3/2}\left\|\xi_2^{m-1}\frac{\p
v}{\p\xi_2}\right\|_{L_2(\mathbb{R}_+)}.
\end{equation*}
Integrating this inequality over $\Pi^{(j)}\cap\{\xi:
\xi_2>R\}$, we get:
\begin{equation*}
\|v\|_{L_2(\Pi^{(j)}\cap\{\xi: \xi_2>R\})}\le
\frac{R^{-m}}{\sqrt{(2m-3)(2m-4)}} \left\|\xi_2^{m-1}\frac{\p
v}{\p\xi_2}\right\|_{L_2(\Pi^{(j)}\cap\{\xi: \xi_2>R\})}.
\end{equation*}
Taking
\begin{align*}
{}&v=X_\eta,\quad v=\xi_2^k\frac{\p X_\eta}{\p\xi_i},\quad
v=\xi_2^{k+1}\frac{\p^2 X_\eta}{\p\xi_i\p\xi_2},\quad
v=v_2^{odd},\quad v=\frac{\p v_2^{odd}}{\p\xi_i},\quad i=1,2,
\end{align*}
in this inequality we arrive at the statement of the lemma. The
proof is complete.

\begin{lemma}\label{lm5.5} The functions $\l_1(\e)$ and
$\psi_1(x,\e)\in C^\infty(\overline{\Om})$ are uniformly bounded
on $\e$:
\begin{equation*}
|\l_1|\le C,\quad \|\psi_1\|_{C^3(\overline{\Om})}\le C.
\end{equation*}
\end{lemma}

\PF{Proof.} The boundedness of $\l_1(\e)$  follows from
assumption \ref{cnA0} and formula (\ref{5.10}). The smoothness
of the function $\psi_1$ is obvious. By well-known estimates for
solutions of elliptic boundary value problems we have:
\begin{equation*}
\|\psi_1\|_{H^2(\Om)}\le C\left(|\l_1|\|\psi_0\|_{L_2(\Om)}+
\|\psi_0^\nu/\th'_\e\|_{C^2(\p\Om)}\right)\le C,
\end{equation*}
where $C$ is independent on $\e$, what by Theorem on embedding
$H^2(\Om)$ into $C(\Om)$ implies:
$\|\psi_1\|_{C(\overline{\Om})}\le C$ with independent on $\e$
constant $C$. Employing now Schauder estimates (see
\cite[Chapter III, \S 1, formual (1.11)]{Ld}) and taking into
account the boundedness of norm
$\|\th'_\e\|_{C^{3+\mathsf{d}}(\p\Om)}$, we deduce:
\begin{equation*}
\|\psi_1\|_{C^3(\overline{\Om})}\le C\left( |\l_1|
\|\psi_0\|_{C^2(\overline{\Om})}+ \|\psi_1\|_{C(\overline{\Om})}
+\|\psi_0^\nu/\th'_\e\|_{C^{3+d}(\p\Om)}\right)\le C,
\end{equation*}
where $C$ is independent on $\e$. The proof is complete.

We denote:
\begin{align*}
\widehat\l_\e=&\l_0+\e\ln\sin\eta\l_1,
\\
\widehat\psi_\e(x)=&\psi_0(x)+\e\ln\sin\eta\psi_1(x,\e)+
\chi(\tau/c_0)\psi_\e^{bl}(\xi,s,\eta)+R_\e(x),
\\
R_\e(x)=&\e^2\ln^2\sin\eta\chi(\tau/c_0) \psi_0^\nu/\th'_\e,
\end{align*}
where $\l_1$ is from (\ref{5.10}), $\psi_\e^{bl}$ is from
(\ref{5.3}) with $v_1$ and $v_2$ from (\ref{5.7}) and
(\ref{5.25}).

Next statement is an analogue of Lemma~\ref{lm2.5}.

\begin{lemma}\label{lm5.6} The function $\widehat{\psi}_\e\in
C^\infty(\Om\cup\g_\e\cup\G_\e)\cap H^1(\Om)$ converges to
$\psi_0$ in $H^1(\Om)$ and satisfies to the boundary value
problem (\ref{2.79}) with $u_\e=\widehat{\psi}_\e$, $\l=\l_\e$,
$f=f_\e$, where for $f_\e$ the uniform estimate
\begin{equation*}
\|f_\e\|_{L_2(\Om)}\le
C\e^{3/2}\left(|\ln\eta|^{3/2}+1\right)\left(\frac{\pi}{2}-
\eta\right),
\end{equation*}
takes place, constant $C$ is independent on $\e$ and $\eta$. For
the function $R_\e\in C^\infty(\overline{\Om})$ a uniform on
$\e$ and $\eta$ estimate
\begin{equation*}
\|R_\e\|_{C^2(\overline{\Om})}\le C\e^2\ln^2\sin\eta
\end{equation*}
is valid.
\end{lemma}

\PF{Proof.} The smoothness $\widehat{\psi}_\e$ and $R_\e$ are
direct implication of definitions of these functions. Maintained
boundary condition for $\widehat{\psi}_\e$ follows from
(\ref{5.5}), (\ref{5.6}), (\ref{5.8}), (\ref{5.9}),
(\ref{5.11}). The proof of the estimate for  $R_\e$ is based on
Lemma~\ref{lm5.5} and the assumption \ref{cnA0}:
\begin{equation*}
\|R_\e\|_{C^2(\overline{\Om})}\le
C\e^2\ln^2\sin\eta\|\psi_1\|_{C^3(\overline{\Om})}\le
C\e^2\ln^2\sin\eta.
\end{equation*}
Let us prove the estimate for $f_\e$. This function can be
represented as
\begin{align*}
{}&f_\e=-(\D_x+\h\l_\e)\h\psi_\e=-\sum\limits_{i=1}^3
f^{(i)}_\e,
\\
{}&f_\e^{(1)}=\e^2\ln^2\sin\eta\left(\l_1\psi_1+\left(\D+
\h\l_\e\right)\chi(\tau/c_0) \psi_0^\nu/\th'_\e\right),
\\
{}&f_\e^{(2)}=\chi(\tau/c_0)(\D_x+\h\l_\e)\psi_\e^{bl},
\\
{}&f_\e^{(3)}=2\left(\nabla_x\psi_\e^{bl}, \nabla_x \chi(\tau/
c_0)\right)+\psi_\e^{bl}\D_x\chi(\tau/c_0),
\end{align*}
The function $f_\e^{(1)}$ is easily estimated owing to
Lemma~\ref{5.5}:
\begin{equation*}
\|f_\e^{(1)}\|_{L_2(\Om)}\le C\e^2\ln^2\sin\eta,
\end{equation*}
where $C$ is independent on $\e$ and $\eta$. Since
$\nabla_x\chi(\tau/c_0)$ and $\D_x\chi(\tau/c_0)$ are nonzero
only for $c_0/4<\tau<3c_0/4$, taking into account the definition
of the variables $\xi$ and using Lemma~\ref{5.4} with $m=3$ and
$R=c_0 c_1/(4\e)$ (here $c_1$ is from \ref{cnA0}), we arrive at
an estimate:
\begin{equation*}
\|f_\e^{(2)}\|_{L_2(\Om)}\le C \e^{7/2} |\ln\sin\eta|^{1/2},
\end{equation*}
where $C$ is independent on $\eta$ and $\e$. Employing the
harmonicity $X$ and the equation for $v_2^{odd}$, we obtain a
representation for the function  $f_\e^{(3)}$:
\begin{align*}
f_\e^{(3)}&=\e\sum\limits_{k=0}^2\sum\limits_{i=1}^2
\left((\ln\sin\eta)^{\left[\frac{3-k}{2}\right]}
\mathsf{p}_{4k+2i-1}+\xi_2\mathsf{p}_{4k+2i}\right)
\xi_2^{2\left[\frac{k+1}{2}\right]}
\frac{\p^{k+1}X}{\p\xi_i\p\xi_2^k}+
\\
{}&+\e(\e^2\ln^2\sin\eta\mathsf{p}_{13}+
\e\ln\sin\eta\mathsf{p}_{14}+\mathsf{p}_{15})X +
\e\sum\limits_{k=0}^1\sum\limits_{i=1}^2\xi_2^k
\mathsf{p}_{2k+i+15}\frac{\p^{k+1}v_2^{odd}}{\p\xi_i\p\xi_2^k}+
\\
{}&+\e^2\left(\e\ln\sin\eta\mathsf{p}_{20}+\mathsf{p}_{21}\right)
v_2^{odd},
\end{align*}
where $\mathsf{p}_i=\mathsf{p}_i(\xi_2;s,\e)$ are polynomials on
$\xi_2$ whose coefficients depending on $s$ and $\e$ are
estimated uniformly on $s$ and $\e$ by Lemma~\ref{lm5.5},
$[\bullet]$ indicates the integral part of number. Bearing in
mind these estimates and using
Lemmas~\ref{lm5.2}~and~\ref{lm5.3}, we conclude that
\begin{equation*}
\|f_\e^{(2)}\|_{L_2(\Om)}\le C
\e^{3/2}\left(|\ln\eta|^{3/2}+1\right)
\left(\frac{\pi}{2}-\eta\right),
\end{equation*}
where $C$ is independent on $\e$ and $\eta$. Here we have also
used obvious relationships: $\ln\sin\eta=O(\eta)$, $\eta\to0$;
$\ln\sin\eta=O((\pi/2-\eta)^2)$, $\eta\to\pi/2$. The proof is
complete.

The justification of the asymptotics constructed is carried out
by analogy with one from the second section.

The formal construction of asymptotics in the case of multiply
limiting eigenvalue does not differ in general from the case of
simple limiting eigenvalue. The only difference is that we
simultaneously construct asymptotics of all eigenvalues
converging  to multiply limiting eigenvalue; in whole the formal
construction reproduces the arguments given above word for word.
The justification of asymptotics  in the case of multiply
limiting eigenvalue is similar to the second section, too. The
proof of Lemma~\ref{lm5.1} is complete.

\sect{Estimates for perturbed eigenvalues}

In this section we will prove Theorems~\ref{th1.2}--\ref{th1.3}.
Their proof will be based on the following auxiliary statement.

\begin{lemma}\label{lm4.1}
Suppose sets $\g_1(\e), \g_2(\e)\subseteq\p\Om$ are such that
$\g_1(\e)\subseteq\g_2(\e)$, $\l_{\e,1}^k$, $\l_{\e,2}^k$ are
eigenvalues of the perturbed problems with $\g_\e=\g_1(\e)$ and
$\g_\e=\g_2(\e)$, respectively, taken in ascending order
counting multiplicity. Then for each  $k$ the inequalities
\begin{equation*}
\l_{\e,1}^k\le\l_{\e,2}^k
\end{equation*}
hold true.
\end{lemma}
Lemma~\ref{lm4.1} is a standard statement about variational
properties of eigenvalues for elliptic boundary value problems,
the proof is based on the minimax property of eigenvalues and an
obvious inclusions of functional spaces:
$H^1(\Om,\g_2(\e))\subseteq H^1(\Om,\g_1(\e))$, where
$H^1(\Om,\g_i(\e))$, $i=1,2$  is a set of function belonging to
$H^1(\Om)$ and vanishing on $\g_i(\e)$.

\smallskip
\PF{\indent Proof of Theorem~\ref{th1.2}}. In accordance with
Lagrange theorem, the functions $a^j$ and $b^j$ introduced in
the second section, can be represented by $a_j$ and $b_j$ as
follows:
\begin{equation*}
a^j=\th'_\e(M_{j,\e}^{(3)})a_j,\qquad
b^j=\th'_\e(M_{j,\e}^{(4)})b_j,
\end{equation*}
where $M_{j,\e}^{(3)}\in(s^\e_j-\e a_j,s^\e_j)$,
$M_{j,\e}^{(4)}\in(s^\e_j,s^\e_j+\e b_j)$ are midpoints. By
representations obtained and the
assumptions~\ref{cnA0}~and~\ref{cnA6} we deduce that
\begin{equation*}
a^j\ge c_1 a_j,\qquad b^j\ge c_1 b_j.
\end{equation*}
These estimates, the assumptions~\ref{cnA0}~and~\ref{cnA6} and
the disjointness of sets $\g_{\e,j}$ yield:
\begin{equation*}
2\eta(\e)\le \min\limits_j a^j(\e)+\min\limits_i b^i(\e)\le \pi,
\end{equation*}
i.e., the function $\eta$ is bounded above by the number
$\pi/2$. Moreover, last inequalities imply the existence of
functions $a_*(\e)$ and $b_*(\e)$ such that $a_*+b_*=2\eta$, and
for a set
\begin{equation*}
\g_{\e,*}=\{x: x\in\p\Om, -\e
a_*(\e)<\th_\e(s)-\th_\e(s_j^\e)<\e b_*(\e), j=0,\ldots,N-1\}
\end{equation*}
the inclusion $\g_{\e,*}\subseteq\g_\e$ holds. By $\l_{\e,*}^k$
we indicate the eigenvalues of the perturbed problems with
$\g_\e=\g_{\e,*}$, taken  in ascending order counting
multiplicity. The set $\g_{\e,*}$ obeys the hypothesis of
Lemma~\ref{lm5.1} with the function $\eta$ from
assumption~\ref{cnA6}. According with Lemma~\ref{lm5.1}, the
eigenvalues $\l^k_{\e,*}$ converge to eigenvalues $\l_0^k$ of
problem (\ref{1.4}) and satisfy the asymptotics from this lemma.
Applying Lemma~\ref{lm4.1} twice: with $\g_1(\e)=\g_{\e,*}$,
$\g_2(\e)=\g_\e$ and $\g_1(\e)=\g_\e$, $\g_2(\e)=\p\Om$, we
establish double-sided estimates:
\begin{equation*}
\l_{\e,*}^k\le\l_\e^k\le\l_0^k.
\end{equation*}
Now we replace $\l_{\e,*}^k$ by their asymptotics from
Lemma~\ref{lm4.1}, what implies, first, convergence of $\l_\e^k$
to $\l_0^k$, and, second, needed double-sided of differences
$(\l_\e^k-\l_0^k)$. The proof of Theorem~\ref{th1.2} is
complete.

\smallskip
\PF{\indent Proof of Theorem~\ref{th1.4}}. We deduce from the
first estimate of Lemma~\ref{lm2.6} and the assumption
\ref{cnA5} that
\begin{equation*}
2\eta_0\eta\le a^j+b^j\le 2\eta.
\end{equation*}
These inequalities imply that, first, the function $\eta_0$ is
bounded above by one, and, second, there exist nonnegative
bounded functions $a^j_*(\e)$, $b^j_*(\e)$, $a^{j,*}(\e)$,
$b^{j,*}(\e)$, such that $a^j_*+b^j_*=2\eta_0\eta$,
$a^{j,*}+b^{j,*}=2\eta$, and sets
\begin{align*}
{}&\g_{\e,*}=\{x: x\in\p\Om, -\e
a^j_*(\e)<\th_\e(s)-\th_\e(s_j^\e)<\e b^j_*(\e),
j=0,\ldots,N-1\},
\\
{}&\g^*_\e=\{x: x\in\p\Om, -\e
a^{j,*}(\e)<\th_\e(s)-\th_\e(s_j^\e)<\e b^{j,*}(\e),
j=0,\ldots,N-1\}
\end{align*}
meet inclusions
\begin{equation}\label{4.9}
\g_{\e,*}\subseteq\g_\e\subseteq\g^*_\e.
\end{equation}
By $\l_{\e,*}^k$ and $\l_\e^{k,*}$ we denote the eigenvalues of
the perturbed problem with $\g_\e=\g_{\e,*}$ and
$\g_\e=\g^*_\e$, taken in ascending order counting multiplicity.
The sets $\g_{\e,*}$ and $\g^*_\e$ obey the assumptions
\ref{cnA0} and \ref{cnA2}: role of the function $\eta$ from
\ref{cnA2} for them is played by the functions $\eta_0\eta$ and
$\eta$ from the assumption~\ref{cnA5}, respectively; the
equality (\ref{1.5}) for these functions holds with the same
$A>0$. The quantities $\d^j(\e)$ for the sets $\g_{\e,*}$ and
$\g^*_\e$ are zero, therefore, by Lemma~\ref{lm2.8} the
eigenvalues $\l_{\e,*}^k$ and $\l_\e^{k,*}$ converge to the
eigenvalues of the problem (\ref{1.6}) and asymptotics
\begin{equation}\label{4.2}
\begin{aligned}
{}&\l_{\e,*}^k=\L_0^k(\t\mu,\e)+\e\int\limits_{\p\Om}
\left(\Psi_0^k(x,\mu,\e)\right)^2\ln
\mathsf{f}_\e(\th_\e(s))\th'_\e(s)\,\mathrm{d}s+O(\e^{3/2}),
\\
{}&\l^{k,*}_\e=\L_0^k(\mu,\e)+\e\int\limits_{\p\Om}
\left(\Psi_0^k(x,\t\mu,\e)\right)^2\ln
\mathsf{f}_\e(\th_\e(s))\th'_\e(s)\,\mathrm{d}s+O(\e^{3/2}),
\end{aligned}
\end{equation}
hold, where $\mu=\mu(\e)=-\left(\e\ln\eta(\e)\right)^{-1}-A$,
\begin{equation*}
\t\mu=\t\mu(\e)=-\left(\e\ln\eta_0(\e)\eta(\e)\right)^{-1}-A=
\mu(\e)+\frac{(A^2-\mu(\e)^2)\e\ln\eta_0(\e)}{1+(A+\mu(\e))\e\ln\eta_0(\e)}.
\end{equation*}
Lemma~\ref{lm2.1} yields an estimate
$\|\Psi_0^k\|_{L_2(\p\Om)}\le C$ with constant $C$ independent
on $\e$ and $\mu$. This estimate, (\ref{2.104}) and the
assumption~\ref{cnA0} allows to estimate the integrals in
(\ref{4.2}):
\begin{equation}\label{4.16}
-C\le \int\limits_{\p\Om}
(\Psi_0^k)^2\ln\mathsf{f}_\e(\th_\e)\th'_\e\,\mathrm{d}s\le 0,
\end{equation}
where $C>0$ is independent on $\e$ and $\mu$. Lemma~\ref{lm4.1}
due to inclusions (\ref{4.9}) maintains the validity of
estimates
\begin{equation*}
\l_{\e,*}^k\le \l_\e^k\le \l^{k,*}_\e,
\end{equation*}
those, the asymptotics (\ref{4.2}), (\ref{3.17}) and the
inequalities (\ref{4.16}) imply the convergence
$\l_\e^k\to\l_0^k$ and needed double-sided estimates for the
quantities $(\l_\e^k-\l_0^k)$. The proof of Theorem~\ref{th1.4}
is complete.

\smallskip
\PF{\indent Proof of Theorem~\ref{th1.3}}. The main idea of
proof is same with one in Theorems~\ref{th1.2},~\ref{th1.4}.
From the first estimate of Lemma~\ref{lm2.6} and the
assumption~\ref{cnA4} it follows the existence of nonnegative
functions $a^{j,*}(\e)$ and $b^{j,*}(\e)$, such that
$a^{j,*}+b^{j,*}=2\eta$, and a subset of the boundary $\p\Om$
\begin{equation*}
\g^*_\e=\{x: x\in\p\Om, -\e a^{j,*}(\e)<\th_\e(s)-\e\pi j<\e
b^{j,*}(\e), j=0,\ldots,N-1\}
\end{equation*}
satisfies $\g_\e\subseteq\g^*_\e$. Let $\l_\e^{k,*}$ be
eigenvalues of the perturbed problem with $\g_\e=\g^*_\e$. The
set $\g^*_\e$ meets the hypothesis of Theorem~\ref{th1.6} with
the function $\eta$ from \ref{cnA4}. Thus,
\begin{equation}\label{4.5}
\l_\e^{k,*}=\l_0^k+\mu\int\limits_{\p\Om}\left(\psi_0^k\right)^2
\th' \mathrm{d}s+o(\mu),
\end{equation}
where, we recall, $\l_0^k$ are eigenvalues of the problem
(\ref{1.6}) for $A=0$. The inclusions
$\emptyset\subseteq\g(\e)\subseteq\g^*(\e)$ by Lemma~\ref{lm4.1}
imply the inequalities:
\begin{equation*}
\l_0^k\le \l_\e^k\le \l_\e^{k,*},
\end{equation*}
from those and the asymptotics (\ref{4.5}) it arises the
statement of the theorem. The proof of Theorem~\ref{th1.3} is
complete.


\sect{Appendix}

In this section we will prove the formulae (\ref{3.17}) for the
eigenvalues of the problem (\ref{2.6}), (\ref{2.25}). Let
\begin{equation*}
\h\L_0^k=\l_0^k+\mu\l_1^k,\quad
\h\Psi_0^k=\psi_0^k+\mu\psi_1^k+\boldsymbol{\psi}^k, \quad
\l_1^k=\int\limits_{\p\Om}(\psi_0^k)^2\th'_0\,\mathrm{d}s.
\end{equation*}
The functions $\psi_0^k$ associated with multiply eigenvalue are
additionally chosen to be orthogonal in $L_2(\p\Om)$ weighted by
$\th'_0$. The functions $\psi_1^k$ and $\boldsymbol{\psi}^k$ are
defined as solutions of the problems:
\begin{gather*}
(\D+\l_0^k)\psi_1^k=-\l_1^k\psi_0^k,\quad x\in\Om,
\qquad\left(\frac{\p}{\p\nu}+A\th'_0\right)\psi_1^k=-\th'_0\psi_0^k,\quad
x\in\p\Om,
\\
(\D-1)\boldsymbol{\psi}^k=-\mu^2\l_1^k\psi_1^k,\quad x\in\Om,
\\
\left(\frac{\p}{\p\nu}+(A+\mu)\th'_\e\right)\boldsymbol{\psi}^k
=-(\th'_\e-\th'_0)((A+\mu)\psi_0^k+A\mu\psi_1^k)
-\mu^2\th'_\e\psi_1^k,\quad x\in\p\Om.
\end{gather*}
The problem for $\psi_1^k$ is solvable, the formula for $\l_1^k$
and the assumption for $\psi_0^k$ mentioned above are exactly
the solvability condition. The functions $\psi_1^k$ are selected
to be orthogonal to all eigenfunctions associated with $\l_0^k$.
Clear, the problem for $\boldsymbol{\psi}^k$ is uniquely
solvable. General properties of solutions of elliptic boundary
value problems yield that $\psi_1^k$ and $\boldsymbol{\psi}^k$
are infinitely differentiable on $x$ functions, for those the
estimates
\begin{equation*}
\|\psi_1^k\|_{H^1(\Om)}\le C,\qquad
\|\boldsymbol{\psi}^k\|_{H^1(\Om)}\le C(\mu^2+(A+\mu)\si),
\end{equation*}
hold, where the constants $C$ are independent on $\e$ and $\mu$.
Employing these estimates and the definition of $\l_1^k$,
$\psi_1^k$ and $\boldsymbol{\psi}^k$ one can check that the
functions $\h\L_0^k$ and $\h\Psi_0^k$ converge to  $\l_0^k$ and
$\psi_0^k$ and  satisfy a problem
\begin{gather*}
(\D+\h\L_0^k)\h\Psi_0^k=\widehat{F}_k,\quad x\in\Om, \qquad
\left(\frac{\p}{\p\nu}+(A+\mu)\th'_\e\right)\h\Psi_0^k=0, \quad
x\in\p\Om,
\\
\|\widehat{F}_k\|_{L_2(\Om)}\le C(\mu^2+(A+\mu)\si),
\end{gather*}
where the constant $C$ is independent on $\e$ and  $\mu$. Let
$\l_0=\l_0^q=\ldots=\l_0^{q+p-1}$ be a $p$-multiply eigenvalue.
By the problem for $\h\Psi_0^k$ and the estimate for the right
side $\widehat{F}_k$ employing results \cite{Kt}, it is easy to
show that for  $k=q,\ldots,q+p-1$ the representation and uniform
on $\e$ and $\mu$ estimate
\begin{equation*}
\h\Psi_0^k=\sum\limits_{i=q}^{q+p-1}\frac{\Psi_0^i}
{\L_0^i-\h\L_0^k}\int\limits_{\Om}\Psi_0^i\h F_k\,\mathrm{d}x+
\h u_k,\quad \|\h u^k\|_{H^1(\Om)}\le C(\mu^2+(A+\mu)\si)
\end{equation*}
take place. By analogy with the justification from the second
section on the base of last assertions we get the estimates
\begin{equation*}
|\L_0^k-\h\L_0^k|\le C(\mu^2+(A+\mu)\si),
\end{equation*}
those prove the equalities (\ref{3.17}).

\medskip

In conclusion we thank R.~R.~Gadyl'shin for permanent attention
to the paper, discussion of the results and useful remarks.

\renewcommand{\refname}





\end{document}